\documentclass[preprint2]{pp7}

\usepackage[bookmarks = true, bookmarksnumbered = true, pdfpagemode =None, pdfstartview = FitH, pdfpagelayout = SinglePage, colorlinks = true, urlcolor = blue, citecolor = blue]{hyperref}
\usepackage{textcomp}
\usepackage[normalem]{ulem}
\usepackage[switch]{lineno}

\newcommand{\rev}[1]{#1} 
\newcommand{\rrm}[1]{} 

\def\aap{A\&A}

\input{pp7.h}

\begin{document}

\title{\textbf{\LARGE Giant Planets from the Inside-Out}}

\author {\textbf{\large Tristan Guillot}}
\affil{\small\it Universit\'e C\^ote d'Azur}
\author {\textbf{\large Leigh N. Fletcher}}
\affil{\small\it University of Leicester}
\author {\textbf{\large Ravit Helled}}
\affil{\small\it University of Zurich}
\author {\textbf{\large Masahiro Ikoma}}
\affil{\small\it National Astronomical Observatory of Japan}
\author {\textbf{\large Michael R. Line}}
\affil{\small\it Arizona State University}
\author {\textbf{\large Vivien Parmentier}}
\affil{\small\it University of Oxford}

\begin{abstract}
\baselineskip = 11pt
\leftskip = 1.5cm 
\rightskip = 1.5cm
\parindent=1pc
{\small 
Giant planets acquire gas, ices and rocks during the early formation stages of planetary systems and thus inform us on the formation process itself. Proceeding from inside out, examining the connections between the deep interiors and the observable atmospheres, linking detailed measurements on giant planets in the solar system to the wealth of data on brown dwarfs and giant exoplanets, we aim to provide global constraints on interiors structure and composition for models of the formation of these planets. 

New developments after the Juno and Cassini missions point to both Jupiter and Saturn having strong compositional gradients and stable regions from the atmosphere to the deep interior. This is also the case of Uranus and Neptune, based on available, limited data on these planets. Giant exoplanets and brown dwarfs provide us with new opportunities to link atmospheric abundances to bulk, interior abundances \rev{and to link these abundances and isotopic ratios to formation scenarios. Analysing the wealth of data becoming available} will require new models accounting for the complexity of the planetary interiors and atmospheres. 
 \\~\\
\noindent {\it Revised chapter submitted to Protostars and Planets VII, Editors: Shu-ichiro Inutsuka, Yuri Aikawa, Takayuki Muto, Kengo Tomida, and Motohide Tamura\\ 1 April 2022}
~\\~\\}
\end{abstract}


\section{INTRODUCTION}

According to the classical picture, giant planets form from planetary cores made of ices and rocks that grew beyond a critical mass of about 10 times the mass of the Earth, at which point they began accreting hydrogen and helium and increased their mass rapidly \citep{Perri+1974,Mizuno1980,Bodenheimer+86}. \rev{Their large mass implies a large internal energy reservoir that is} released slowly, implying that they are hot, fluid, and mostly convective \citep{Hubbard1968,Stevenson+Salpeter1977a}.  In parallel, observations \rev{of} tropospheric temperatures in Jupiter and Saturn \rev{show} very limited equator-to-pole gradients, despite having a pattern of insolation that is strongly variable from equator to pole, pointing to efficient convective mixing redistibuting energy \citep{Ingersoll+Porco1978}. This led to today's models of giant planets (extended to giant exoplanets) made of a central dense core surrounded by a hydrogen-helium envelope of nearly uniform composition, upon which planetary rotation imposes a latitudinally-banded structure in the atmosphere.

Recent observations, in particular by \rev{Juno \citep{Bolton+2017SSRv} and Cassini \citep{Spilker2019}}, show that Jupiter and Saturn depart -- potentially strongly -- from that ideal picture. Both their interiors show signs of strong variations in composition and potentially extended stably stratified regions. The temperatures, composition, and aerosol properties of their atmospheres are non-uniform to great depths (including the deep troposphere) and variable over a variety of timescales linked to seasonal climates and local meteorology.  In parallel, the analysis of observations of exoplanets and brown dwarfs also shows signs of this complexity, linked to the presence of clouds, strong horizontal variations in temperature and probably chemical composition and time variability. Linking Solar-System giant planets to exoplanets and brown dwarfs is crucial to better characterize these objects and understand planet formation, and planetary environments, in general. 
\rev{The previous PPVI review chapters on planetary interior structures and giant planet formation \citep{2014prpl.conf..763B,2014prpl.conf..643H} focused on evidence available on solar system planets and moons, and on exoplanets based on their mass-radius properties. We now have refined measurements of Jupiter and Saturn's properties, and are beginning to truly characterise exoplanetary atmospheres. The present chapter is based on this new evidence.}

We first present the current understanding of the interiors and atmospheres of Jupiter, Saturn, Uranus and Neptune in \S~2.  Next, we review constraints on brown dwarfs and giant exoplanets in \S~3. In \S~4 we apply these findings to the history of the formation planets. In each subsection, we organise our discussion `\textit{from the inside out}', starting with the deep interior and moving to the regions more readily accessible to remote sensing - the atmospheres.  We provide our conclusions in \S~5.

\section{THE SOLAR SYSTEM GIANT PLANETS}

In this section, we discuss our present understanding of the interiors and atmospheres of the giant planets, following the Galileo, Juno, and Cassini orbital remote sensing of Jupiter and Saturn, as well as Voyager-2 observations (and three decades of ground- and space-based astronomy) of Uranus and Neptune. We focus on questions that may have relevance to the characterisation of exoplanets and brown dwarfs in \S~3. In particular, how are heat and elements transported in the atmosphere and interior of giant planets? What can be inferred from the observation of their atmosphere? What are the consequences for their evolution?  We structure this section by starting in the deep interior, moving upwards into the banded troposphere and the cloud-forming `weather layer,' and finally into the stably-stratified middle atmosphere above the clouds.

\subsection{Interior Structure}
\label{sec:GP_interior}

\subsubsection{The core-envelope structure}
A lack of constraints and Ockham’s razor have meant that giant planet interiors have long been thought to be relatively simple: A well-defined central core, leftover from the planet’s formation, and a convective mostly homogeneous hydrogen-helium envelope (except for a transition in helium content due to a phase separation in Jupiter and Saturn) on top \citep[see][and references therein]{Guillot2005,Fortney+Nettelmann2010,Helled2014}. Yet, the question of their inherent complexity, including the presence of possible important compositional gradients, was raised already in the 1980's \citep{Stevenson1985}. New data show that this complexity must be accounted for.   

Spacecraft measurements of the planets' gravity fields have thus far been the main constraints used for interior models. These measurements have been improved by more than two orders of magnitude thanks to the close-in, polar orbits of Juno at Jupiter \citep{Iess+2018, Durante+2020} and of the Cassini Grand Finale at Saturn \citep{Iess+2019}. These measurements led to models showing that Jupiter's interior is clearly inhomogeneous, with a deep metallic envelope that must be enriched in heavy elements (all elements heavier than hydrogen and helium) compared to its upper molecular envelope, possibly requiring the presence of a fuzzy core that instead of being well-defined extends in the overlaying envelope \citep{Wahl+2017,Debras+Chabrier2019,Ni2019, Nettelmann+2021, Miguel+2022}. In Saturn, a milestone has been reached thanks to the detection of the normal modes of oscillation of the planet \citep[e.g.,][]{Hedman+Nicholson2013, Hedman+2019}. Seismology has thus provided evidence of the presence of a deep stable region \citep{Fuller2014}, of substantial helium differentiation leading to the formation of a helium-rich core \citep{Mankovich+Fortney2020}, and of a gradient in the distribution of heavy elements similar to that inferred in Jupiter and extending to about 60\% of Saturn's radius \citep{Mankovich+Fuller2021}. In comparison, Uranus and Neptune, which have been briefly visited only by Voyager~2 in 1986 and 1989, respectively, remain much less well characterised \citep[][and references therein]{Helled+2020}. 

\begin{figure}[tb!]
    \centering
    \includegraphics[width=\linewidth]{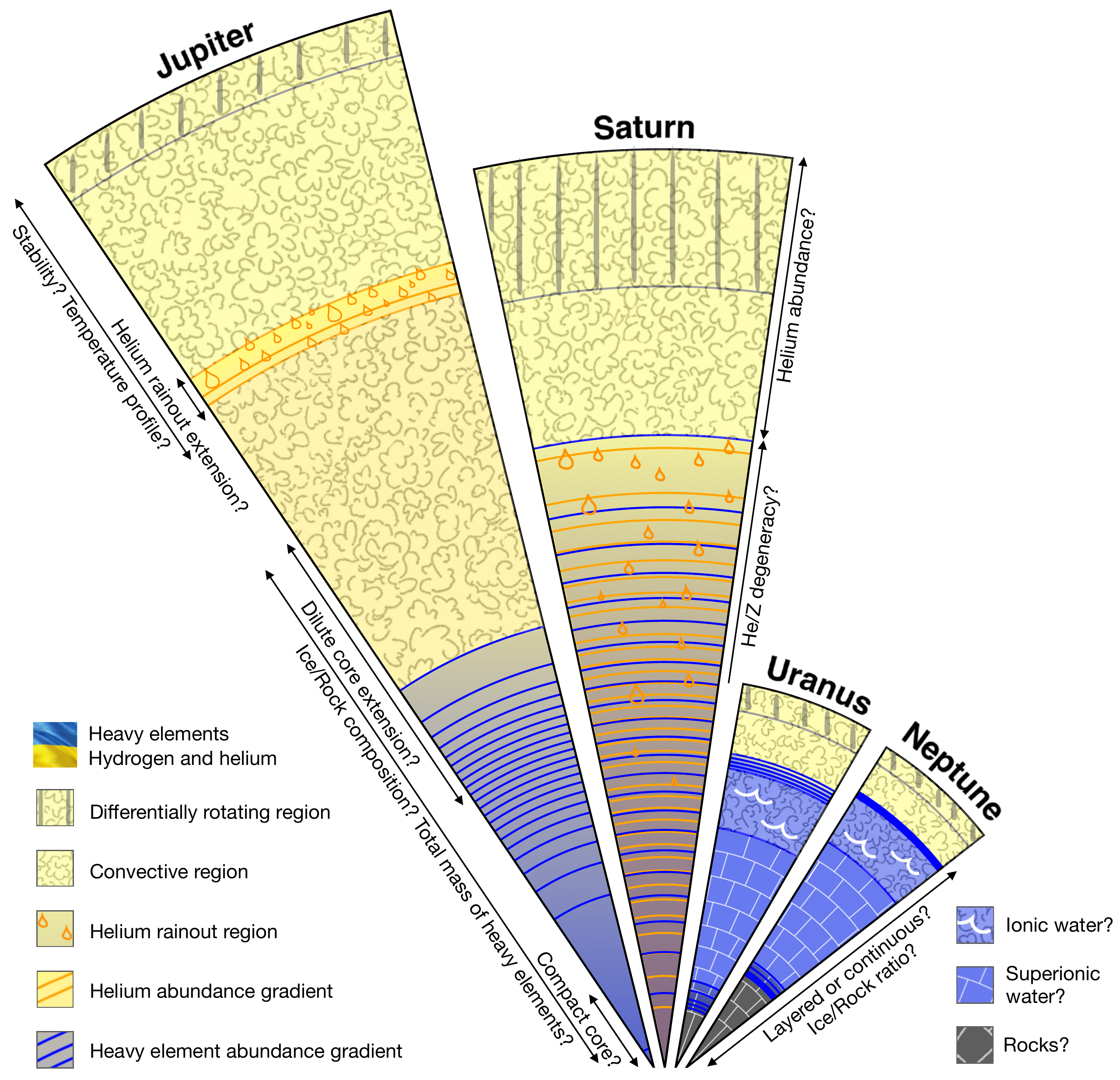}
    \caption{\rev{Slices of the internal structures of Jupiter, Saturn, Uranus and Neptune highlighting compositional gradients in helium (orange) and heavy elements (blue). Jupiter and Saturn are characterized by a phase separation of helium in metallic hydrogen starting near Mbar pressures. Juno and Cassini measurements also indicate the presence of a dilute heavy element core in these planets, implying that heavy elements are partially mixed with hydrogen and helium rather forming separate pure layers. In these planets, each layer correspond to a 2\% increase of the helium/hydrogen ratio (orange) or of the heavy element mass fraction $Z$ (blue). For Uranus and Neptune, a structure dominated by well-separated layers, including a solid superionic water layer is presented but solutions with compositions that evolve more continuously are also highly plausible (see text). }}
    \label{fig:interiorslices}
\end{figure}

Fig.~\ref{fig:interiorslices} sketches the interior structures of the four giant planets, as envisioned from the latest available publications. 
Jupiter is thus made of a central dense compact core of small mass (less than about 6$\,\rm M_\oplus$, possibly zero), probably a dilute core extending to possibly a high fraction ($\sim 50\%$) of the planet's radius, an inner envelope of high helium abundance and an outer envelope with a low helium abundance \rev{\citep{Miguel+2022}}. A region of variable helium abundance in which helium droplets should form is sandwiched between the inner and outer envelope. It extents to up to 8\% of the planet in radius \citep{Mankovich+Fortney2020} but could be smaller, depending on the temperature profile in the region. The dilute core is believed to have a variable composition, from up to perhaps $50\%$ of heavy elements in the deep regions down to $\approx 8\%$ when merging with the inner envelope (and the complement in hydrogen and helium). Although Jupiter has a high intrinsic luminosity \citep{Li_Liming+2018} and should be largely convective \citep[e.g.,][]{Guillot+2004}, these zones of variable compositions should be either stable to convection or characterised by double-diffusive convection \citep[e.g.][]{Rosenblum+2011,Leconte+Chabrier2012,Wood+2013,Nettelmann+2015}. The nature of heavy elements in terms of the fraction of ices, rocks and iron that they contain is unknown.   

Saturn has a structure that is qualitatively similar to Jupiter but with important differences. The work of \cite{Mankovich+Fuller2021} which integrates constraints from both gravimetry and seismology assumes no central compact core. The same work provides evidence for the existence of a dilute core with a gradient in both heavy elements and helium abundance extending to 60\% of the planet's radius. The mass of heavy elements involved in this dilute core is constrained to values in the range $15.5-20.8\rm\,M_\oplus$. Helium demixing is much more pronounced than in Jupiter, possibly leading to an inner region with less than 5\% hydrogen \citep{Mankovich+Fortney2020}. The upper envelope is homogeneous with a mass fraction of heavy elements $Z=0.028-0.084$.

Uranus and Neptune are of comparatively much smaller mass, with an interior dominated by heavy elements and an even larger variety of possible structures. Uncertainties are large, due both to the loose constraints on the planets' gravity fields, but also to uncertain rotation rates \citep{Helled+2010, Helled+2011}. This allows a full range of solutions for the core ($11-13.3\rm\,M_\oplus$ for Uranus, $13-15.5\rm\,M_\oplus$ for Neptune). Fully differentiated solutions include a central rock core, a shell of solid \rrm{(but low viscosity)} super-ionic water and an overlaying layer of fluid ionic water \citep{Redmer+2011, Nettelmann+2013}. \rev{Recent studies have shown super-ionic water to possess a significant shear modulus which should allow only slow convective motions \citep{Millot+2019} and imply that Uranus and Neptune would progressively solidify with time \citep{Stixrude+2021}. However } solutions are also possible \rev{in which} with rocks, ices and even a small fraction ($\lesssim 10\%$) of gas \rev{are mixed, potentially modifying these conclusions. In all cases, }their overlaying hydrogen-helium-dominated envelopes are relatively small, being only $1.25-3.5\rm\,M_\oplus$ for Uranus and $1.6-4.15\rm\,M_\oplus$ for Neptune \citep{Helled+2011, Nettelmann+2013,Helled+2020}. 

Thus, strong compositional gradients occur in all four giant planets. The gradient in helium composition in Jupiter and Saturn is linked to a phase separation of helium in metallic hydrogen (see \S~\ref{sec:EOS} hereafter). The increase in molecular weight thus formed is strongly stabilising to convection, but it is not yet clear whether this region is stable, diffusive-convective or fully convective (see \S~\ref{sec:non-adiabatic}). Jupiter and Saturn's dilute core regions seem at this point to be unrelated to any phase transition, and would rather be a leftover from formation processes (see \S~\ref{sec:formation}). These regions should therefore either be diffusive-convective or convectively stable. The existence of an extended region which is stable to convection is demonstrated in Saturn, thanks to seismology and the discovery of oscillation modes mixing fundamental waves (f-modes) and gravity waves (g-modes) \citep{Fuller2014}, as the latter can only propagate in stable regions. In Uranus and Neptune, the transition from a hydrogen and helium-dominated envelope to a much more dense interior should lead either to diffusive convection or to a stable region in which heat is mainly transported by conduction.

\subsubsection{Equations of State and Phase Diagrams}\label{sec:EOS}

\begin{figure*}[htb!]
    \centering
    \includegraphics[width=\linewidth]{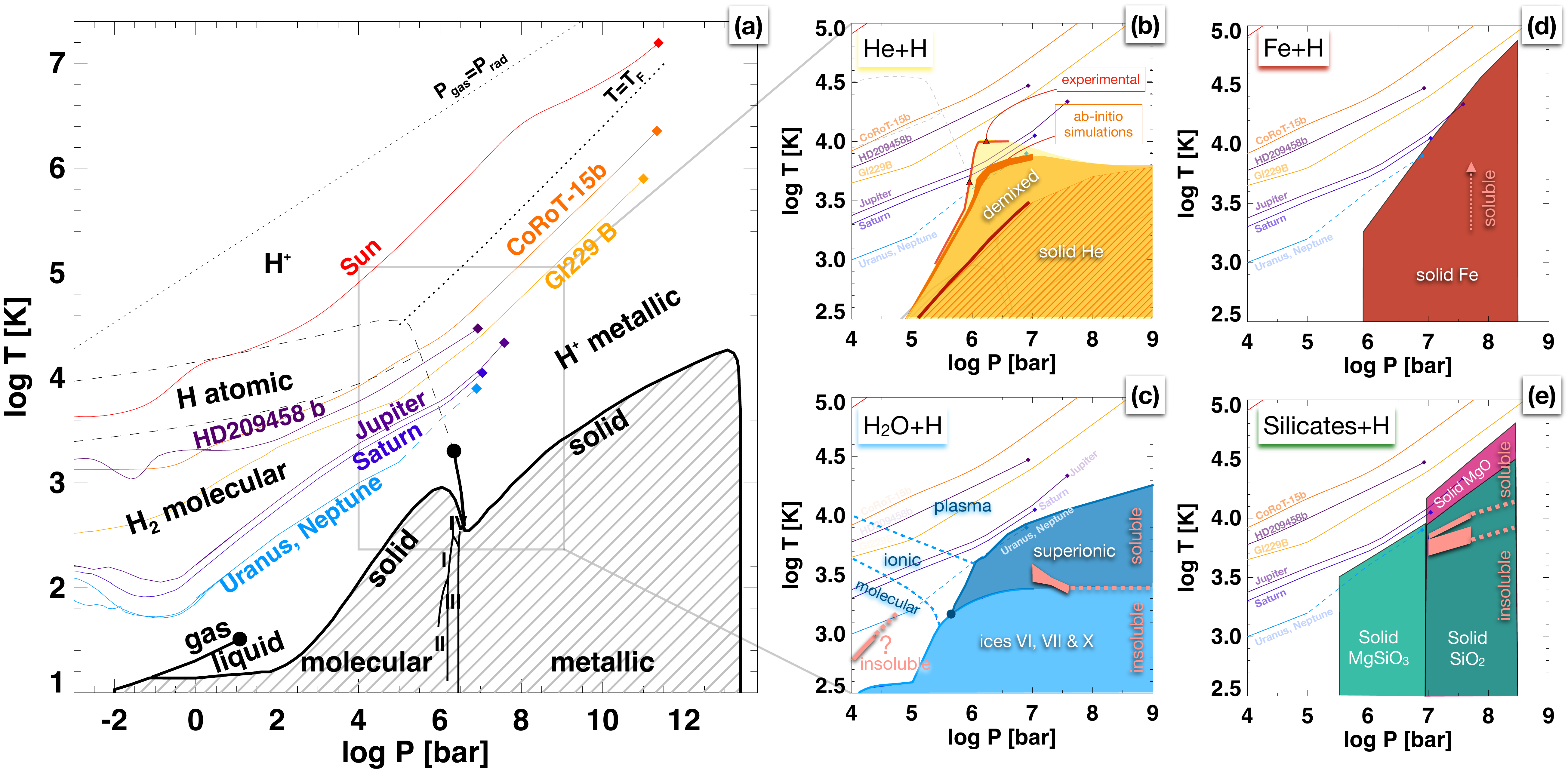}
    \caption{Phase diagrams of key elements with pressure-temperature profiles of relevant astrophysical objects (the Sun, brown dwarfs CoRoT-15b and Gl229~B, giant exoplanet HD~209458~b, and Jupiter, Saturn, Uranus and Neptune). Pressures are in bars ($\rm 1\,bar=10^5\,Pa=10^6\,dyn/cm^{2}$). (a): Phases of hydrogen, with solid phases shown as a hashed region, critical points as black circles, thick lines indicating a first order (discontinuous) phase transition, and dashed lines a continuous transition \citep[from][]{Guillot+Gautier2015}. (b): Location of the hydrogen-helium phase separation leading to helium rain-out in Jupiter and Saturn (yellow). The critical demixing temperatures obtained for an 11 mol\% He fraction from ab-initio simulations \citep{Schottler+Redmer2018} are shown in orange. Those obtained from high-pressure experiments are shown as two triangles connected by a red curve \citep{Brygoo+2021}. The demixing region is extended to high pressure using the fully pressure-ionised limit \citep{Stevenson1975}.   (c): Phase diagram of H$_2$O, including the fluid molecular, ionic and plasma phases, the solid ice phases and the superionic phase, as labelled \citep{Redmer+2011,Salzmann2019,Mazevet+2019a}. The H$_2$O transitions are mostly relevant for Uranus and Neptune and ice giants. The solubility of H$_2$O in metallic hydrogen at $P\sim 10-100$\,Mbar is from \cite{Wilson+Militzer2012}. A region of water insolubility in molecular hydrogen is speculated at $P\sim 10-100\,$kbar \citep{Bali+2013, Bailey+Stevenson2021} but controversial \citep{Soubiran2015}. (d): Region of iron solidification \citep[from][]{Mazevet+2019}, relevant for the central regions of Jupiter and possibly Saturn. In planetary interiors, iron is believed to be entirely soluble in metallic hydrogen \citep{Wahl+2013} (e): Regions of silicate solidification \citep[from][]{Mazevet+2019}, relevant in the interiors of Uranus, Neptune, Saturn and Jupiter. Note that MgSiO$_3$ decomposes at high pressures into MgO (which should be solid in Jupiter and Saturn) and SiO$_2$ (which may still be liquid in the deep interior of Jupiter and Saturn). These two phases have slightly different solubilities but should be both mostly soluble in metallic hydrogen in planetary interiors \citep{Gonzalez-Cataldo+2014}.}
    \label{fig:phases}
\end{figure*}

For a large part, our knowledge of the interiors of giant planets is inferred from hydrostatic models and equations of state governing the behaviour of matter at pressures reaching $\sim 70\,$Mbar ($\sim 7$\,TPa) in Jupiter and even much more in selected brown dwarfs (see Fig.~\ref{fig:phases}a). In this regime, hydrogen transitions from a weakly conducting molecular fluid to a metallic fluid by pressure ionization. In giant planets and brown dwarfs, this transition occurs smoothly around pressures $\sim 1\,$Mbar \citep{Sano+2011, Loubeyre+2012}. At temperatures below 4000\,K and pressures between 1.5 and 2.5\,Mbar, this transition occurs discontinuously \citep{McMahon+2012, Knudson+2015}. This regime, as that of the solidification of hydrogen, occurs at too low temperatures to be of relevance for giant planets and brown dwarfs \citep[see][]{Guillot2005}. 

The calculation of an equation of state \rev{is difficult, in particular where} electrons are partially degenerate and Coulombian interactions are important. The pioneering work of \cite{SCvH95} included the possibility of a "Plasma Phase Transition" (PPT) for giant planets and led to a high compressibility of hydrogen, now ruled-out by the experiments. Several equations of state of hydrogen or of the hydrogen-helium mixture solving these issues have become available \citep{Militzer+Hubbard2013, Becker+2014, Miguel+2016, Miguel+2018, Chabrier+2019}. While relying on ab-initio calculations, the difference between them amounts to up to $\sim 8$\% in density and $\sim 10$\% in temperature on a Jupiter adiabat, at pressures between $\sim 0.03$ to 10\,\rev{Mbar}. This is a significant source of uncertainty which must be accounted for, especially when seeking to reproduce the tight constraints obtained for Jupiter and Saturn.

Helium, shown in Fig.~\ref{fig:phases}b, undergoes a phase separation from metallic hydrogen at temperatures of relevance for Jupiter and Saturn. The formation of helium-rich droplets and their sinking under the action of gravity has strong consequences both for the cooling and the structure of these planets \citep[e.g.,][]{Stevenson+Salpeter1977a,Mankovich+Fortney2020}. Considerable progress has been made on this issue: Calculations based on first-principle are available and provide the full phase diagram as a function of pressure, temperature and composition \citep{Morales+2013, Schottler+Redmer2018}. High-pressure experiments provide direct evidence for immiscibility at Jupiter-interior conditions \citep{Brygoo+2021}. Uncertainties remain however: Ab-initio simulations predict that phase separation should occur in Saturn but not necessarily in Jupiter, leading \cite{Mankovich+Fortney2020} to arbitrarily offset the critical temperatures from \cite{Schottler+Redmer2018} upward by 540\,K in order to account for the depleted atmospheric helium abundance (see \S~\ref{sec:atmos_comp}). 
On the other hand, high-pressure experiments using laser-driven shock compression of H$_2$-He samples that have been pre-compressed in diamond-anvil cells indicate a much higher critical temperature reaching 10,500\,K at 1.5\,Mbar \citep{Brygoo+2021}. Given the importance of the issue for Jupiter and Saturn, further models and experiments are warranted. 

Given its high abundance in the Universe, water deserves special consideration. Its phase diagram is shown in Fig.~\ref{fig:phases}c. Importantly, when embedded inside the envelope of a giant planet, water is found to be easily soluble in metallic hydrogen \citep{Wilson+Militzer2012}, allowing for the possibility of an erosion of such a core (see discussion in \S~\ref{sec:re-distribution}). In Uranus and Neptune, below the hydrogen-helium-dominated envelope, water should be present in the form of a ionic fluid, but could then transition into superionic water \citep{French+2009, Redmer+2011}. This phase is a combination of a proton fluid and an oxygen lattice and is thus hybrid between a fluid and a solid, \rev{but experiments and ab-initio calculations with density functional theory indicate that it should behave as a solid \citep{Millot+2019}}. \rrm{\cite{Redmer+2011} find that is is characterised by a viscosity which is only a factor $\sim 2$ lower than that of warm dense fluid water, thus behaving more like a fluid.} The question of how other elements may affect the picture is not yet clear \citep[e.g.,][]{Guarguaglini+2019}, especially given that water can dissolve important amounts of MgO \citep{Kim+2021, Nettelmann2021}. 

Interestingly, water may also separate from molecular hydrogen at low pressures: Experiments indicate immiscibility at $17-26$\,kbar and temperatures lower than $750-1000$\,K \citep{Bali+2013}, leading to the possibility that Uranus and Neptune may possess liquid water oceans \citep{Bailey+Stevenson2021}. This issue is open however for several reasons: The experimental data have to be extrapolated to higher pressures and temperatures in order to meet the conditions relevant for Uranus and Neptune. But more importantly, ab-initio calculations do not find immiscibility in these conditions \citep{Soubiran2015}, raising the possibility that the immiscibility found by \cite{Bali+2013} may result from a contamination by the silicates used in the experiments. This issue which has significant consequences for the structures of Uranus, Neptune and ice giants must be investigated further. 

Iron, shown in Fig.~\ref{fig:phases}d, has a relatively high melting temperature at high pressures \citep{Mazevet+2019}, and thus should be in solid form in the central regions of Jupiter and Saturn. Nonetheless, ab initio simulations indicate that it is highly soluble in metallic hydrogen \citep{Wahl+2013} and therefore could be eroded if energetically possible. 

Finally, Fig.~\ref{fig:phases}e shows a potential phase diagram for silicates, or more precisely, MgSiO$_3$, expected to transform into MgO and SiO$_2$ above $\sim 10\,$Mbar \citep{Mazevet+2019}. The melting temperature is relatively high, indicating that MgO (but not SiO$_2$) should be solid in Jupiter and Saturn's cores. For the conditions expected in the deep interiors of these planets, both MgO and SiO$_2$ are expected to be soluble in metallic hydrogen, assuming abundances consistent with a solar composition \citep{Gonzalez-Cataldo+2014}. At lower pressures, MgSiO$_3$ should be solid in Uranus and Neptune. 

In order to understand the formation of giant planets and the fate of their primordial cores, one must consider that they initially formed with significantly higher temperatures ($\sim 30,000K$ or more at Jupiter's center). Figure~\ref{fig:phases} shows that during a significant fraction of their evolution, heavy elements in giant planets were entirely fluid and soluble. This does {\em not} imply that they were necessarily mixed efficiently, as this depends on whether this was energetically possible (see \S~\ref{sec:re-distribution}). Depending of the elements considered, insolubility or solidification occurred first. For example, helium separation should have occurred first in Saturn after $\sim 1.5\,$Gyr of evolution, then in Jupiter, $\sim 4\,$Gyr after its formation \citep{Mankovich+Fortney2020}, but helium solidification is beyond reach. Silicates in Jupiter and Saturn's core may have partially solidified, without becoming insoluble. Water in Uranus and Neptune may have become super-ionic (at high pressures) and may have become insoluble as well (at low pressures).  

Representative exoplanets shown in Fig.~\ref{fig:phases} \rev{are fully miscible because of their} high entropies, either because they are close to their star, or because they are massive and retained a large fraction of the internal energy. \rrm{These are the planets that we can best characterise presently.} We are however beginning to have the possibility to characterise giant planets with lower entropies. This should enable an extremely useful comparison with solar system giant planets. For example, the issue of helium phase separation should concern planets with masses between about $\sim40$ and $400\rm\,M_\oplus$ and effective temperatures below about $\sim 150\,$K, i.e., weakly irradiated and sufficiently old planets \rev{\citep{Fortney+Hubbard2004}. With the possibility to measure atmospheric compositions (see \S~\ref{sec:GP_atmos}), the existence of phase separations of other elements and of the link between atmospheric and interior composition will become highly significant.  }

\subsubsection{Departures from an Isentropic Interior}
\label{sec:non-adiabatic}
Traditionally, models of giant planets have been built with the assumption of an isentropic structure, with the idea that (1) the relatively high intrinsic luminosities coupled to the high radiative and conductive opacities should imply convective interiors and that (2) the superadiabaticity needed to transport the observed heat fluxes is small \citep[see][and references therein]{Guillot+2004}. Since efficient (nearly adiabatic) convection must lead to a uniform composition, this assumption must break down in the presence of compositional gradients, as was in fact recognised early-on for Uranus and Neptune \citep{Podolak+1991,Hubbard+1995}.  

In the presence of a gradient of mean molecular weight $\nabla_\mu\equiv d\log \mu/d\log P$ (where $\mu$ is the mean molecular weight and $P$ the pressure), the temperature gradient $\nabla_T\equiv d\log T/d\log P$ (where $T$ is temperature) should satisfy the convective stability criterion \citep{Ledoux1947, Kippenhahn+Weigert1990}:
\begin{equation}
\nabla_T \le \nabla_{\rm ad}+(\varphi/\delta)\nabla_\mu,\label{eq:Ledoux}
\end{equation}
where $\nabla_{\rm ad}\equiv (\partial\log T/\partial\log P)_{S,\mu}$ is the adiabatic gradient assuming uniform composition, $S$ is specific entropy, and $\varphi$ and $\delta$ are dimensionless thermodynamical quantities that are equal to $1$ for a perfect gas ($\varphi=-(\partial \log\rho/\partial\log\mu)_{P,T}$ and $\delta=-(\partial \log\rho/\partial\log T)_{P,\mu}$). 

In such a region, in the absence of a phase transition, several situations can be envisioned: 
\begin{itemize}
    \item Convection is shut down completely, heat is transported by conduction or radiation. This implies that $\nabla_T=\nabla_{\rm rad}(L)$, where $\nabla_{\rm rad}$ is the radiative/conductive gradient, a function of the properties of the medium and proportional to $L$, the intrinsic luminosity to be transported. 
    \item Double-diffusive convection sets in: an oscillatory instability due to the different coefficients of diffusion for heat and elements leads to the formation of a time-variable series of convective layers sandwiched in-between small stratified diffusive interfaces \citep{Rosenblum+2011,Wood+2013}. Globally, the temperature gradient is higher than in the absence of a compositional gradient, i.e., $\nabla_{\rm ad}<\nabla_T$ and it of course satisfies eq.~\eqref{eq:Ledoux}. 
    \item \rev{Eq.~\eqref{eq:Ledoux} is not satisfied. Rapid convective overturning occurs and leads to a fast mixing of the region. This generally creates a highly stable interface with a steep compositional gradient. This interface may evolve, either through double-diffusive convection, or because the adjacent region cools} \citep[see][and \S~\ref{sec:re-distribution}]{Vazan+2018}. \rrm{The composition gradient (and thus temperature gradient) in this interface. This situation is rapidly evolving with a with a timescale that is set by the cooling of the region above.} 
\end{itemize}
The presence of such a stable layer or of double-diffusive convection \rev{thus complexifies the analysis with a deep planetary entropy that} can be higher or smaller than that obtained for a pure adiabat \citep[see][]{Debras+2021}.  

\rev{In addition, }the presence of a phase transition (e.g., water or rock condensation, helium demixing) changes the picture in several ways. First, in the presence of convection, the release of latent heat generally favours heat transport, leading to a temperature gradient that can be smaller than the standard adiabat calculated without including this effect (so-called the {\em dry} adiabat). This process is known as moist convection \citep[e.g.,][]{Emanuel1994}. However, when the abundance of condensates is high, in hydrogen atmospheres, the molecular weight gradient starts having a dominant effect, leading to a possible inhibition of moist convection. For a perfect gas, this occurs when the mass mixing ratio of the condensing species $q$ exceeds a critical value \citep{Guillot1995}:
\begin{equation}
    q_{\rm crit}=\left(1-\frac{m_{\rm d}}{m_{\rm v}}\right)\frac{RT}{m_{\rm v}L_{\rm v}},
    \label{eq:qcrit}
\end{equation}
where $m_{\rm d}$ is the mass of the dry gas, $m_{\rm v}$ that of the condensing species, $L_{\rm v}$ the latent heat released upon condensation and $R$ the gas constant. For condensing species such as methane, water or iron, $q_{\rm crit}\approx 0.05$ to $0.1$, implying that moist convection inhibition should occur when the condensing species account for more than 5\% to 10\% of the mass of the mixture. Importantly, \cite{Leconte+2017} and \cite{Friedson+Gonzales2017} show that, in such a case, double-diffusion is also inhibited, raising the possibility that heat can only be transported locally by radiation or conduction,  with a very high temperature gradient which may even possibly violate eq.~\eqref{eq:Ledoux}. However, the moist convection inhibition criterion is derived locally, assuming full saturation. Moist convection and storms generated above, in regions such that $q<q_{\rm crit}$ (see eq.~\eqref{eq:qcrit}) may lead to important rainfall and consequently generate strong downdrafts \citep[see, in a slightly different context,][]{Guillot+2020a}. \rev{Their role for chemical transport and on the final temperature gradient should be investigated by dedicated simulations}.

The growth to sinkable $\sim 0.1$\,mm-size helium droplets should be fast, of order $0.1\,$s \citep{Stevenson+Salpeter1977a, Mankovich+2016}. It is not clear whether this may allow convection to proceed unhindered, whether diffusive-convection should set-in \citep{Nettelmann+2015,Mankovich+2016}, or whether it will be inhibited as well \citep{Guillot1995, Leconte+2017, Friedson+Gonzales2017}. 

Altogether, the presence of large compositional gradients ($\Delta\mu/\mu\gtrsim 1$) imply a high uncertainty on the temperature gradient, so that the interior temperatures (and consequently entropies) may exceed those calculated assuming adiabaticity by up to $\Delta T/T\approx\Delta\mu/\mu$, leading to a possibly significant underestimation of the total amount of heavy elements present in giant planets \citep[see][]{Leconte+Chabrier2012}.

\subsubsection{Gravitational and Seismological Sounding}\label{sec:sounding}

Gravity sounding provides an essential way to probe the planetary interior structure, especially when the measurements are extremely accurate \citep{Iess+2018, Iess+2019}. However, with only a handful of gravitational moments available and weighting functions peaking near external regions of the envelope \citep{Guillot2005}, the constraints are limited. The variety of possible structures, compositions, intrinsic uncertainties on the equations of state (see \S~\ref{sec:EOS}) and deep entropies (see \S~\ref{sec:non-adiabatic}) imply that the degeneracy in possible solutions remains high. 

Advances in our understanding of Saturn's interior \citep{Fuller2014,Mankovich+2019,Mankovich+Fuller2021} demonstrate the power of seismology. The ability to probe giant planet interiors through seismology was postulated already in a pioneering work by \cite{Vorontsov+1976} and applied to Saturn's rings \citep{Marley1991}, more than twenty years before their discovery by \cite{Hedman+Nicholson2013}. 
The discovery of normal modes in Saturn's rings proves that a mechanism, yet unidentified \citep[see][]{Bercovici+Schubert1987,Markham+Stevenson2018}, is capable of exciting these to detectable amplitudes. Saturn's rings are powerful amplifiers of f-mode planetary oscillations \citep{Marley1991, Marley+Porco1993}, but unfortunately this technique is difficult to apply to other planets, and the f-modes seen in the rings represent a small subset of all the possible modes. Experience from solar seismology shows that f-modes have amplitudes that are 1 to 2 orders of magnitude smaller than p-modes \citep[e.g.,][]{Christensen-Dalsgaard2002}. The analysis of Cassini gravity data indeed provides evidence for p-modes oscillations at frequencies of 500 to 700$\,\mu$Hz and with relatively high amplitudes of several m/s \citep{Markham+2020}. Ground-based searches using Doppler imaging have also provided evidence for the presence of p-modes in Jupiter, at frequencies of 1000 to 1500\,$\mu$Hz and amplitudes of up to 40\,cm/s \citep{Gaulme+2011}. Observations aimed at confirming these measurements are under way \citep{Goncalves+2019}. 

\cite{Jackiewicz+2012} show that, given the detection of a wide-enough variety of modes, one can probe the entire planet, from the atmosphere to the deeper interior. Having the capability to measure normal modes, from p-modes to f-modes in all four giant planets would provide us with the ability to truly constrain the structure and deep compositions of these planets.

\subsection{Magnetic Fields and Interior Rotation}

\begin{figure*}[htb!]
    \centering
    \includegraphics[width=\linewidth]{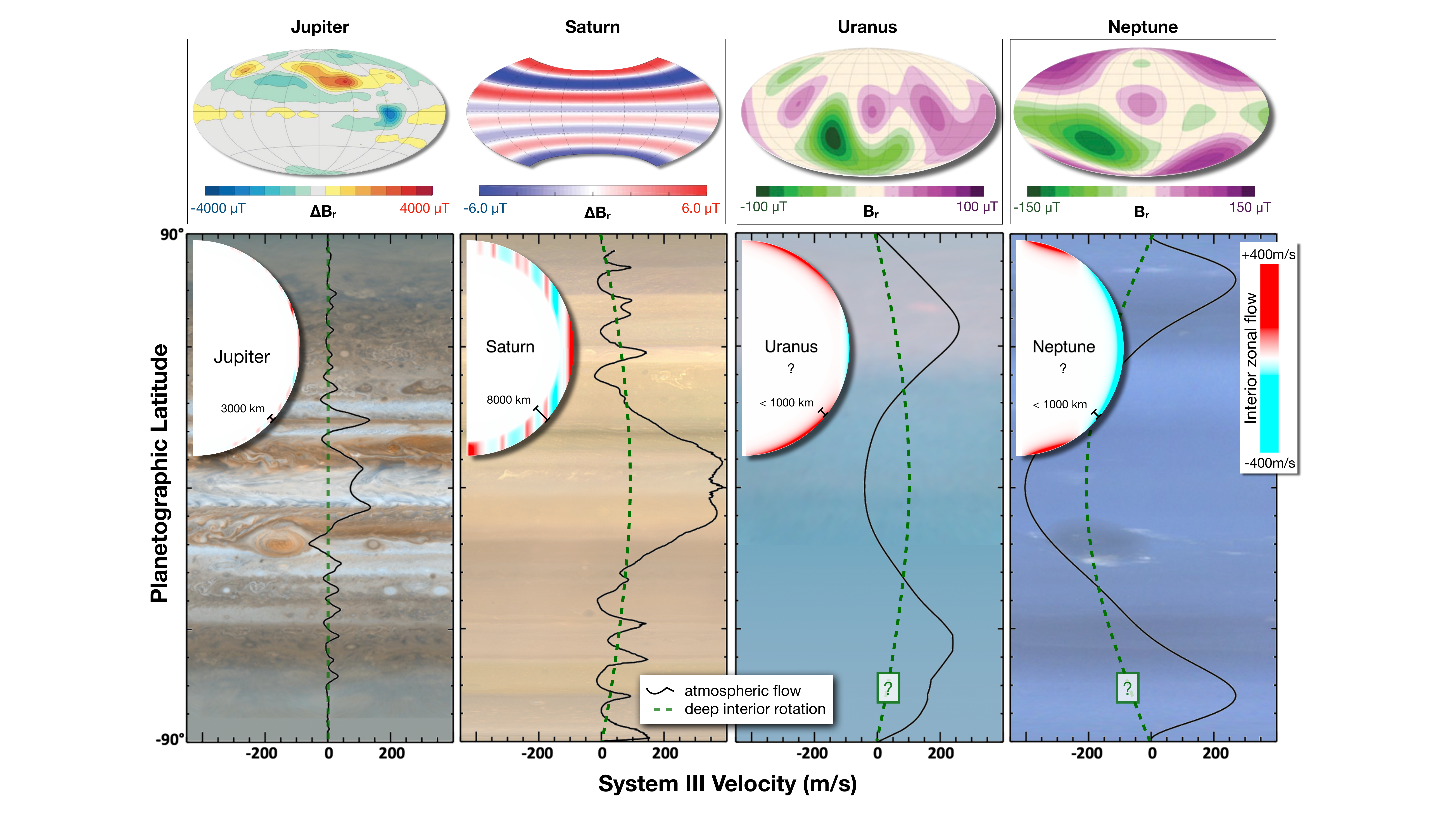}
    \caption{Magnetic fields, deep rotation, zonal winds and cloud bands on Jupiter and Saturn. The top panels show the non-dipole radial magnetic fields, for Jupiter at $0.9\rm\,R_J$ \citep[from][]{Moore+2018}, for Saturn at $0.75\rm\,R_J$ \citep[from][]{Dougherty+2018}, and the full radial fields for Uranus and Neptune at $1.0\rm\,R_{\rm tot}$ \citep[from][]{Soderlund+Stanley2020}. Cloud bands, zonal winds, and deep rotation on the outer planets.  The wind profiles are shown at the same scale for each planet and show zonal jets reaching 200-400 m/s; Jupiter and Saturn exhibit prograde equatorial jets, while Uranus and Neptune’s are retrograde. Deep rotation speeds (dashed lines) are determined from magnetic field rotation in Jupiter, seismology in Saturn \citep{Mankovich+2019}, and estimated from planet shape arguments in Uranus and Neptune \citep{Helled+2010}.  Interior wind profiles (insets) are inferred from gravity data \citep{Kaspi+2018, Guillot+2018}. In Uranus and Neptune only an upper limit to zonal flow depth can be determined \citep{2013Natur.497..344K}. Image credits: NASA/ ESA/ CICLOPS/ Bjorn Jonnson; winds from \cite{03porco}, \cite{11garcia}, \cite{15sromovsky}, \cite{11karkoschka} / T. Guillot, M. Hedman, L. Fletcher, A. Simon}
    \label{fig:flows_magnetic}
\end{figure*}


The magnetic fields of giant planets all differ in surprising ways. Jupiter's field is dipole-dominated with a $\sim 10^\circ$ tilt between the magnetic and rotation axis, with a non-dipolar part of the field which is confined almost entirely to the northern hemisphere and peaks at 3\,mT, almost three times more than the peak dipolar field \citep{Connerney+2018, Moore+2018}. Saturn's magnetic field is also dipole-dominated, with a nearly axisymmetric field, and a magnetic axis aligned with the spin axis to within $0.007^\circ$ \citep{Dougherty+2018,Cao+2020}. The fields of Uranus and Neptune are constrained only from flyby measurements by {\it Voyager~2} in 1986 and 1989, respectively. They are highly multipolar, with a dipole component tilted by $59^\circ$ at Uranus and $47^\circ$ at Neptune \citep[][and references therein]{Soderlund+Stanley2020}.

The reason for Jupiter's magnetic field strong north-south asymmetry (see Fig.~\ref{fig:flows_magnetic}) is not clear but it appears to preclude a dynamo operating in a thick, homogeneous shell. Rather, it has been proposed that this field morphology could arise from a dynamo operating in a thin layer, due to rapid variations in density or electrical conductivity \citep{Dietrich+Jones2018, Moore+2018}. This again points to the presence of stably-stratified layers in the interior \citep{Wicht+Gastine2020}. 

Saturn's highly axisymmetrical field shown in Fig.~\ref{fig:flows_magnetic} remains a mystery. A possibility is that the components of the fields are filtered out by a differentially-rotating stable conductive region above the dynamo region \citep[][]{Stevenson1982}. Examination of the field measured by Cassini leads \cite{Cao+2020} to estimate that this stable region must be at least 2500\,km thick, i.e. 4\% of Saturn's radius. This \rev{seems incompatible} with Saturn's interior model as derived by \cite{Mankovich+Fuller2021} which predicts a stable region at depth but none above the dynamo region. 

Reproducing the complex yet weakly constrained multipolar magnetic fields of Uranus and Neptune is clearly a challenge, especially given the many unknowns on the planets' interiors (see \S~\ref{sec:GP_interior}). One possibility is that these fields result from thin-shell dynamos overlying a region of stable stratification, near $R/R_{\rm tot}\sim 0.7$ \citep{Stanley+Bloxham2006}. This would be qualitatively consistent with either the strong stratification due to a water-rich layer or the presence of a solid superionic shell around this location \citep[][and \S~\ref{sec:GP_interior}]{Redmer+2011}. But other possibilities are that the multipolar dynamos result from the planets' relatively slow rotation rates and strong inertial effects \citep{Soderlund+2013} or from the complex interplay between moderate electrical conductivity and density stratification \citep{Gastine+2012}. 

Between the conducting interior where the dynamo originate and the atmosphere, rotation must change from being close to uniform \citep{Cao+Stevenson2017} to being strongly dependent on latitude. As shown in Fig.~\ref{fig:flows_magnetic}, the gravity field measurements from Juno and Cassini enabled a determination of the depth of the transition region, at about 3000\,km below the cloud tops ($R/\rm R_J \sim 0.96$) in Jupiter \citep{Kaspi+2018,Guillot+2018} and about 8000\,km ($R/\rm R_S\sim 0.86$) in Saturn \citep{Iess+2019,Galanti+2019}. The transition corresponds to an increase of hydrogen's conductivity and may be attributed to magnetic field drag. In Jupiter, a secular variation of the magnetic field was detected and shown to be compatible with advection of the field by the zonal flow near 93-95\% of the radius \citep{Moore+2019}.  

For Uranus and Neptune, the penetration depths of the winds are less constrained. Based on gravity data, the depths of the winds are estimated to be confined to a thin weather layer no more than $\sim$1,000 km ($\sim 96\%$ of the planetary radius) in both planets \citep{2013Natur.497..344K}, consistent with estimates of the interior conductivity and Ohmic dissipation arguments \citep{2020MNRAS.498..621S}. The rotation rate of their magnetic field, supposedly inferred from Voyager~2 data is also in question \citep{Helled+2010}, with consequences for the interior structures of these planets \citep{Nettelmann+2013}.

\subsection{Atmospheres: Spatial \& Temporal Variability}
\label{sec:GP_atmos}

The atmospheres of giant planets represent the lens through which we glimpse the deep, hidden properties of the planetary bulk, and the transitional domain between the convective interior and the external charged-particle environment of the magnetosphere. As seen in Fig.~\ref{fig:flows_magnetic} they are characterized by spatial variability, strong zonal winds and a remarkable banded structure. They also exhibit temporal variability on multiple timescales.    


\subsubsection{Banded Structure}
\label{atmos_bands}

Rotating fluid planets develop a system of planetary bands due to the dominance of the Coriolis force in the momentum balance, the injection of energy and momentum from small scales (eddies and storms) to larger scales (zonal jets), and the conservation of angular momentum and potential vorticity as differentially-heated air moves with latitude.  These bands manifest in the `visible' troposphere as a series of alternating prograde and retrograde zonal jets, separating bands of different temperatures, cloud opacity, and chemical composition.  These contrasts are thought to be representative of vertical and meridional circulations on the scale of the bands, as we discuss below.  The atmospheres accessible in our Solar System naturally fall into two categories:  
\begin{itemize}
\item Rapid rotators ($\sim10$ hours) with 5-8 bands in each hemisphere and a relatively uniform distribution of key condensables away from the equator (Jupiter and Saturn).
\item Intermediate rotators ($\sim17$ hours) with one equatorial retrograde jet, and a single prograde jet in each hemisphere, and strong equator-to-pole gradients in condensables (Uranus and Neptune).
\end{itemize}

\begin{figure*}[htb!]
    \centering
    \includegraphics[width=\linewidth]{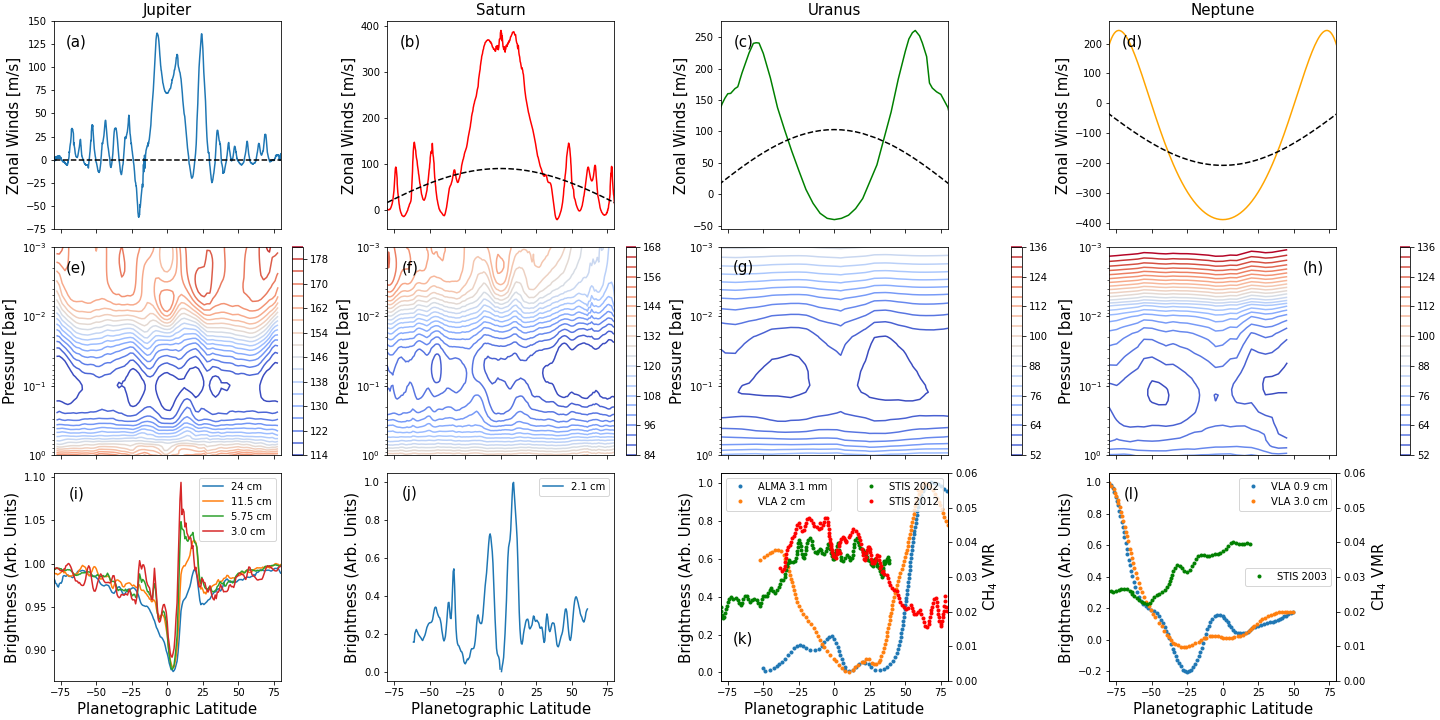}
    \caption{Latitudinal gradients in winds (top row), temperatures (middle row), and condensable volatiles (bottom row) for each of the giant planets.  Cloud-tracked winds for (a) Jupiter in 2000 \citep{03porco} and (b) Saturn in 2009 \citep{11garcia} from Cassini; (c) Uranus from a combination of Voyager (1986) and Keck data in 2012-13 \citep{15sromovsky}, and (d) Neptune from Voyager in 1989 \citep{11karkoschka}.  Tropospheric and stratospheric temperatures from Cassini/CIRS for (e) Jupiter in 2000 \citep{16fletcher_texes} and (f) Saturn between 2004-2009 \citep{18fletcher_hex}, and from Voyager/IRIS for (g) Uranus in 1986 \citep{15orton} and (h) Neptune in 1989 \citep{14fletcher_nep}.  Microwave brightnesses in the bottom row have been normalised for ease of comparison (for Jupiter, we normalise to latitudes beyond $20^\circ$). \rev{Microwave observations broadly sense NH$_3$ gradients for Jupiter (i) and Saturn (j); H$2$S gradients for Uranus (k) and Neptune (l).  CH$_4$ does not condense and is well-mixed on Jupiter and Saturn (not shown); but condenses and displays latitudinal gradients on Uranus and Neptune (k, l).} For (i) Jupiter, we show nadir brightness in four channels as measured by Juno in 2016-18 \citep{20oyafuso}.  For (j) Saturn, we show 2.1 cm from passive radiometry with the Cassini radar 2009 \citep{13laraia}.  For (k) Uranus, we show ALMA (2017) and VLA (2015) brightness scans from \citet{Molter+2021}, plus CH$_4$ from \citet{14sromovsky} from Hubble STIS observations (their Fig. 12, referenced to the right-hand axis); and for (l) Neptune, we show VLA (2015) brightness scans from \citet{Tollefson+2021}, \rev{alongside the CH$_4$ mole fraction estimated by multiplying the latitudinal gradients reported by \citet{Karkoschka+Tomasko2011} from Hubble STIS observations (their Fig. 8) by the vertical profile in their Fig. 10, with a maximum deep mole fraction of 0.04.  The mole fraction (referenced to the right-hand axis) should therefore be representative of abundances in the 1.8-3.0 bar region.}}
    \label{fig:GPTwinds}
\end{figure*}

These two groups are shown in Fig. \ref{fig:GPTwinds}, displaying their measured winds, tropospheric and stratospheric temperatures ($p<1$ bar, above the main clouds), and microwave brightness distributions.  Microwave-dark regions exhibit significant absorption from condensing species (NH$_3$, H$_2$S, H$_2$O) in the few-tens-of-bars domain.  Conversely, microwave-bright regions (e.g., at the poles of the ice giants) are depleted in volatiles.  Hemispheric contrasts in temperatures and winds, driven by seasonal insolation and superimposed onto the smaller-scale banded structure, will be discussed in Section~\ref{atmos_temporal}.  

Fig. \ref{fig:GPTwinds} demonstrates that giant planet atmospheres are latitudinally heterogeneous, with different latitudinal domains exhibiting somewhat different climatologies.  A `classical' picture of the bands of Jupiter and Saturn divides them into anticyclonic zones (with prograde jets on their poleward sides, retrograde jets on their equatorward sides) of low temperature, enhanced cloud opacity due to condensation, and enhanced abundances of species like NH$_3$, PH$_3$, para-H$_2$ due to upwelling \citep[e.g., see reviews by][]{04ingersoll, 09delgenio}.  
Conversely, cyclonic belts are regions of warmer temperatures, cloud-free conditions, and depletions in gaseous species. The implied temperature gradients are in geostrophic balance with the zonal winds, implying winds decaying with altitude from the cloud-tops into the stably-stratified upper troposphere and stratosphere \citep[e.g.,][]{83conrath}.  The belt/zone contrast is most apparent in the tropical regions in Fig. \ref{fig:GPTwinds}, featuring enhanced NH$_3$ and cool temperatures in the equatorial zones, and depleted NH$_3$ and warm temperatures in the equatorial belts \citep{86gierasch, 06achterberg, 13janssen, 16fletcher_texes, Li+2017}.  
At mid-latitudes on Jupiter and Saturn, the correspondence between temperatures and winds remains clear, implying vertical decay of the winds, but the connection to gaseous abundances and aerosols becomes weaker \citep[e.g.,][]{Fletcher+2011, Giles+2017, 19antunano, Grassi+2020}. 


This picture of the banded structure was called into question by findings from the Galileo \citep{00ingersoll}, Juno \citep{21fletcher_mwr}, and Cassini missions \citep{Fletcher+2011}.  Several lines of evidence support deeper motions in the \textit{opposite} sense to those described above \citep[see][for a full review]{20fletcher_beltzone}: lightning was found to be prevalent in Jupiter's belts (suggesting uplift by moist convective plumes) \citep{00ingersoll}; eddy-momentum flux convergence into the zonal jets required a compensating meridional circulation with rising motion in belts, sinking in zones \citep{06salyk, 12delgenio}; and some chemical contrasts in Saturn's deep troposphere appeared to oppose those seen in the upper troposphere \citep{Fletcher+2011}.  Adding to this, Jupiter's microwave brightness gradients appear to flip in sign as we probe deeper into the atmosphere \citep{21fletcher_mwr}:  belts are microwave-bright in the upper troposphere but microwave-dark below 5-10 bars (vice versa for zones).  This could imply a series of stacked meridional circulation cells on the scale of the belts and zones, \rev{analogous to Earth’s Ferrel circulation cells} (\citeauthor{Duer+2021} \citeyear{Duer+2021}, \citeauthor{21fletcher_mwr} \citeyear{21fletcher_mwr}, see also \citeauthor{17ingersoll} \citeyear{17ingersoll}).  \rev{The cells would be responsible for} advecting NH$_3$ (and other gaseous species) in opposite directions above and below a \rev{transitional layer somewhere in the 1-10 bar region, creating ammonia-depleted belts at shallow depths, and ammonia-enriched belts at greater pressures.}


Observations of Uranus and Neptune, from Voyager-2 in the 1980s, through three decades of ground- and space-based remote sensing, reveal atmospheres rather unlike the Gas Giants \citep[e.g., see reviews by][]{18mousis, 19hueso, 20moses, 20fletcher_icegiants}.  Fig. \ref{fig:GPTwinds} reveals the same geostrophic balance between temperatures and winds, implying decay of the zonal winds with altitude \citep{98conrath}. 
Finer-scale albedo banding is observed on both planets, reminiscent of the bands of Saturn \citep[e.g.,][]{15sromovsky}, but to date no temperature or wind contrasts have been observed on these scales.  However, the key difference from the Gas Giants is the strong equator-to-pole gradient in CH$_4$ \citep{Karkoschka+Tomasko2011, 14sromovsky} and H$_2$S \citep{Molter+2021, Takasao+2021}.  The former manifests as bright poles in the visible and near-IR (due to the dearth of CH$_4$ absorption), the latter manifests as bright poles in the radio.  These gradients may imply larger-scale circulation within and below the clouds, with air rising at the equator, moving polewards, and descending at high latitudes \citep{14depater, 14sromovsky}.  

\subsubsection{Deep Vertical Structure}

We have few measurements of vertical structure in a planetary atmosphere below the 1\,bar pressure level. In 1995, the Galileo probe plunged into Jupiter's atmosphere with an entry speed of 47 km/s, at a 6.5$^\circ$ North latitude, at the edge of a hot spot. The region was found to be relatively devoid of clouds \citep{Ragent+1998}, with a relatively low abundance of condensates at higher levels: The abundance of NH$_3$ increased rapidly to $\sim 200$\,ppmv near 2 bar and then slowly to its maximal value (between $400$ and $800$\,ppmv) near 7 bar. That of H$_2$S increased progressively from $\sim 10$\,ppmv near 9 bar to about 100\,ppmv near 16 bar. Water was found to have a very low mixing ratio $\sim 50$\,ppmv near 10 bar and reaching only $\sim 500$\,ppmv near 20 bar, significantly less than the solar value \citep{Wong+2004}. The probe also measured a temperature profile that was nearly dry adiabatic, except for regions with a static stability of $0.1$ to $0.2$\,K/km at $0.5-1.7$\,bar, $3-8.5$\,bar, and $14-20$ bar \citep{Magalhaes+2002}. It reached a pressure of 22\,bar for a temperature of 427.7\,K. 

\begin{figure}[htb!]
    \centering
    \includegraphics[width=\linewidth]{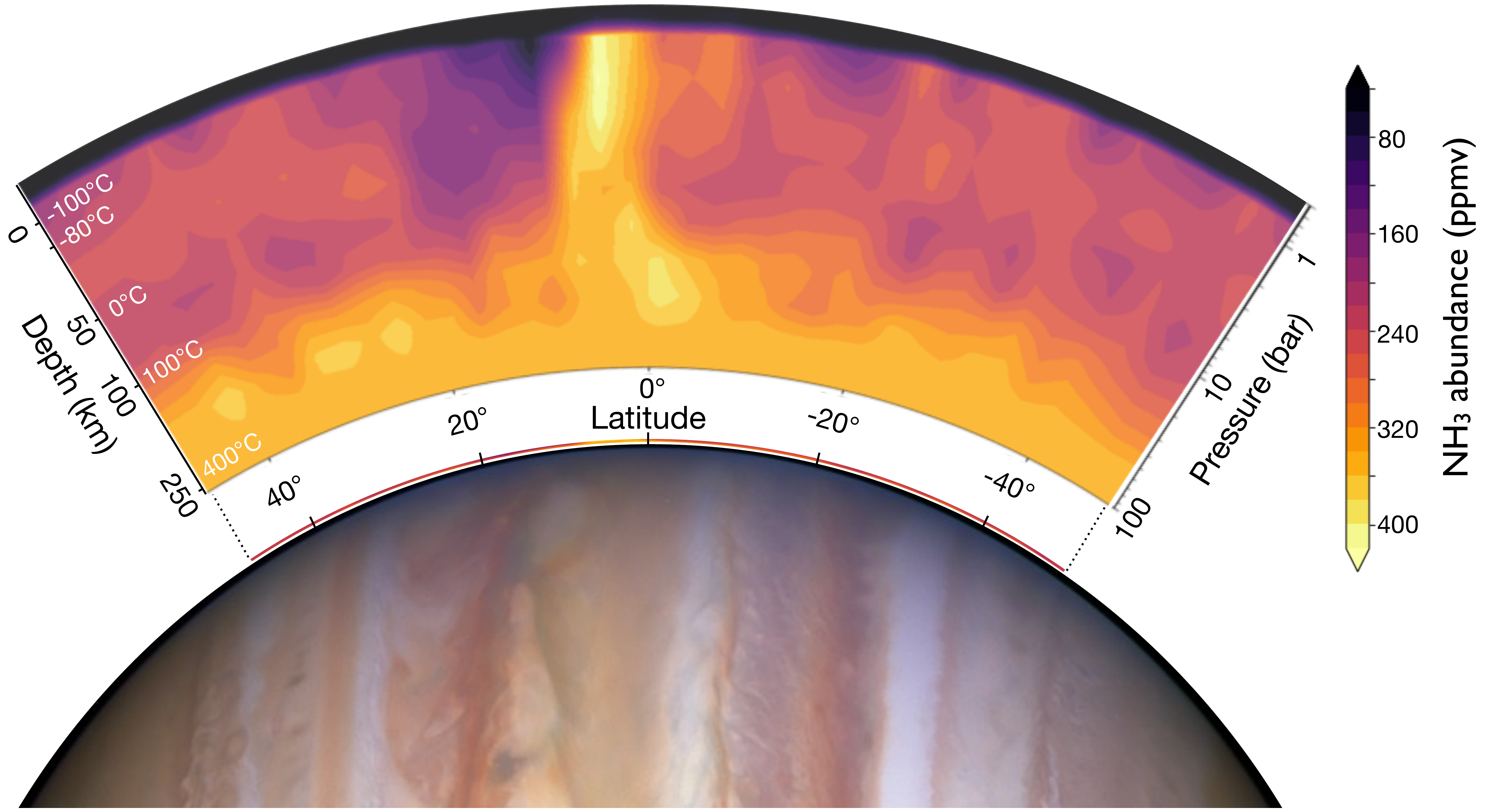}
    \caption{Ammonia abundance measured by Juno MWR in the deep atmosphere of Jupiter. The map corresponds to a mean abundance in parts per million by volume as a function of latitude and pressure level (or depth), obtained by a median average from 10 Juno flybys (PJ1 to PJ11) \citep[from][]{Li+2017,Guillot+2020b}. The Jupiter visible image is from HST/WFC3/Mike Wong.}
    \label{fig:jupiterdeepnh3}
\end{figure}

Ground-based observations from the Very Large Array have allowed probing Jupiter's atmosphere, providing first evidence for a depletion of ammonia except near its equatorial zone \citep[see][and references therein]{dePater+2016}. But the Juno Microwave Instrument measurement, operating inside Jupiter's synchrotron radiation belt at wavelengths of $3$ to $24$\,cm (see Fig.~\ref{fig:GPTwinds}) was able to probe the deep atmosphere, down to more than $100$\,bar \citep{17bolton}. As shown in Fig.~\ref{fig:jupiterdeepnh3}, an inversion of the MWR data revealed that Jupiter’s ammonia has a variable abundance as a function of
depth and latitude down to at least 200 km below the cloud tops,
far beneath the expected cloud base \citep{Li+2017}. \rev{The analysis of individual vortices also point to deep structures with variable depths, with both the microwave and gravity data showing that Jupiter's Great Red Spot extends to about 300 to 500\,km deep \citep{Bolton+2021, Parisi+2021}.} 

The ammonia depletion is correlated with enhanced lightning activity, as measured through the flash rate as a function of latitude \citep{Brown+2018}. It has been proposed that during strong storms powered by water condensation in the $\sim 5$\,bar region, ammonia vapor enables the melting of water ice crystal at very low temperatures $\sim -85^\circ$C, thus leading to the efficient formation of water-ammonia hailstones called mushballs \citep{Guillot+2020a}. The downward transport of ammonia in stormy regions by mushballs and associated cold and dense downdrafts accounts for the Juno measurements \citep{Guillot+2020b}. Importantly, the Juno results reveal that Jupiter's deep atmosphere must be on average stable down to great depths. The abundance variations seen with ammonia could also apply to water, providing at least part of the explanation for the results of the Galileo probe. Finally, the same process at play on Jupiter should also apply to Saturn, Uranus and Neptune. 

\subsubsection{Temporal Variability}
\label{atmos_temporal}

\begin{figure*}[htb!]
    \centering
    \includegraphics[width=\linewidth]{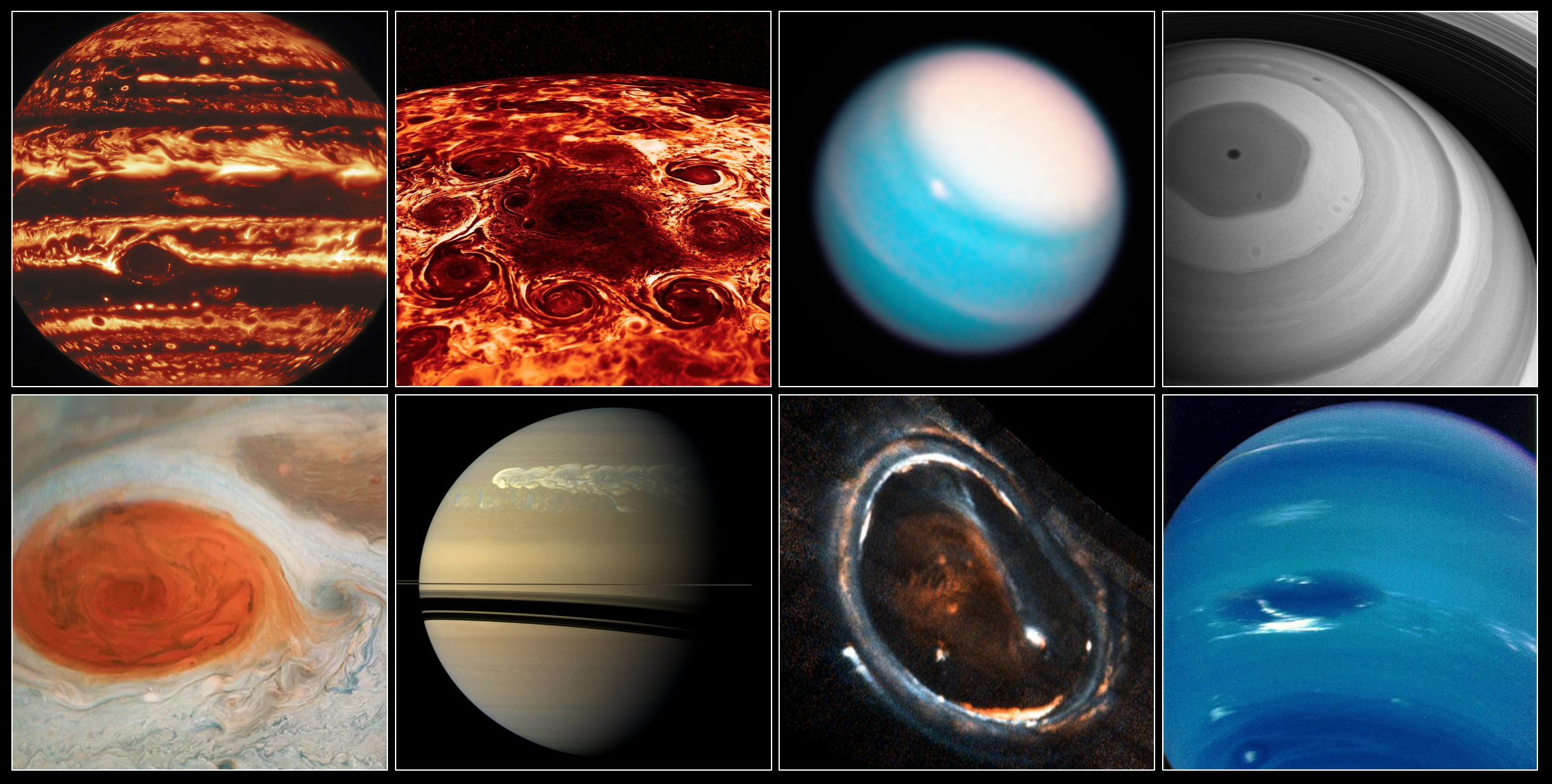}
    \caption{Examples of dynamic phenomena on the Giant Planets. \textit{Top row, left to right:} Jupiter at 5 $\mu$m seen by Gemini (credit:Gemini Observatory/NOIRLab/NSF/AURA/M.H. Wong); and the north polar cyclone and `octagon' of circumpolar cyclones observed at 5 $\mu$m by Juno \citep{18adriani}; Uranus' seasonal reflective polar cap of aerosols observed by Hubble (credit:NASA/ESA/A. Simon/M.H. Wong/A. Hsu); and Saturn's north polar hexagon and bands observed by Cassini (credit:NASA/JPL-Caltech/SSI). \textit{Bottom row, left to right:}  Jupiter's Great Red Spot observed by JunoCam (credit: NASA/JPL-Caltech/SwRI/MSSS/Gerald Eichstädt/Seán Doran); Saturn's 2010-11 storm system from Cassini (credit:NASA/JPL-Caltech/SSI); Jupiter's north polar auroral oval observed in the UV by Juno \citep{17mauk}; and Neptune and its Great Dark Spot observed by Voyager 2 (credit:NASA/JPL-Caltech). }
    \label{GPmontage}
\end{figure*}

Two further sources of variability have not yet been discussed:  changes to the atmosphere over time, and longitudinal variability associated with large-scale disturbances (vortices, plumes, waves, etc.).  Indeed, long-term monitoring of the four giant planets reveals changes over timescales from years (seasons), to months (planet-wide disturbances), to days (storms), and even minutes (auroras and impacts).  These phenomena could modulate light-curves seen from afar, so we discuss temporal variability in this section.  On our journey through a giant planet \textit{from the inside out,} this represents the top-most level that is most readily accessible to remote sensing:  the upper troposphere and middle-atmosphere above the clouds, where reflected light and thermal emission in Fig. \ref{GPmontage} can reveal dynamic and ephemeral phenomena.  

Absorption of short-wave sunlight by methane and aerosols dominates radiative heating, to be balanced by thermal emission and radiative cooling by the collision-induced hydrogen-helium continuum, and from stratospheric acetylene (13 $\mu$m), ethane (12 $\mu$m), and to a lesser extent methane (7 $\mu$m).  Planets with Earth-like axial tilts (Saturn and Neptune), or extreme obliquities (Uranus), might therefore be expected to display seasonal variability.  The radiative time constant characterises an atmosphere's inertia to temperature response to seasonal change - this is shortest in the upper and middle-atmosphere, but lengthens at higher pressures so that temperatures in the troposphere tend to follow the annual-averaged insolation.  Observations of the upper troposphere and stratospheres, however, tend to see seasonal asymmetries in temperatures,
stratospheric chemicals (from a combination of temperature- and sunlight-dependent reactions and seasonally-dependent vertical mixing), and aerosols/hazes (from seasonally-dependent growth of aerosols).  The former has been revealed via 13 years or orbital remote sensing of Saturn by the Cassini spacecraft, whereas the latter is exemplified by the slow seasonal growth of tropospheric aerosols at the Uranian poles, producing reflective polar caps \citep{Sromovsky+2019}.  These slow seasonal variations can also be influenced by periodic variations on shorter timescales, such as those associated with equatorial stratospheric oscillations on Jupiter \citep{91leovy} and Saturn \citep{08fouchet, 08orton}, or those associated with seasonally-reversing stratospheric circulations \citep[e.g.,][]{12friedson}.  \rrm{See \citet{20fletcher_saturn} for a review of these processes pertinent to Saturn.}  Thus any disc-averaged observations sensing domains in the radiatively-controlled middle atmosphere could find conditions biased towards particular seasons or phases of seasonally-dependent phenomena.

Besides the seasons, planetary emission and reflection are observed to vary over shorter, but still periodic, cycles.  Entire bands of Jupiter can be seen to fade (i.e., whiten over) and then spectacularly revive (i.e., regain their red-brown colour) due to localised convective plumes \citep{17fletcher_seb}.  Other bands expand and contract in latitude with predictable, but poorly understood timescales \citep{19antunano}, although the influence of these changes on light curves may be minimal \citep{19ge}.  The brightness of Uranus and Neptune has been monitored over many decades \citep{19lockwood}, appearing to show some relationship with the solar cycle \citep{16aplin}, exemplifying planet-star interactions as a driver for the brightness of an atmosphere.  Saturn's 2010 storm is an extreme example of a storm modulating temperature and composition variations with longitude and time \citep{18sanchez-lavega}, creating a new cloud-free and volatile-depleted band that persisted for many years after the storm, and producing huge changes to stratospheric temperatures and chemistry known as the `beacon' \citep{12fletcher}.  These storms appear to occur on seasonal timescales, potentially as a result of a radiatively influenced charge-recharge cycle associated with the overcoming of convective inhibition near the water-cloud \citep{Li+Ingersoll2015}.      

Water is key to understanding the giant planets, playing a crucial role in their meteorology.  Moist convection may be the most important mode of transport for internal heat on Jupiter and Saturn \citep{Gierasch+2000}, although water storms are observed to be highly intermittent, localised, but widespread \citep{Hueso+2002, Sugiyama+2014, Brown+2018}.  In the case of Jupiter's planet-wide disturbances, immense clusters of water-driven plumes are readily visible and reflective in certain longitude domains.  Disturbances in Jupiter's North Temperate Belt generate vigorous plumes within a few days of one another, but at widely separated longitudes, indicating some degree of connection over many thousands of kilometres \citep[e.g.,][]{08sanchez}.  Like Saturn's storm, these sporadic water plumes could create ephemeral changes to a rotational light curve, whilst associated precipitation could also transport condensed volatiles down to great depths \citep{Guillot+2020a, Guillot+2020b}.  Understanding the occurrence of these storms remains a significant challenge, given the stabilising effects of the molecular weight of moist air (i.e., convective inhibition), balanced by the buoyant effects of latent heat release \citep{Guillot1995, Leconte+2017}.  Unfortunately, the depth of this water-driven convection is hard to access on Jupiter and Saturn, but methane-driven convection on the Ice Giants may provide a vital means for testing convection in hydrogen-rich planets at higher, more accessible altitudes \citep{20hueso}.

Periodic variations in planetary atmospheres, in addition to sporadic storms, will cause a planet's disc-averaged emission to change with time.  In addition, giant planet atmospheres host vortices with a variety of scales \citep[e.g.,][]{04ingersoll,05vasavada}, from large-scale anticyclones (typically cool and cloudy with white or red aerosols on Jupiter and Saturn, or darker ovals on Uranus and Neptune with their associated `orographic' white clouds), to smaller scale and often elongates cyclonic structures (typically cloud-free but prone to outbursts of convection creating turbulent filamentary patterns).  Small vortices are unlikely to be accessible in the disc-average, but the largest anticyclones (Jupiter's Great Red Spot and Neptune's Great Dark Spot) could modulate the light curve at the rotation period.  Indeed, rapidly-evolving storm features on Neptune were seen to modulate the lightcurve measured by Kepler as the planet rotated \citep{16simon}.  These vortices evolve with time, such as the shrinking of Jupiter's Great Red Spot \citep{18simon}, or the equatorward drifting and eventual disappearance of Neptune's Great Dark Spot \citep{01stratman}.

Finally, giant planet atmospheres are influenced by external processes, including cometary impacts and interplanetary dust \citep{17moses}, and auroral heating \citep{21odonoghue}.  Fig. \ref{GPmontage} shows an example of UV emission from Jupiter's northern aurora, where energy injection is known to heat the thermosphere and ionosphere to significantly higher temperatures than would be expected from solar heating alone.  
\rrm{The auroral heating is visible throughout the infrared (emission from ionospheric H$^+_3$ and stratospheric hydrocarbons), and as Jupiter's magnetic axis is tilted, we would therefore see a modulation of the rotational lightcurve at these wavelengths.  The strength of the auroral UV emission, and the ionospheric heating, is seen to vary with time in connection with processes both internal to the magnetosphere and external, in the solar wind. Auroras can also have other consequences, such as influencing aerosol production through ion-neutral chemistry, which subsequently influence the balance between radiative heating and cooling, and once again meaning that auroral domains may display unusual environmental conditions compared to other latitudes.}


\subsubsection{Implications for the Characterisation of Exoplanets} 
With the caveat that the environmental conditions encountered on the present-day census of exoplanets and brown dwarfs will be significantly different to our own giant planets \rrm{(warmer temperatures, faster rotation)}, the discussion above suggests latitudinal and vertical variability could influence the interpretation of disc-averaged observations: 
(i) Certain latitude domains will be significantly brighter in the thermal than colder, cloudier, volatile-enriched regions - the belts of Gas Giants, or the poles of Ice Giants.  These brighter domains may contribute more to the disc average, meaning a bias towards volatile-depleted and cloud-free regions.  An extreme case is Jupiter at 5 $\mu$m, where radiance is only observed from cloud-free belts and compact features, like Galileo's infamous `hotspot.'  Furthermore, if a giant exoplanet is viewed from an oblique angle (e.g., consider the appearance of Uranus as it approaches solstices), a volatile-depleted polar domain may dominate the disc average.
(ii) The `climate domain' being characterised depends on the depth of penetration at the wavelength of interest - observations sounding the deeper troposphere would reveal conditions rather different to those sensing the upper troposphere (or even the stratosphere, see \S~\ref{atmos_temporal}).  It is also clear that cloud condensation \rev{sometimes} deviates from the equilibrium expectations:  aerosol layers \rev{rarely} occur where they are expected, and deep volatiles show considerable spatial variability and vertical gradients, rather than being well-mixed. Put simply:  vertically-layered climate domains with different compositions and circulations are likely adding complexity to interpretation of disc-averaged spectra. 

In summary, inferring the properties of the deep interior and bulk composition `\textit{from the outside in},' when it is hidden beneath such a highly variable and complex atmospheric layer, should remain a significant challenge for any giant planets.

\subsection{Composition}

\subsubsection{Bulk composition}
\label{sec:bulk_comp}
The bulk chemical composition of each planet reflects the proportion of rocks, ices, and gases accreted by the forming giants from the surrounding nebula (see \S~\ref{sec:formation} hereafter). 
Unfortunately, we presently cannot distinguish between rocks and ices, something that would be highly desirable in order to constrain the history of the formation of the solar system \citep[see][]{Kunitomo+2018}. Instead, we have to treat them together as heavy elements. 

Jupiter is the planet for which the uncertainties on interior composition have been the largest, owing mostly to the fact that a large fraction of its interior lies in this 0.1 to 10$\,$Mbar region for which the EOSs of hydrogen and helium are the most uncertain (see \S~\ref{sec:EOS}). Three-layer models inferred a total mass of heavy elements between $10$ and $42\rm\,M_\oplus$  for a core smaller than about $10\rm\,M_\oplus$ \citep{Saumon+Guillot2004}. 
Recent models \rev{based on the Juno measurements (\S~\ref{sec:sounding})} have not narrowed down this uncertainty, with a total mass of heavy elements ranging from about \rev{8 to $46\rm\,M_\oplus$ \citep{Wahl+2017,Debras+Chabrier2019,Ni2019,Nettelmann+2021,Miguel+2022}.
The central compact core is smaller than $7\rm\,M_\oplus$ \citep{Miguel+2022}, with most of the heavy element mass being held in a dilute core that may encompass a limited region to most of the metallic hydrogen envelope (see Fig.~\ref{fig:interiorslices}). Most of these uncertainties are linked to uncertainties in the equations of state of hydrogen and helium \citep[e.g.][]{Mazevet+2020}. In addition, a significant tension exists between interior models which favor low metallicities for the outer envelope, in contradiction with atmospheric constraints. Possible solutions include a heavy element abundance that would decrease with depth in the molecular envelope region \citep{Debras+Chabrier2019} or higher temperatures in the deep atmosphere \citep{Nettelmann+2021, Miguel+2022}, both raising further questions in terms of the long-term stability and formation of such an inverted Z-gradient and in terms of possible latitudinal temperature variations in Jupiter, given the Galileo probe constraint.} \rrm{The values of the core masses range from about zero to up to $32\rm\,M_\oplus$ including the dilute part that can encompass a large fraction of Jupiter's metallic region. Not all of these solutions fully match all the available constraints. Work is under way that should provide a clearer picture.} 

On Saturn, three-layer models were predicting a total mass of heavy elements between $16$ and $30\rm\,M_\oplus$ and a compact core mass between $8$ to $25\rm\,M_\oplus$ \citep{Saumon+Guillot2004}. The \rev{gravitational and seismological constaints from Cassini} allow \cite{Mankovich+Fuller2021} to derive much tighter constraints, a total mass of heavy elements $19.1\pm 1.0\rm\,M_\oplus$ and a compact+dilute core containing $17.4\pm 1.2\rm\,M_\oplus$ of rocks and ices. Interestingly, the amount of rocks and ices in the envelope is relatively limited corresponding to an envelope metallicity $Z_{\rm out}=0.041\pm 0.009$. \rev{Again, as for Jupiter, this may be in tension with the generally high metallicity of Saturn's atmosphere (\S~\ref{sec:atmos_comp})}. 

\rev{Unfortunately, the lack of accurate constraints on Uranus and Neptune's gravity fields translates into considerable uncertainties for these planets, with in particular the impossibility to determine whether they are formed of discrete layers (as in Fig.~\ref{fig:interiorslices}) or instead mixed regions with progressive compositional gradients}. Interior models suggest a metallicity of $\sim 76\%$ to 90\% for Uranus and 77\% to 90\% for Neptune \citep{Helled+2011, Nettelmann+2013}. While they are called "ice giants", we presently have no information on their rock-to-ice ratio \citep{Helled+2020}.


\subsubsection{Atmospheric Composition}
\label{sec:atmos_comp}

\begin{figure*}[tb!]
    \centering
    \includegraphics[width=\linewidth]{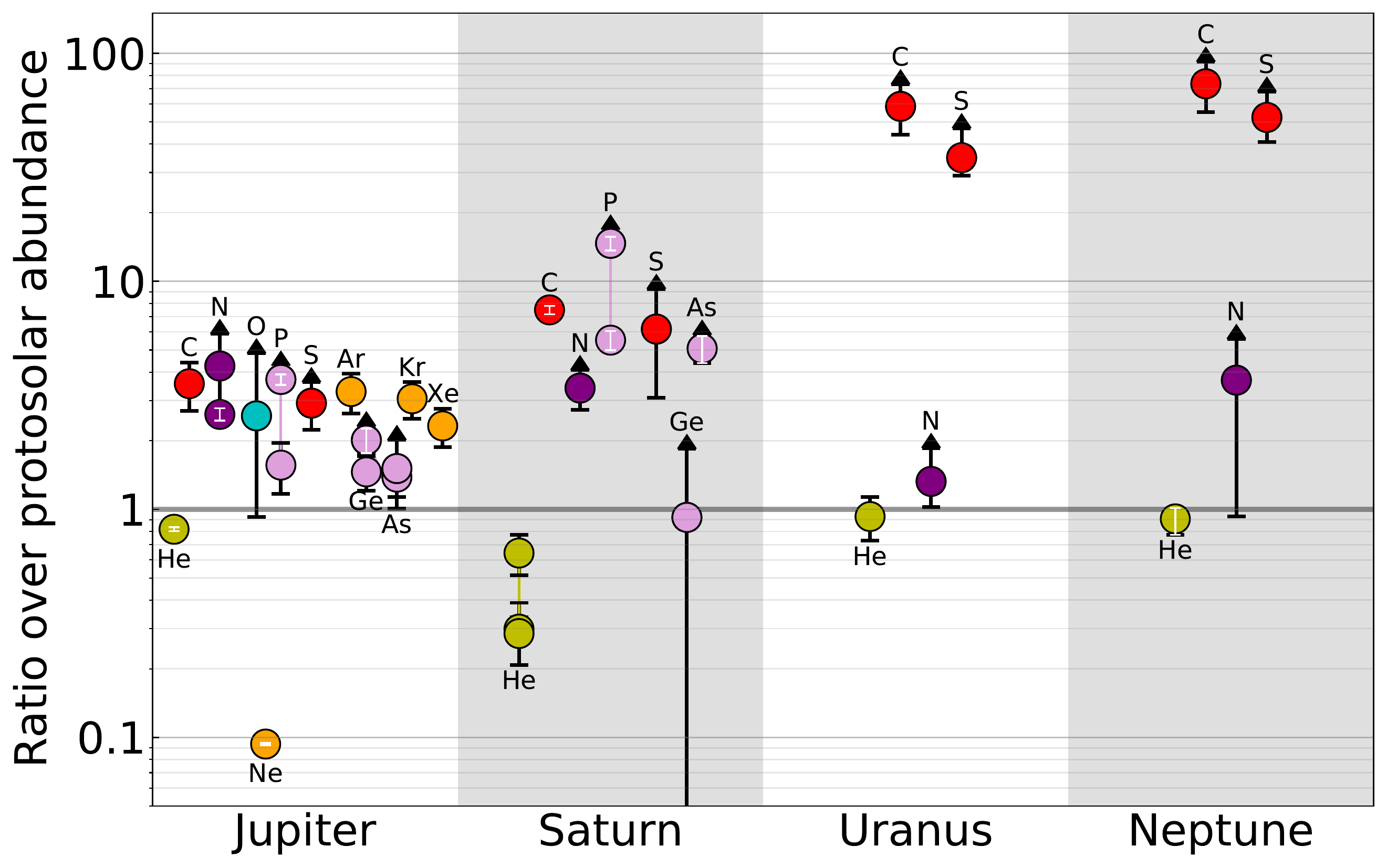}
    \caption{\rev{Elemental abundances of He, C, N, O, P, S, Ar, Ge, As, Kr and Xe in protosolar units in Jupiter, Saturn, Uranus and Neptune. Elements carried by chemical species which are condensing in the atmosphere (e.g., C, which is mainly carried as CH$_4$ which condenses in Uranus and Neptune) are indicated by a top arrow. For these species, when the abundance is variable (horizontally and vertically), the maximum value was chosen. Several points are sometimes provided for a given species when several measurements are available. The colours are green for helium, orange of noble gases, red for carbon and sulfur, purple for nitrogen, blue for oxygen and pink for elements carried by disequilibrium species. See Table~\ref{tab:comp} for values. }}
    \label{fig:compositions_JSUN}
\end{figure*}

The extent to which the observed atmospheric composition (from the stratosphere to the deeper troposphere at a few tens of bars) reflects that of the deeper interior remains an open question.  Atmospheric remote sensing, in addition to \textit{in situ} measurements by the Galileo probe, reveal that cosmogonically-common elements are present in their reduced/hydrogenated forms, either as condensable volatiles (methane, ammonia, H$_2$S, water), disequilibrium tracers (e.g., PH$_3$, AsH$_3$, GeH$_4$, CO), photochemical products of UV photolysis (e.g., tropospheric hazes, stratospheric hydrocarbons), or as externally-sourced contaminants to the upper atmosphere.  The elemental abundances measured on each planet is provided in Table \ref{tab:comp}, and shown in Fig. \ref{fig:compositions_JSUN}. However, as the vertical distribution of each molecular species can influenced by chemical sinks (condensation, photochemistry) and sources (e.g., vertical mixing), accessing well-mixed `bulk' reservoirs is a distinct challenge.  

Classical thermochemical `equilibrium' models predict the formation of condensate clouds from key species (methane, ammonia, hydrogen sulphide, water) in relatively well-defined layers, and uniform mixing below \citep{Atreya+2003}.  For the top-most clouds on the Gas Giants (NH$_3$ ice and solid NH$_4$SH), Cassini and Juno measurements of the abundances of NH$_3$ \citep{13laraia, Fletcher+2011, dePater+2016, Li+2017} below the clouds has shown to be highly variable, and even less is known about H$_2$S, with estimates of elemental abundances being subject to considerable uncertainty owing to (i) remote-sensing degeneracies between abundance, temperature, and aerosols; and (ii) spatial variability associated with dynamics and meteorology.  The latter implies that it remains unclear how representative the Galileo \textit{in situ} measurements were for Jupiter's atmosphere.  The top-most clouds of the Ice Giants (CH$_4$ ice and H$_2$S ice) also limit the accessibility of the bulk carbon and sulphur enrichments:  estimates of CH$_4$ and H$_2$S rely on precise separation of gaseous absorption from cloud reflectance in the near-infrared \citep{09karkoschka, Karkoschka+Tomasko2011, 14sromovsky, 18irwin}, or on millimetre-centimetre-wave sounding \citep{Molter+2021, Tollefson+2021}, both of which are further hindered by strong equator-to-pole gradients in condensables (see \S \ref{atmos_bands}).  Only methane on the Gas Giants has been reliably constrained as temperatures are too warm for condensation and, in the absence of carbon sinks, the derived CH$_4$ is expected to be representative of the bulk \citep{Wong+2004, Fletcher+2009ch4}.  

In summary, extrapolating atmospheric abundances of condensibles (and disequilibrium species) to the deeper interior is fraught with difficulty, and this is before we reach a vitally-important molecule for understanding planet formation:  water.  The importance of water ice as a carrier of elements to forming planets means that it remains an essential comparative measurement on all four giants.  On Uranus and Neptune, it is locked away at such great depths that we can only infer its abundance indirectly by means of its chemical reactions with measured disequilibrium species like CO, \rev{after having separated stratospheric CO \citep[which can arise from external sources, such as cometary impacts or interplanetary dust,][]{Bezard+2002} from the tropospheric reservoir. \citep{17cavalie, 20venot, 20moses} derive an upper limit to the O/H ratio of 250 times solar in Uranus and a value between 250 and 650 times solar in Neptune, but these values are highly uncertain due to our poor understanding of CO transport.} On Jupiter and Saturn the well-mixed water could be shallower and more readily accessible, but infrared spectral signatures of H$_2$O are only seen over limited domains and oft associated with vigorous dynamics \citep{13sromovsky, Bjoraker+2018, Grassi+2020}.  Hopes to derive Jupiter's bulk water abundance via the Galileo probe were dashed when it entered a dry and desiccated region near the equator \citep[known as a 5-$\mu$m hotspot,][]{98orton}, where water was found to be significantly sub-solar and still increasing at 20 bars \citep{Wong+2004}.  Juno's Microwave Radiometer \citep{17bolton} is capable of measuring water at tens of bars, provided its opacity can be disentangled from the horizontally- and vertically-variable distribution of NH$_3$ \citep{Li+2017}, something which has only been accomplished at the equator so far \citep{Li+2020}, where it is at least solar in abundance near its 6-bar condensation level. Determinations of giant planet bulk-water content remains an active area of research.

Despite the challenges associated with many of the species in Table~\ref{tab:comp}, we note that the noble gases are chemically inert, meaning that the Galileo probe measurements \citep{Atreya+2003, Wong+2004} are expected to be representative of Jupiter's bulk. A notable exception is neon, predicted to dissolve into helium-rich droplets inside Jupiter and Saturn \citep{Roulston+Stevenson1995, Wilson+Militzer2010} and indeed depleted in Jupiter \citep{Wong+2004}. Unfortunately, the absence of \textit{in situ} measurements for any other giant planet hampers attempts at comparative planetology, so noble gas measurements should be a key goal for future giant planet exploration.  
Finally, we note that comparisons of isotope ratios \rev{(see Table~\ref{tab:isotopes})} in common molecules provides further constraint on planetary accretion. These are discussed in \S~\ref{sec:isotopes} hereafter. 
\rrm{The D/H ratio has been determined on all four giants, being close to solar on Jupiter and Saturn \citep{00mahaffy, 01lellouch, Fletcher+2009ch4, Pierel+2017}, and enriched on Uranus and Neptune \citep{13feuchtgruber, 10fletcher_akari, 14orton}.  The $^{13}$C/$^{12}$C ratio has been measured to be uniform (at the terrestrial ratio) across all the giants, whereas the $^{15}$N/$^{14}$N ratio on Jupiter, \rev{$(2.3\pm0.3)\times 10^{-3}$} (and an upper limit for Saturn \rev{$< 2.0\times 10^{-3}$}) is low \citep{Wong+2004, 14fletcher_texes}, implying that $^{15}$N-enriched ammonia ices could not have been a substantial contributor to the bulk nitrogen inventory of either planet, instead favouring accretion of primordial N$_2$ \citep{Owen+2001}. Comparative \textit{in situ} measurements of these isotopic ratios, including those in water, would be an invaluable constraint on planetary accretion models.}

\section{CHARACTERIZING BROWN DWARFS AND GIANT EXOPLANETS}

%

Currently, thousands of brown dwarfs \citep{Best18} and \rev{nearly a thousand\footnote{According to the Transiting Exoplanet Catalogue (TEPCat), there are 874 planets with measured masses and radii, 674 with mass/radius precision better than 20\%.}} exoplanets \citep{Thompson2018} are known. The boundaries between brown dwarfs, self-luminous directly imaged young planets, and close-in transiting planets are mostly artificial as these objects occupy a continuum in irradiation ($0$ to $10^4$ that of the Earth), mass (M$_\oplus$ to $\sim 80$\,M$_{\rm jup}$), rotation periods (hours to days), internal heat flux (none up to 1000s of K), and composition (solar to $\sim$100s $\times$ solar).   Understanding these  atmospheres is key to understanding this continuum and providing clues as to their possible formation avenues.   Most of the focus to date has been on determining the abundances of the major carbon, oxygen, and nitrogen bearing species as well as metals like iron, magnesium, and the alkali's, \rev{both because these were the most easily observable with the current instruments and because} they are thought to be the primary ``heavy element" tracers of planet formation
(see \S\ref{sec:formation}). Below we summarise how we obtain abundance information from the these objects and how we can leverage abundances across this diverse population to provide insight into planet formation processes.

\subsection{Constraining Interior Structure \& Composition}
\subsubsection{Evolution Models}\label{sec:gp_evolution}

As seen in Fig.~\ref{fig:phases}, giant planets and brown dwarfs cover a range of pressures and temperature in which hydrogen is fluid and therefore, compressible. After a first phase of accretion and rapid contraction in which their interior heats up, their contraction continues through a cooling of the interior. For a planet like Jupiter, about half of the planet's gravitational potential energy corresponds to the intrinsic luminosity and is radiated away. This loss also corresponds to a decrease of the thermal energy of the protons. However, the planet's internal energy still increases as it should because the energy of the degenerate electron gas increases by a larger amount \citep{Guillot2005}. For ice giants, coulombian effects and phase changes imply that the compressibility is reduced, but still significant. 

\begin{figure}[tbh]
    \centering
    \includegraphics[width=\linewidth]{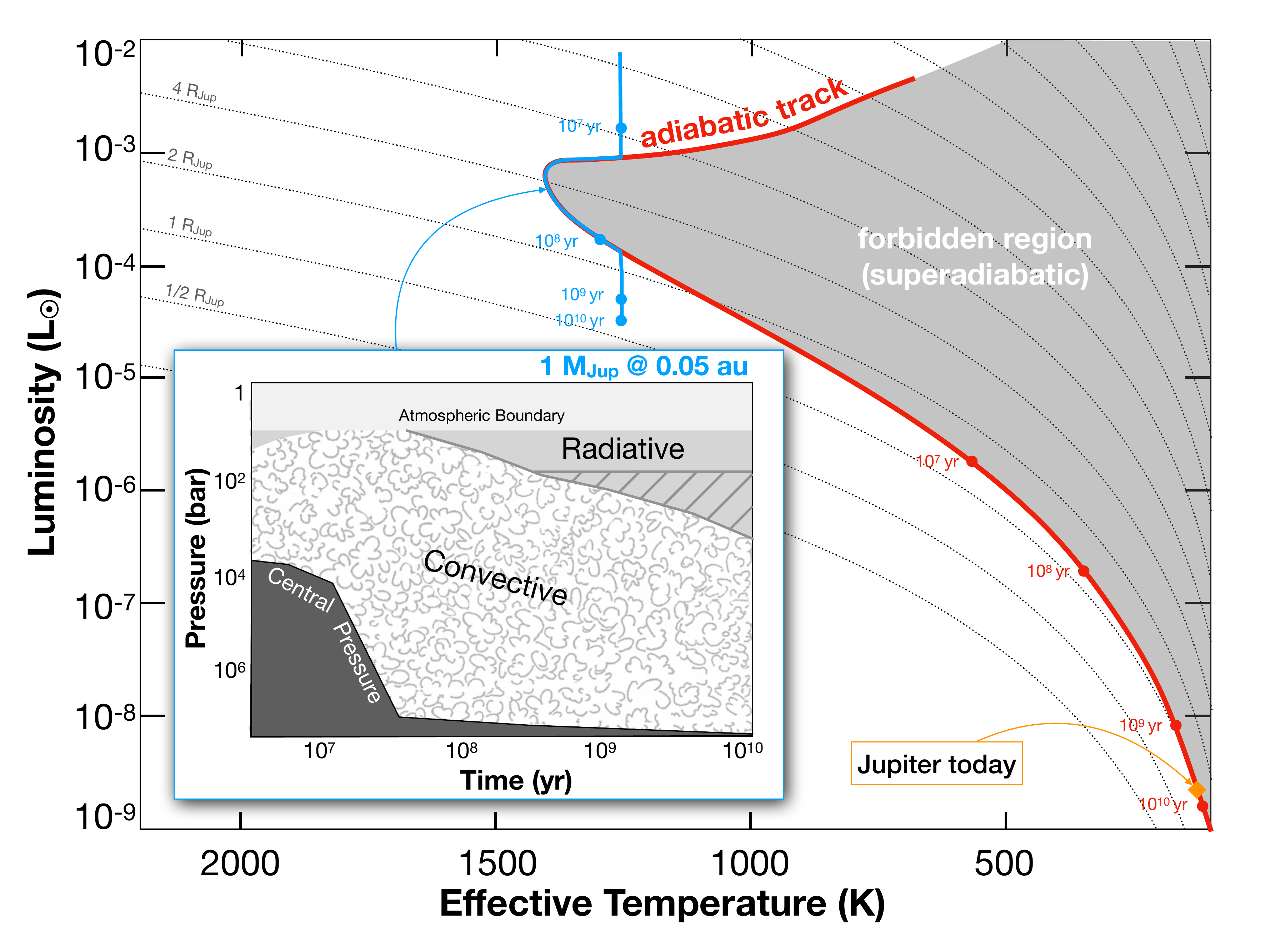}
    \caption{Theoretical Hertzprung-Russell diagram for a $1\rm\,M_{Jup}$ planet in isolation (red) and at 0.05\,au from a solar-type star (blue) \citep[from][]{Guillot+1996}. The effective temperature corresponds to the sum of the planet's intrinsic and absorbed stellar flux. Evolution times are indicated in years (from $10^7$ to $10^{10}\,$yr after an arbitrary initial condition). Dotted curves indicate planetary radius. {\it Inset:} Kippenhahn diagram \citep[see][]{Kippenhahn+Weigert1990} showing the evolution of the interior structure of the $1\rm\,M_{Jup}$ at 0.05\,au from its star and the growth of the inner radiative zone \citep[from][]{Guillot+Showman2002}. \rev{The radiative-convective zone boundary may be limited because of heat dissipation, in which case the hashed region should remain convective. Alternatively, the radiative zone may be extended but less statically stable due to wind-driven downward energy advection (see text).} }
    \label{fig:hrdiag}
\end{figure}

In this regime, the opacities are high, the release of primordial heat due to the loss of internal entropy is progressive, implying that, in isolation, substellar objects are fully convective \citep{Burrows1997, Baraffe+1998}. But irradiation effects can alter this picture: When the intrinsic luminosity becomes of the same order or smaller than the irradiation luminosity, as shown in Fig.~\ref{fig:hrdiag}, a radiative zone must develop below the photosphere in order to allow the planetary interior to continue to cool \citep{Guillot+1996}. 

Given an approximate age, and a measurement of the mass and radius, given appropriate evolution models, it is therefore possible in principle to provide constraints on the bulk composition of giant planets and brown dwarfs. Cooling tracks for these objects \citep[e.g.,][]{Fortney+2007} are essentially defined by the EOSs, atmospheric properties, and radiative opacities. The presence of a central heavy element core leads to a planet being smaller compared to the same planet without a core. Enriching the envelope uniformly however leads to opposing effects: On one hand, the increase in mean molecular weight tends to increase the planetary density. On the other hand, this is opposed by an increase in the opacity in the radiative zone which slows the cooling. It also leads to a slightly change in atmospheric properties. Models show that for highly irradiated planets, the opacity effect dominates at early times, in the first Gyr or so, and then at later times, molecular weight effects begin to dominate \citep{Guillot2005,Baraffe+2008}.

An outstanding issue however is that a significant fraction of known hot-Jupiters (i.e., with orbital periods closer than about 10 days around solar-type stars) are oversized compared to theoretical predictions \citep{Bodenheimer+2001,Guillot+Showman2002}. Many possibilities have been proposed to solve this, including, tides \citep{Bodenheimer+2001}, downward kinetic energy transport \citep{Guillot+Showman2002}, convective inhibition \citep{Chabrier+Baraffe2007}, \rev{enhanced atmospheric opacities \citep{Burrows+2007},} thermal tides \citep{Arras2010}, ohmic dissipation~\citep{Batygin2010} or wind-driven downward advection of energy~\citep{Youdin+Mitchell2010,Tremblin+2017ApJ}. \rev{Statistical studies show that the effect is more pronounced for planets with equilibrium temperatures around $T_{\rm eq}\sim$2000\,K \citep{Thorngren+Fortney2018,Sarkis+2021}. This dependence on equilibrium temperature rules out tides, convective inhibition, and enhanced opacities, favouring instead mechanisms that tap of order $\epsilon\sim 1\%$ energy from the stellar irradiation reservoir \citep[see][]{Guillot+Showman2002}. The fact that models with $\epsilon$ increasing from $\sim$0 for $T_{\rm eq}\sim 1000$\,K to $\sim$3\% for $T_{\rm eq}\sim 1500-2000$\,K and decreasing past that value are strongly favoured statistically \citep{Thorngren+Fortney2018,Sarkis+2021} points to the importance of magnetic drag that leads to slower atmospheric winds on highly irradiated planets \citep{Perna+2010,Menou2012,Ginzburg+Sari2016}. Thus, mechanisms involving Ohmic dissipation \citep{Batygin2010}, wind-driven downward advection of energy \citep{Youdin+Mitchell2010,Tremblin+2017ApJ, Sainsbury-Martinez2019} or thermal tides \citep{Arras2010} appear favoured.}

\rev{The downward energy transport that results has consequences for the extent and structure of the interior radiative zone in hot-Jupiters.} While standard models predict it to grow to reach $\sim $kbar levels in a few Gyr (see Fig.~\ref{fig:hrdiag}), \rev{evolution models aimed at reproducing observable constraints} imply that the extent of this radiative zone should be \rev{limited to a few hundred bars for $T_{\rm eq}\lesssim 1000$\,K to a few bars for $T\gtrsim 2000$\,K \citep[][]{Thorngren+2019b,Sarkis+2021}. However, for wind-driven downward energy advection, the entire temperature profile of the radiative region would be affected \citep{Tremblin+2017ApJ}, with the possibility of a smaller static stability but deeper radiative-convective boundary than for a heat deposition at greater depth}.  

For temperate giant planets, the most important effects to be considered in light of the new developments for solar system planets are (1) changes to the atmospheric structure and (2) consequences of possible compositional gradients. As seen from Fig.~\ref{fig:phases}, phase changes in the deep interior should affect only a small fraction of the known population of giant exoplanets, those with relatively small masses and low irradiation temperatures. However, the condensation of elements should be taken into account, particularly for objects with effective temperatures below 375\,K which begin condensing water \citep{Morley+2014}. Finally, deep compositional gradients are acting in two ways: By suppressing convection, they can store heat and release it at a later time \citep[e.g.][]{Chabrier+Baraffe2007,Leconte+Chabrier2012}. By leading to upward mixing (core erosion) they also affect the energy balance \citep{Guillot+2004,Moll+2017}. These effects can potentially modify the planetary evolution, something that is still to be examined. 

\subsubsection{Inferring Bulk Abundances}

Given an evolution model, the total (bulk) amount of heavy elements in a giant planet or a brown dwarf can be inferred from the knowledge of its age, mass, and either radius or luminosity. In practice, measurements of luminosities can only be made for brown dwarfs and bright planets which therefore must be massive and/or young. In these cases, such constraint is presently very difficult to obtain becasue of the small relative mass of a potential core, the uncertainties on the initial entropy and/or on the atmospheric properties.  

Transiting planets have therefore been the target of choice to obtain bulk metallicities. For simplification purposes, models have usually assumed that heavy elements are entirely embedded into a central dense core (see \S~\ref{sec:gp_evolution} for details). The technique was first applied to hot Jupiters, by arbitrarily transporting a fraction $\sim 1$\% of the irradiated energy to the planetary interior in order to account for the inflation problem \citep{Guillot+2006}. This assumption however can be lifted by limiting the ensemble to the cooler transiting planets for which the inflation mechanism becomes negligible \citep{Thorngren2016}. The results consistently highlight a great diversity of bulk compositions, with planets which can have very low to very high masses in heavy elements \citep[e.g.,][]{Guillot+2006,Burrows+2007,Moutou+2013,Thorngren2016}. \rev{Whereas a correlation between the heavy element content of planets and the metallicity of their host star was found for hot Jupiters \citep{Guillot+2006, Moutou+2013}, it was not confirmed in a sample of 24 well-characterized warm Jupiters \citep{Teske+2019}. }

\begin{figure}[tbh]
    \centering
    \includegraphics[width=\linewidth]{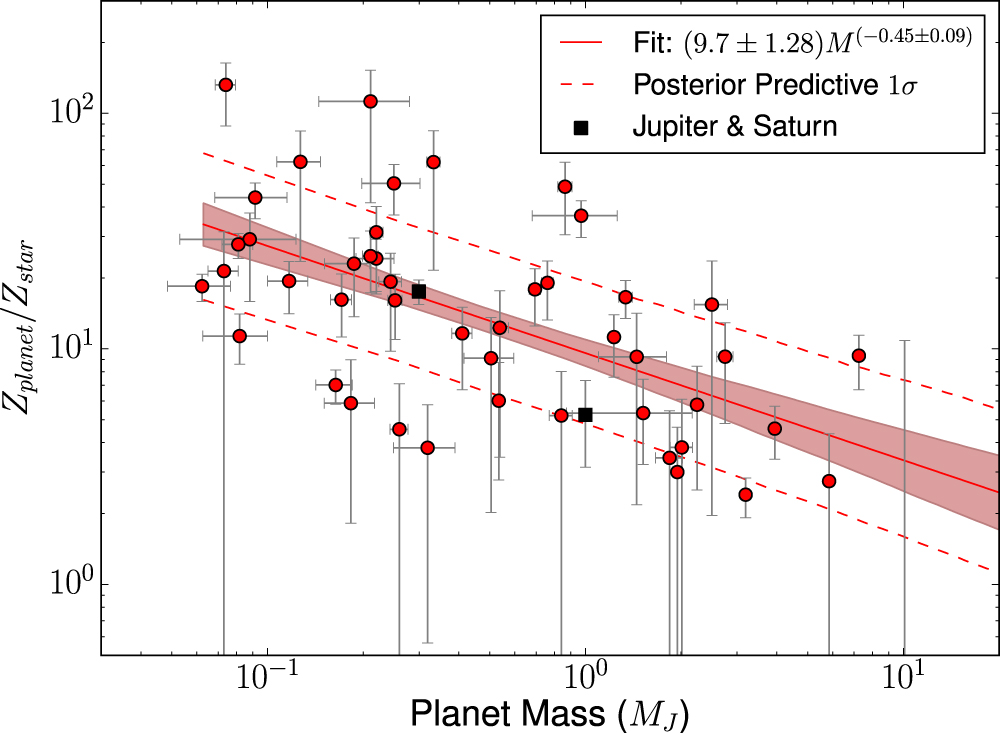}
    \caption{Heavy element enrichment of non-inflated transiting planets relative to their parent stars as a function of mass. The line is the median fit to the distribution from bootstrapping, with 1, 2, and $3\sigma$ error contours. Jupiter and Saturn are shown in blue, from Guillot (1999). [From \citet{Thorngren2016}]}
    \label{fig:thorngren+2016}
\end{figure}

A robust inverse correlation shown in Fig.~\ref{fig:thorngren+2016} links the giant planets heavy element content and their mass \citep{Thorngren2016}. This inverse relation, expected in the framework of core-accretion formation models, 
indicates a bulk enrichment of $\sim 30$ for $0.1\rm\,M_{Jup}$ planets decreasing to $\sim 3$ for $10\rm\,M_{Jup}$ planets. It is important to caution that the spread in that relation is very large: In Fig.~\ref{fig:thorngren+2016}, some $0.1\rm\,M_{Jup}$ planets end up much less enriched than some $\sim 1\rm\,M_{Jup}$ planets. While massive planets tend to be relatively less heavy-element rich than their lighter counterparts, the dominant feature remains the very large diversity of bulk properties for any given planetary mass. 

\subsubsection{Towards Exoplanetary Core Masses}

In very special circumstances, one can obtain a constraint on the interior structure of an exoplanet. This is the case when the Love number of a planet, i.e., the proportionality relation between an applied tidal potential and the induced field at the surface of the planet, can be measured through its apsidal precession \citep{Ragozzine+Wolf2009}. This link requires a system of two planets orbiting a central star in a so-called fixed-point eccentricy configuration. \cite{Batygin+2009} and \cite{Mardling2010} show that the second Love number $k_2$ of the inner planet can be determined if (i) the mass of the inner planet is much smaller than the mass of the central body, (ii) the
semimajor axis of the inner planet is much less than the
semimajor axis of the outer planet, (iii) the eccentricity of the
inner planet is much less than the eccentricity of the outer
planet, (iv) the planet is transiting, and (v) the planet is
sufficiently close to its host star, such that the tidal precession is
significant compared to the precession induced by relativistic
effects \citep{Buhler+2016}. 

The HAT-P-13 system is the first and only currently known system to fulfill these criteria. It consists of a central $1.3\rm\,M_\odot$ G-type star, and two planets, HAT-P-13~b with a mass of $0.9\rm\,M_{Jup}$, a radius of $1.5\rm\,R_{Jup}$ and an orbital period of 2.91 days, and the outer HAT-P-13~c which has a minimum mass of $14.2\rm\,M_{Jup}$, an orbital period of 446 days and an eccentricity of 0.66 \citep{Bakos+2009, Winn+2010, Southworth+2012}. An outer massive companion lies between 12 and 200\,au \citep{Winn+2010}. Observations of secondary eclipses of HAT-P-13~b lead to an eccentricity $e_{\rm b}=0.007\pm 0.001$, a value of the Love number $k_2=0.31_{-0.05}^{+0.08}$ leading to a core mass that is $<25\rm\,M_\oplus$ (68\% confidence interval) \citep{Buhler+2016}. 
For comparison, on Jupiter, gravity fields measurements indicate a Love number associated with Io's tide $k_{22}=0.565\pm 0.006$ \citep{Durante+2020}, slightly lower than static model predictions \citep{Wahl+2016}, but in agreement with theory when accounting for the Coriolis acceleration \citep{Idini+Stevenson2021}. In Saturn, astrometric constraints indicate that $k_2=0.390\pm 0.024$ \citep{Lainey+2017}, in line with model predictions \citep{Wahl+2017b}. 

The possibility to measure core masses is extremely interesting. The difficulty with the fixed-point eccentricity configuration is that it is rare and an extremely accurate determination of the eccentricity of the inner planet is needed. 
Future determinations may rely instead on extremely accurate ($<50$\,ppm/min photometric precision) determination of planetary shape \citep[see][]{Akinsanmi+2019}. This method is possible with space-based photometry only, but is not limited to systems in a fixed-point eccentricity configuration.

\subsection{Atmospheric Abundances from Spectra}

\begin{figure*}[htb!]
    \centering
    \includegraphics[width=\linewidth]{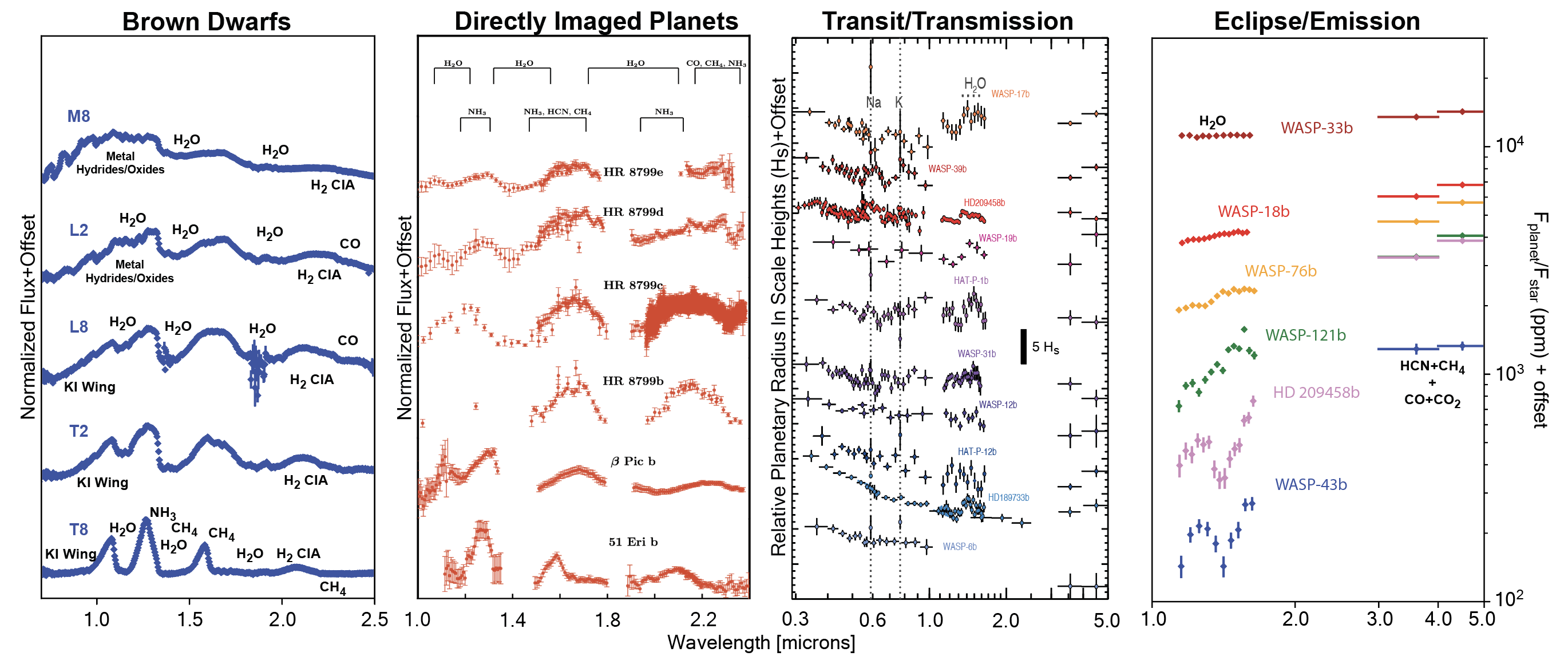}
    \caption{Summary of typical spectra obtained, from left to right, for brown dwarfs (from the SpeX Prism Library), directly imaged self-luminous exoplanets (from \cite{Madhu2019}), and transiting planets in transmission \citep[from][]{Sing2016} and in eclipse/thermal emission \citep{Mansfield2021}.  The spectra of all objects are carved out by various atoms, molecules, and grain opacity. We can leverage these opacity sources to extract meaningful abundance information like atmospheric metal enrichment and elemental abundance ratios.}
    \label{fig:spectra_summary}
\end{figure*}

Classic stellar spectroscopy involves measurements of the self-luminous thermal emission from an isolated (field) star.  Equivalent-width analyses or data-model spectral synthesis tools are used to extract precise constraints ($<$0.1 dex) on elemental abundances while accounting for uncertainties in the \rev{stellar} effective temperature and gravity \citep{Asplund2005}.  Like stars, spectroscopic characterisation (Figure \ref{fig:spectra_summary}) is the primary tool by which we can answer fundamental questions regarding the intrinsic composition of these objects and their possible formation history \citep{Madhu2019, Kirk05}. However, most sub-stellar objects are typically of temperatures much lower than stars ($<$3000K)--abundance analyses must leverage both atomic and molecular opacity sources--classic stellar abundance analyses cannot be used.  

Instead, model dependent parameter estimation methods (``atmospheric retrievals") are employed to extract the pertinent, often degenerate, information. These entail a forward model that takes in as parameters the vertical temperature-pressure profile, numerous abundance parameters, and any other nuisance/process parameters. This forward model is combined with a parameter estimator, like Markov chain Monte Carlo (see \cite{Madhu18} for a review of retrieval methods) to fit the data and to find the optimal parameters and corresponding uncertainties. The resulting ``posterior-probability distribution" is where any abundance information is extracted and processed for subsequent analyses. 

\begin{figure}
    \centering
    \includegraphics[width=\linewidth]{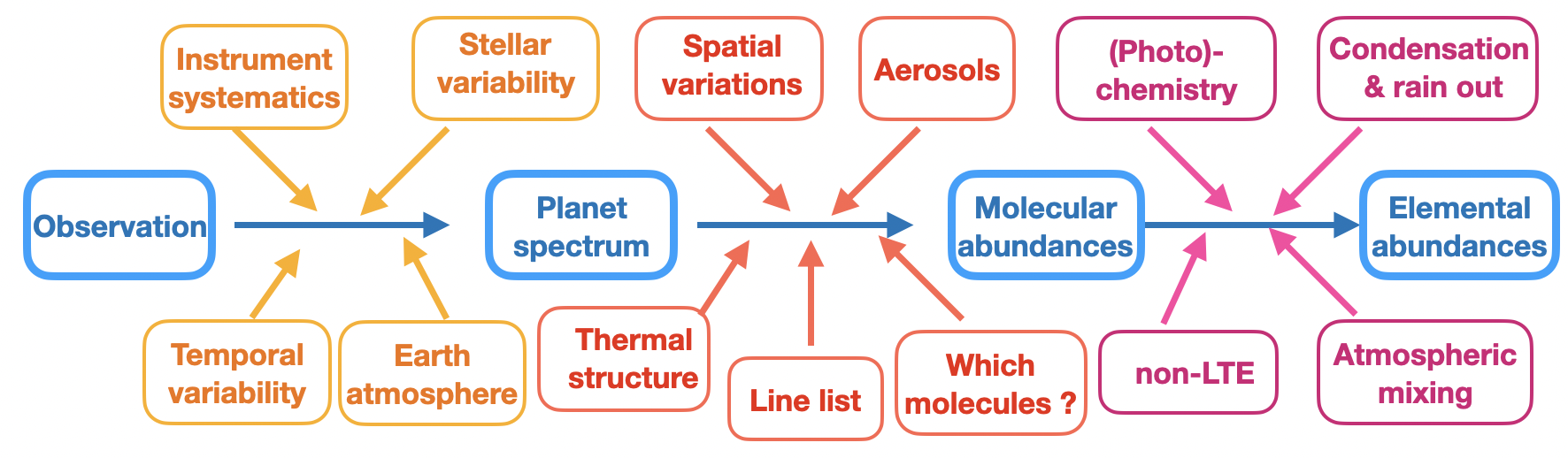}
    \caption{Flow chart showing the different steps needed to go from an exoplanet observation to an atmospheric elemental abundance measurement.}
    \label{fig:complicating_effects}
\end{figure}

The cooler temperatures and broad diversity in bulk properties of the planet/substellar population drives numerous transitions in atmospheric chemistry, dynamics, chemistry, and radiative processes. This diversity, and without the luxury of in-situ orbiters or probes, provides numerous challenges in decoupling the effects of planetary processes from intrinsic properties like envelope composition.  Disentangling these effects (Figure \ref{fig:complicating_effects}) within these retrieval methods is critical to ascertaining unbiased elemental abundances and their ratios--the key quantities needed to infer a planets formation history. Below we summarise the key challenges and recent results in determining brown dwarf and giant exoplanet atmospheric abundances.   

\subsubsection{Brown Dwarf Abundances}
Brown dwarfs (M$<$80M$_J$) provide a control sample for understanding the transition from ``stellar" to ``planetary" atmospheres.  They are presumed to form like stars \citep[see][e.g.,]{Whitworth2007}, but unlike stars, they do not fuse H into He and they possess molecular dominated atmospheres (Figure \ref{fig:spectra_summary}, left) like planets. Under this assumption, it is expected that field brown dwarf elemental abundances should reflect the local stellar population.  In effect, brown dwarfs are like non-irradiated planets, which greatly simplifies their observations (their spectra are not washed out by any host star) and the interpretation of their spectra.  However, due to their molecular dominated atmospheres, extracting the atmospheric metallicities and abundance ratios is challenging. 

\begin{figure}[htb!]
    \centering
    \includegraphics[width=\linewidth]{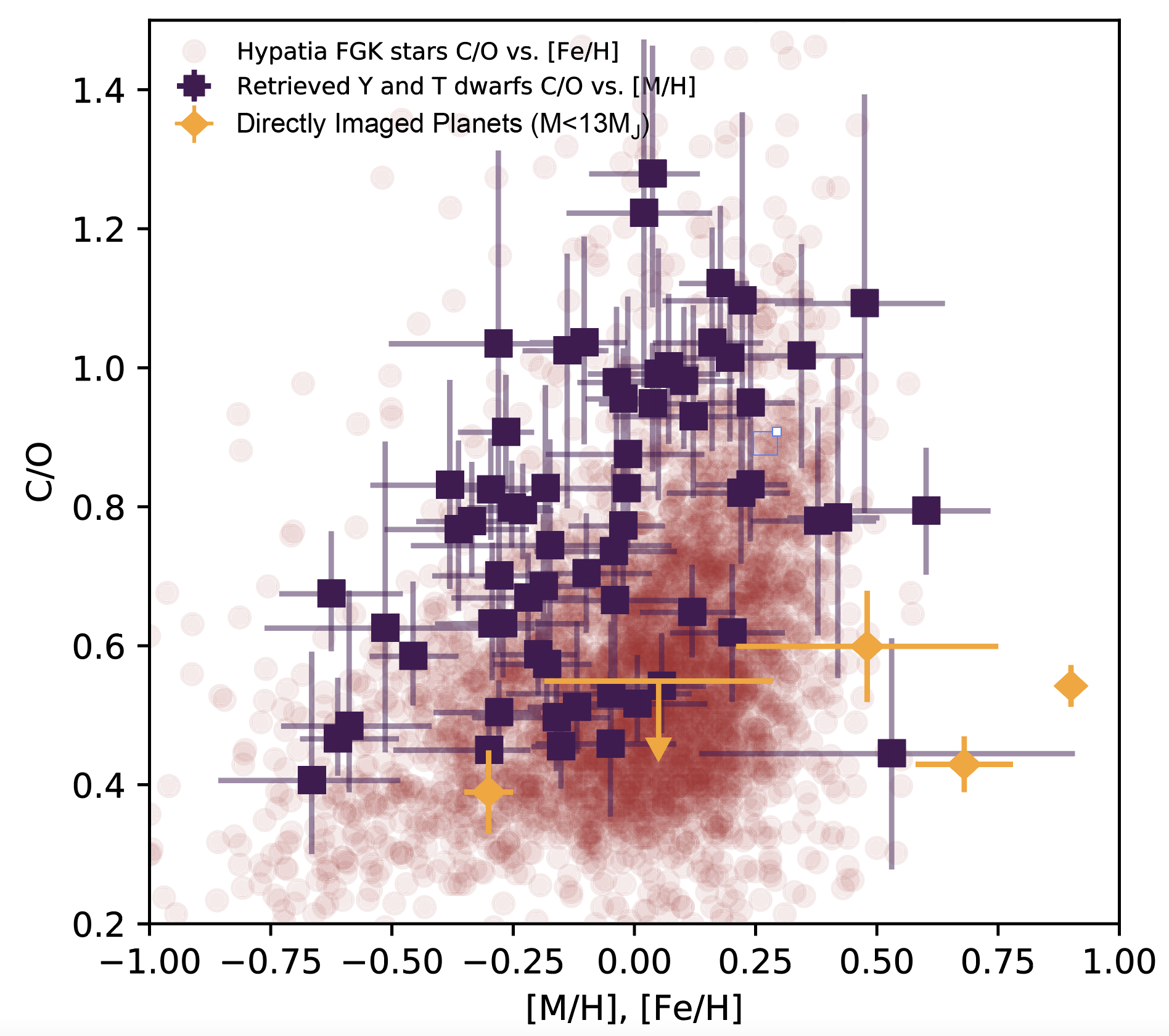}
    \caption{Brown Dwarf  (dark blue) and directly imaged planet abundances (orange) compared to those derived for stars (red, \cite{Hinkel2014}). The brown dwarf abundances here are derived from a uniform atmospheric retrieval analysis \citep{Line17, Zalesky19} applied to $\sim$50 late T and Y-dwarf low-resolution (R$\sim$120) spectra (Zalesky et al. in prep) to derive the atmospheric metallicity ([M/H]) and carbon-to-oxygen ratio. The directly imaged planet abundances are derived from a variety of sources given in the text. Such analyses on brown dwarfs provide a control sample for interpreting the abundances of exoplanets.}
    \label{fig:bd_abund}
\end{figure}

\cite{Line15,Line17} and \cite{Zalesky19} leveraged the aforementioned atmospheric retrieval methods to apply a uniform retriaval/abundance analysis on homogeneous near-infrared spectra\footnote{SpeX Prsim Spectral Library:\\ http://pono.ucsd.edu/{\texttildelow}adam/browndwarfs/spexprism/html/tdwarf.html} like those shown in Figure \ref{fig:spectra_summary}, left, to obtain constraints on the primary molecular constituents while taking into account the correlations/degeneracies between gas abundances, vertical thermal structure, gravity, radius, and instrumental artefacts.  These works specifically focused on the cooler late-T-type brown dwarfs as they are observed (and predicted) to be largely cloud free \citep{Burrows1997, Allard2001, Stephens2009} and the major carbon and oxygen bearing species are fairly homogenised in altitude, enabling a more pristine, less degenerate, abundance analysis.  At these cooler temperatures ($<$800K) water, methane, ammonia, and potassium are the dominant trace gas constituents and hence, sculpt the dominant spectral features.  From the abundance constraints on water and methane--the dominant oxygen and carbon-bearing species at these temperatures--the atmospheric metallicity ((CH$_4$+H$_2$O)/(M/H)$_{\odot}$) and carbon-to-oxygen ratios (CH$_4$/H$_2$O) can be derived.  Figure \ref{fig:bd_abund} summarises these constraints on a sample of over 50 late T- and Y-dwarfs compared to the abundances from the local stellar population. Metallicity precisions between 0.2-0.5 dex and C/O constraints between 0.1 - 0.3 are readily achievable with low spectral resolution, high signal-to-noise, near-infrared spectra.  The spread in abundances is comparable to the stellar population, but with an overall ``offset" to higher C/O and lower [M/H] in the brown dwarf population. \rrm{This offset is possibly due to the influence of refractory condensation deep in the atmosphere which acts to sequester oxygen.}
\rev{The condensation of silicates in the deep layers of the atmosphere provides a good explanation for this offset. Indeed, as shown by~\citep{Burrows1999}, the condensation of enstatite (MgSiO$_3$ and forsterite Mg$_2$SiO$_4$ should take between 2 and 3 oxygen atoms per magnesium atom out of the atmosphere. For solar abundance ratio of magnesium and oxygen, silicate condensation is therefore expected to decrease the number of oxygen atoms by 15 to 20$\%$ and thus increase the C/O ratio from, for example, 0.6 to 0.75. The condensation of other species, such as Fe$_2$O$_3$, if they indeed form, could increase the C/O ratio further to 0.85. Which condensates actually form in brown dwarfs and exoplanet atmospheres and in which quantities they form is, however, complicated to predict from first principles as both kinetics and microphysics processes can alter the predictions from chemical equilibrium. }

The current focus is shifting towards the more data-rich, hotter, L-dwarfs \citep{Burn17,Gonzalez2020} of which present numerous additional absorbers, including CO, and metal hydrides/oxides enabling constraints on more elements.  However, the hotter temperatures present numerous complicating factors including non-uniform vertical abundances and uncertain cloud opacities which can bias abundance determinations.  Nevertheless, once these complications are understood, similar uniform abundance analyses/determinations as show in Figure \ref{fig:bd_abund} can be applied to a much larger sample ($\sim$1000) of objects providing an invaluable control sample for understanding the gradient between star and planet formation.       

\subsubsection{Directly-Imaged Giant Planet Abundances}



``Directly imaged planets" generally refer to self-luminous, young, typically high mass planets ($\sim$few - 13M$_J$) that are observed at wide separations ($\sim$10-100\,au) from their, typically young, host star. Coronographic techniques that suppress the scattered star light are needed in order to detect the oft less than 1 part in 10,000 thermal glow.  Their masses straddle the classic evolutionary defined boundary between “planet” (e.g., low entropy, rocky cored) and “brown dwarf” (high entropy, degenerate) \citep{Burrows1997}. They provide unique tests of planet formation theories as both brown dwarf-like and solar system-like formation mechanisms (see \S \ref{sec:formation}) can explain their masses and orbital properties.  Determining their atmospheric abundances and comparing them to those obtained from the brown dwarf population and the transiting exoplanet population (more below) can shed insight into their possible formation avenues.  We refer the reader to Chapter XX (Currie et al.) for more details on the characterization of these worlds, we discuss them here in the context of their abundances.

Unfortunately, owing to the challenges in obtaining contrasts below 10$^{-6}$ and the very likely low occurrence rates of such types of planets ($\sim$10\% for 5-13M$_J$ at 10-100\,au around $>$1.5M$_{\odot}$ stars, \cite{Nielsen2019}), only about ten have been spectroscopically characterised. Their spectra (Figure \ref{fig:spectra_summary}) are similar to those of brown-dwarfs suggesting that they possess similar opacities.  While several key carbon and oxygen bearing molecules like CO, water, and methane have been been detected (see~Table \ref{Table:Detection_table}), determining their abundances has been challenging owing to their cloudy atmospheres and lower signal-to-noise spectra than their more readily observable brown dwarf cousins. Modelling assumptions tend to influence the reported abundances, usually derived from a combination of atmospheric retrieval modeling that provide direct constraints on the molecular abundances \citep{Lavie2017, Wang2020} and 1D radiative-convective model grid fits \citep{Konopacky2013, Barman2015}, of which typically assume thermochemial equilibrium molecular abundances, and instead fit directly for the metallicity and C/O. Considering only the directly imaged planets with masses less than 13$M_{J}$, there are only 5 objects with reported metallicity and C/O constraints (\cite{GRAVITYCollaboration2020, Molliere2020,Wang2020, Petrus2021} and Chapter XX)).  These constraints are shown in comparison to the stellar and brown-dwarf populations in Figure \ref{fig:bd_abund}. With only five objects, it is difficult to readily discern any clear patterns/differences between the imaged planet population and the brown dwarfs--perhaps the imaged planet population is slightly metal enriched with somewhat lower carbon-to-oxygen ratios, though a much larger sample size is needed to make any confident statements.

However, it is worth being \rev{wary} of the actual uncertainties on these abundance constraints due to the model assumptions used to derive them (e.g., see \cite{Wang2020}). Furthermore, the brown dwarf abundances were largely derived directly from the water and methane abundance constraints where-as most of the directly imaged planet constraints arise from the self-consistent grid fitting, with the latter typically resulting in artificially precise constraints owing to the radiative-convective-thermochemical equilibrium assumption within the models.  

\subsubsection{Transiting Planet Abundances}
Transiting planets \rrm{(about 1000 with measured radii and masses)} make up most of the charcterised exoplanet population, with thousands of Hubble orbits and thousands of Spitzer and ground-based telescope hours spent observing many dozens of atmospheres ranging from small cool terrestrial worlds to planets with hellish star-like atmospheres. It is this planet population that has driven most of the exoplanet-atmosphere literature.

\begin{table}[htb!]
    \centering
    \includegraphics[trim=70 70 70 70,clip,width=1\linewidth]{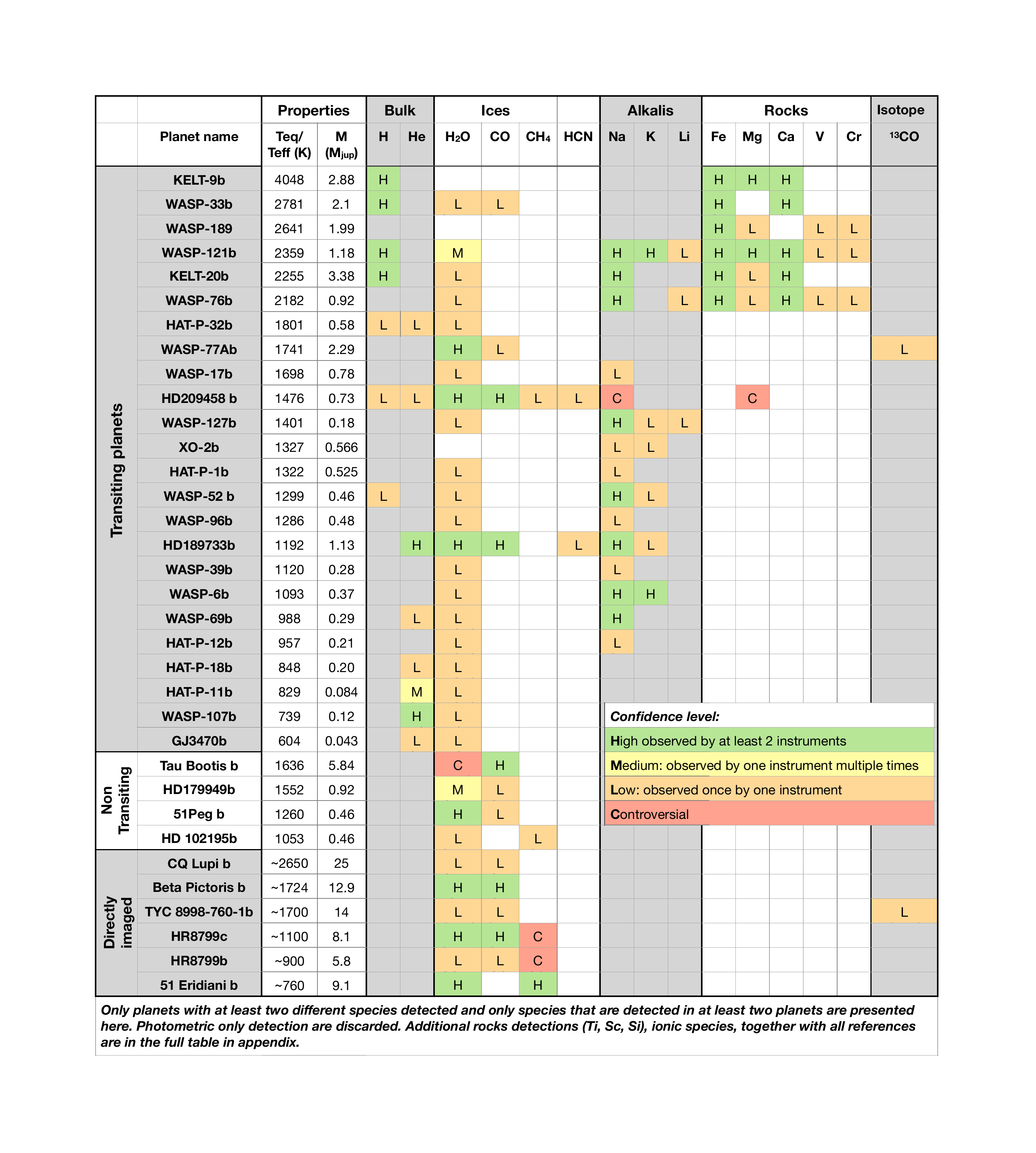}
        \caption{\rev{Chemical species detected in exoplanet atmospheres as of March 2022. We included only planets where more than 2 species have been detected and species that have been detected in at least two planets. We did not include species detected via photometry alone. Additional rock species (Ti,Sc,Si), ionised and non-ionised species and references are provided in the Appendix (see Table~\ref{Table:Detection_table_full})}}\label{Table:Detection_table}
\end{table}

In contrast to brown dwarfs and directly imaged planets, due to their close-in orbits \rrm{(typically much less than 1\,au)},  the spectra of transiting planets cannot be measured directly, but rather, relative to their host stars. The wavelength-dependent opacity structure can be measured either when the planet passes behind its' host star, an eclipse (an emission spectrum, or planet-flux/stellar-flux) or when the planet passes in front of its' host star, a transit (transmission spectrum,  the planet-to-star area or radius ratio) \citep{Deming2019} (Figure \ref{fig:spectra_summary}, right two panels). Most transiting exoplanet atmosphere characterization stems from observations with the Hubble and Spitzer Space Telescopes.  An alternative approach leverages the high orbital velocity of the planets, due to their short periods, which a time-dependent Doppler shift of the planetary molecular/atomic lines relative to the stellar and telluric lines enabling the use of large aperture ($>$ 6 m), high spectral resolving power (R$>$25,000), ground-based platforms \citep{Birkby2018}.  This approach enables the characterisation of short-period non-transiting planets as well.  As with the spectra of brown dwarfs and directly imaged planets, information about the abundances can be ascertained through various atmospheric retrieval \citep{Madhu18} methods applied to the eclipse, transit, or high resolution Doppler spectra \citep{BrogiLine2019}. Below we summarise the current state of species detections, and subsequent abundance determinations within the transiting planet population.

Table ~\ref{Table:Detection_table} summarises the various gaseous species detected in exoplanet atmospheres, with transiting planets, owing to their broad diversity in temperature and size, making up a bulk of the detections. Whether a given chemical species is detected in an atmosphere is determined by both by the intrinsic presence of that species as well as the instrumental set up (wavelength range, sensitivity). We separate the detected species into four categories: rocks, or elements that condense into solids for temperatures $<$1000K, alkalies, which have condensation temperatures between 500 and 800K, and finally, ices, which condense below 300K.

Water is ubuiquitous as it has been detected in most planetary spectra~\citep{Deming2013,Crouzet+2014,Sing2016} observed with the Hubble Space Telescope. Due to the low resolution and relatively narrow wavelength coverage of the HST WFC3 instrument (1.1-1.4$\mu$m), no other molecular species beyond water have been reliably detected.  Through the high-resolution ground based method, which can cover a broader range in wavelengths, carbon monoxide has been seen in only two transiting planets \citep{Snellen2010, deKok2013}. Methane has yet to be reliably detected in all but a single planet, the hot Jupiter HD209458b \citep{Giacobbe2021} and has yet to be found in cooler ($<$1000K) worlds \rev{\citep{Stevenson2010,Kreid18, Benneke2019}
}
where it is expected to be more thermochemically prominent \rev{\citep{Moses2013} (as is seen in similar temperature brown dwarfs, \S 3.2.1)}. Nitrogen bearing species have remained equally as elusive. That is because N$_2$, the main nitrogen bearing species in hot planets, does not have strong spectral lines. In cooler planets, NH$_3$ becomes the main bearer of nitrogen but has so far been detected only in a single hot planet, strongly out of chemical equilibrium\citep{Giacobbe2021}. HCN, thought to be a tracer of high carbon-to-oxygen ratios and/or photochemical processes has also been challenging to detect, but has been suggested to be present in at least one hot Jupiter \citep{MacMadhu2017, Hawker2018, Giacobbe2021}

Alkali metals have broad absorption lines in the optical that makes them observable at low spectral resolution from space \citep{Sing2016} and at both low and high resolution from the ground~\citep{Huitson2017, Seidel2019, Nikolov2016}. They are expected to condense out of the atmosphere around 800K and thus should not be observable in cooler planets. 

\rev{More refractory material (``rocks") can be observed in the hottest ($>$2000 K) of planets. In these ultra-hot Jupiters molecules thermochemically dissociate leaving behind numerous atomic species that can be observed at high spectral resolution~\citep{Lothringer2018,Parmentier2018,Kitzmann2018}.  Iron, magnesium, calcium and scandium have been observed in several planets \citep{Vidal-Madjar2013,Yan2019,Ehrenreich2020,Borsa2021,Kesseli2022} whereas others such as yantium, nickel, titanium and chromium have only been seen in KELT-9b, the hottest known exoplanet~\citep{Hoeijmakers2019}. Some of these species are also transported into the exosphere of the planet by the hydrodynamic escape of hydrogen and helium and get ionized. This leads to strong line-contrast for the observed ioninc species, such as Mg+ or Fe+, that can be observed in the near ultraviolet with HST \citep{Sing2019}.}

Hydrogen and Helium have been seen in numerous planets, the former through the large Lyman-$\alpha$ and H$-\alpha$ absorption~\citep{Vidal-Madjar2003} and the second through the metastable Helium line around $1\mu m$~\citep{Spake2018,Nortmann2018}. These are usually seen very high in the atmosphere, up to the roche lobe limit of the planet, and are indicative of the presence of atmospheric escape \rev{\citep[see][for a review]{Owen2019}}.


Measuring precise and accurate abundances, in contrast to pure detection, from current transiting planet spectra is not straightforward. Whereas the detection of a species is driven by the \emph{strength} of the spectral lines, the measurement of the chemical abundances are driven by the \emph{shape} of the spectral lines, which requires a higher signal-to-noise ratio. Again, retrievals are often used to leverage these spectral shapes to pull out quantifiable abundance constraints. However, at the level of precision needed to quantify abundances, several other nuisance phenomenon need to be accounted for within these retrieval methods. These include the presence of horizontal heterogeneities in temperature, chemistry and cloud coverage in the hottest targets, where the day-to-night thermal contrast can reach hundreds to thousands of degrees. Aerosols are always a cause of uncertainties in planetary spectroscopy, but this is particularly true for exoplanets where an enormous variety of condensates~\citep{Marley2015,Helling2021} and photochemicaly produces hazes~\citep{Horst2018,Kawashima2019} can potentially exist, combined with the long path lengths of starlight through the planetary limbs during transit \citep{Fortney2005}. The presence of aerosols can significantly change the scattering properties of the atmosphere and bias the abundance measurement \citep{Deming2013,Taylor2021}, particularly if they cover only part of the atmosphere \citep{LinePar2016}. For cooler planets (often discovered orbiting M dwarfs due to selection biases) the presence of stellar spots containing the very molecules one is trying to measure in the planet (e.g. water) leaves an imprint in the measured transit spectrum that can significantly bias the abundance measurements~\citep{Desert2011,Rackham2019}.


\begin{figure}[htb!]
    \centering
    \includegraphics[width=\linewidth]{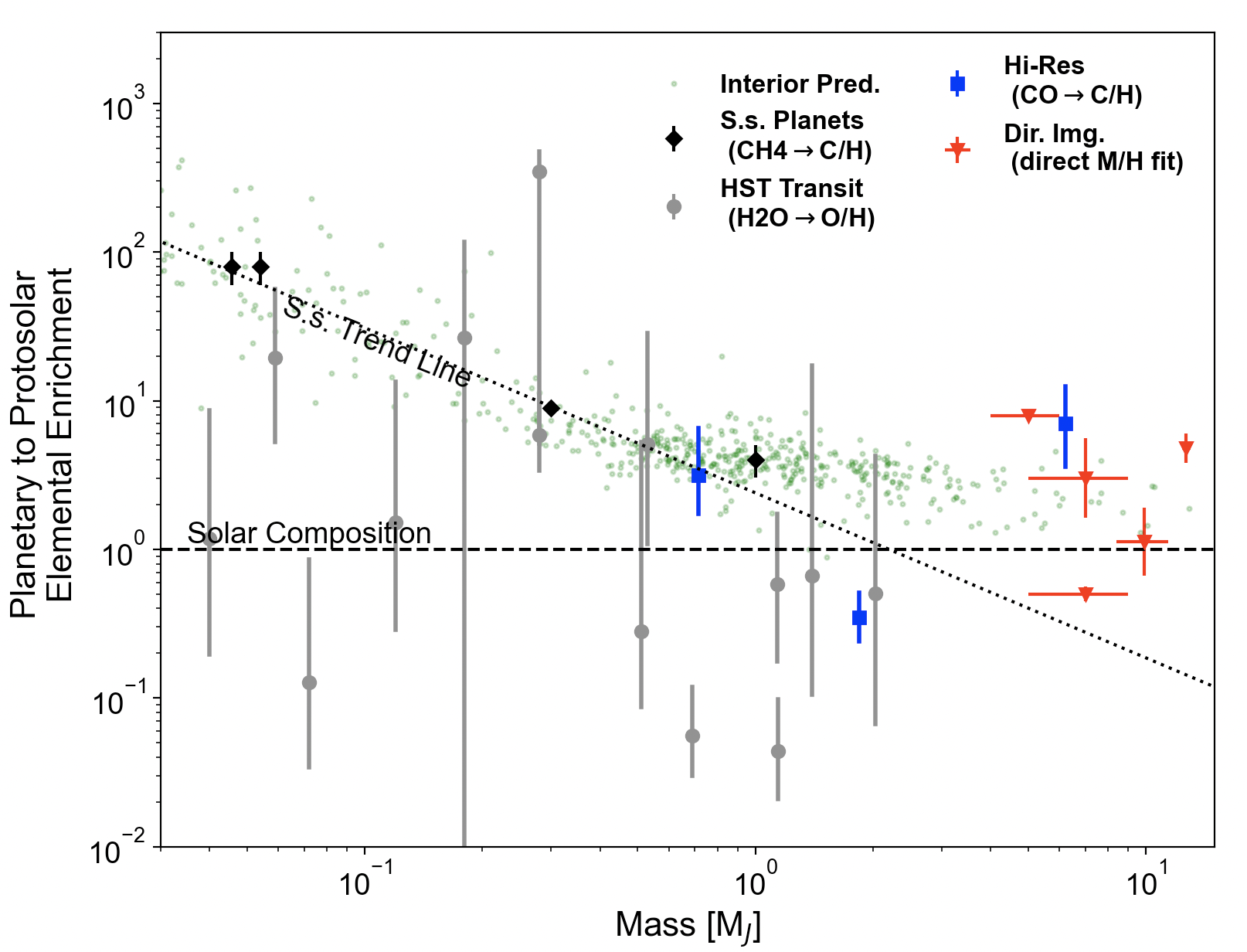}
    \caption{Atmospheric elemental abundance enrichment for select species as a function of planetary mass for the solar system planets (black diamonds), transiting exoplanets observed in transmission with the Hubble Space Telescope (grey circles, based on water/O constraints, \cite{Welbanks2019}), transiting (or near transiting) exoplanets observed with ground-based high resolution spectroscopy (blue squares based on CO/C constraints--\cite[][Line et al. 2021]{BrogiLine2019, Pelletier2021}), and directly imaged self-luminous planets (red triangles, based on direct grid fitting determined enrichments, \cite{GRAVITYCollaboration2020, Molliere2020,Wang2020, Petrus2021} and Chapter XX). The green scatter points are the interior structure based envelope enrichment predictions \citep{Thorngren2016} for the observed exoplanet population assuming a 10/90\% partitioning of metals in the envelope/core--similar to what is assumed for Jupiter. A linear fit to the solar system abundance is shown as the dashed line (see \cite{Kreidberg2014}). As of yet, there is no clear trend in envelope metal enrichment vs. metallicity among the characterised exoplanet population.}
    \label{fig:mass_v_met}
\end{figure}

Despite water being readily detected, there is a large diversity in the signal strength resulting in a very wide range in the retrieved gas mixing ratios (see fig.~\ref{fig:mass_v_met}). Several planets may have sub-solar oxygen abundances \citep{Welbanks2019}, as traced with water, and most planets seem to be less enriched in heavy element species than in the atmospheres of solar-system planets with similar masses, in contrast to what is derived for the population wide bulk planetary enrichments (Fig. \ref{fig:thorngren+2016}). However, whether this inconsistency and the scatter in retrieved abundances is due to a scatter in intrinsic elemental abundances or a uncertainty in some of the other, poorly controlled parameters mentioned above, is not yet understood.

In the near future new instruments such as those on-board the James Webb Space Telescope or high-resolution instruments on the ground (such as IGRINS, ESPRESSO, CRIRES+, MAROON-X, SPIRou) should provide enough spectral information to break some of the common absolute-abundance inhibiting degeneracies. At the same time our theoretical understanding of these atmospheres should increase trust in our prior assumptions, leading to a better mitigation of nuisance parameters and thus more reliable abundance constraints.

\subsubsection{Transiting Planets Abundance Ratios}

Elemental abundance ratios are in theory more readily constrained than absolute abundances due to the lessened sensitivity of these ratios to common degeneracies~\citep{BS2012, Griffith2014}. However, in practice, most instruments have too narrow a wavelength coverage to simultaneously measure the spectral features of multiple molecules and thus rely upon multiple visits with different instruments \citep{Sing2016}. However, as discussed above, the only major volatile species reliably detected with HST is water.  Extracting information on the atmospheric metallicity and other elemental ratios like C/O or N/C, etc., is challenging if only ``O" is being measured owing to the degeneracy between overall metallicity and the ratios themselves (e.g., a high metallicity and high C/O can produce the same O abundance as solar composition, \cite{Moses2013}).

Despite this challenge, there are several broad conclusions, primarily driven by the water/oxygen abundance itself and the lack of presence of expected carbon bearing species (namely, CH4) over the HST wavelength ranges, observed in transmission (where we have the most observations, upwards of two-dozen planets). Firstly, in most hot Jupiters \citep{Tsiaras2018} water (hence oxygen) is abundant enough to present absorption stronger than the obscuring presence of H2-H2/He collision induced opacity, typically requiring values above 1 ppmv. Secondly, the shape of the water absorption feature relative to the collision-induced continuum for many planets is indicative water abundances that are below the stellar (usually solar) values, suggesting a ``depletion" of oxygen \citep{Madhu2014_H2O,Barstow2017, Pinhas2019, Welbanks2019} relative to solar and/or planet formation model atmospheric enrichment predictions (Fig. \ref{fig:mass_v_met}). Whether or not this apparent depletion of O is due to low overall metal enrichment or due to elevated C/O, remains to be seen.  Furthermore, the degeneracy between the water abundance and cloud properties prevents constraints better than $\pm$1dex for most planets.

Nevertheless, inferences of carbon-to-oxygen ratios have been attempted by leveraging the lack-of-carbon species.  Low C/O values have been presumed in two cool ($<$900K) worlds, GJ3470 \citep{Benneke2019} and WASP-107b \citep{Kreid18}, primarily owing to th elack of expected methane over the HST WFC3 pass-band.  These works suggest sub-solar C/O ratios ($<$0.4), however, additional complications in these lack-of-methane based C/O inferences are strongly dependent upon the chemical assumptions (e.g., see \cite{Fortney2020})

Thermal emission/secondary eclipse (which probes the planetary dayside thermal emission) derived inferences are far less common due to the more difficult observational requirements (need hot enough planets to produce an observable spectrum). However, similarly, inferences derived from secondary eclipse observations with Spitzer and HST have provide comparable ambiguity.  The broad-band (3 - 24 $\mu$m) Sptizer photometry observations the warm ($\sim$700K) Neptune, GJ436b \citep{Stevenson2010}, are suggestive of a methane depleted atmosphere (no methane at all detected), interpreted as either being due to disequilibrium \citep{Stevenson2010} chemistry or extreme ($>$300$\times$Solar) metal enrichment \citep{Moses2013}. However, Spitzer photometry alone is not enough to thorughly break this degeneracy. Furthermore, this planet is too cool to be observed in eclipse with HST in the near-IR and the transmission spectrum of this world is relatively featureless \citep{Knutson2014}, providing little further insight into its composition. 

The first evidence of a ``high C/O" planet (C/O$>$1) arose from a combination of a featureless (blackbody) dayside WFC3 emission spectrum and Spitzer photometry of the ultra-hot ($\sim$3000 K) Jupiter, WASP-12b \citep{Madhusudhan2011}. However, this result is entirely dependent upon the reliability of the Spitzer eclipse depths, which are challenging to derive \citep{Cowan2012} due to the oblate shape of the planet, nor did this initial interpretation consider key opacity sources expected to be present at such extreme temperatures \citep{Parmentier2018}, which could also explain the lack of water feature. Furthermore, the HST WFC3 transmission spectrum \citep{Kreidberg2015} largely rules out the high C/O scenario, suggesting a composition consistent with solar. 

Three planets, namely the hot Jupiters WASP-43b \citep{Kreidberg2014}, HD209458b \citep{Line2016}, and HD189733b \citep{Zhang2020} present the most complete emission spectra and produce O 
enrichments that are consistent with solar. Retrievals on the the former two are consistent with a broad range of  oxygen enrichements (0.6-8$\times$Solar, 0.06-10$\times$Solar, respectively), but provide little to no constraint on carbon-species leaving the degeneracy between metallicity and C/O un-broken.  The latter spectrum, HD189733b, is suggestive \citep{Zhang2020} of a metal enrichement of 8-20$\times$Solar and a $0.6<C/O<0.7$ under the assumption of thermochemical equilibrium chemistry (however, see \cite{Lee2012, Line2014} ). Placing unambiguous constraints on the elemental abundance ratios would require the detection of multiple elemental species.

Complimentary to the space-based HST and Spitzer based inferences, high-resolution ground-based observations (see review by \cite{Birkby2018}) have provided an opportunity to detect {\it both} carbon and oxygen, and very recently, to provide constraints on their (3 transiting planets and one non-transiting, see Fig \ref{fig:mass_v_met}) abundances \citep{BrogiLine2019}. Retrievals on the CRIRES K-band dayside emission data for the hot Jupiter's HD~189733b and HD~209458b \citep{BrogiLine2019} produce precise constraints on the CO abundances and stringent upper limits on the water abundance,  \rev{which together point towards C/O$\sim$1. However, these results are potentially at odds with the near-solar constraints obtained from the aforementioned space based HST/Spitzer measurements \citep{Zhang2020, Line2016}.}  

Similarly, recent analysis of the SPIRou \citep{Pelletier2021} observations the non-transiting massive planet, $\tau$ Boo b, placed a stringent upper limit ($\lesssim 10^{-6}$) on water and a very precise ($\pm$0.27 dex) constraint on the CO abundance, resulting in an extremely precise ($\pm$0.01) C/O=1. Very recently, \citep{Line+2021}, using the IGRINS instrument, observed a generic transiting hot Jupiter ($\sim$1700K) WASP-77Ab obtained precise, bounded constraints on both the water and CO abundances, leading to a metallicity of 0.33$\pm$0.11 and a C/O=0.59$\pm$0.08. We anticipate comparable constraints over the next several years as more and more high resolution observations are obtained and analysis procedures developed.




Isotpologue ratios can be measured in \rrm{hot Jupiter} \rev{exoplanetary} atmospheres by resolving the spectral lines of different isotopologues. This has been performed \rrm{uniquely} on carbon monoxide, with the $\rm ^{13}CO/^{12}CO$ ratio measured to be \rev{about} 3 times higher than the solar-system value in both the young \rev{massive planet} TYC 8998-760-1 b \rev{using VLT/SINFONI} ~\citep{Zhang2021} and the hot Jupiter WASP-77Ab~\citep{Line+2021}. \rev{It has also been measured with Keck/NIRSPEC in a young brown dwarf, 2MASS
J03552337+1133437, where it is consistent with the solar value \citep{Zhang2021bd}. }
While there has yet to be a unifying interpretation of these results alone \rev{(see \S~\ref{sec:isotopes})}, isotopic fractionation \rrm{in general is highly sensitive to the temperature within the disk and could potentially} \rev{will} provide an addition dimension to constrain planet formation history. \rrm{ and migration, especially if additional isotopologues (e.g., D/H) are measured.}

Certainly, we cannot discuss all compositional inferences within this chapter, we highlight works that elucidate some of the key issues in the field. Fig. \ref{fig:mass_v_met} is an attempt at summarising the state of the abundances in terms of metal enrichment's of the various planet populations (transiting, directly imaged) and techniques (low res space based, hi-res ground based,direct imaging). We also urge caution when interpreting the ensemble of results as different analysis on different datasets can often result in conflicting conclusions.  As we obtain higher quality data, expected on all fronts, we anticipate a convergence in our overall understanding of atmospheric abundances.

\subsection{Linking bulk and atmospheric abundances}


Whether the observed atmospheric abundances are representative of convective envelope abundances depends on how deep is the radiative/convective boundary situated and how strong is the vertical mixing in the radiative zone. The timescale for mixing elements accross the radiative zone can be estimated as: 
\begin{equation}
    \tau_{\rm mix}=\frac{(nH)^2}{K_{zz}},
    \label{eq::taumix}
\end{equation}
where $nH$ is the distance between the photosphere and the radiative/convective boundary expressed in units of the scale height and $K_{zz}$ is the vertical mixing coefficients.

For Brown Dwarfs and directly imaged planets the photosphere is situated very close to the radiative/convective boundary and no elemental diffferences are expected between what we observe and the convective zone. For close-in planets, however, the strong stellar irradiation leads to the formation of a radiative zone that can extend down to 1000 bars for planets with a Jupiter-like internal flux (\S~\ref{sec:gp_evolution}), which situates the radiative/convective boundary roughly 10 scale heights away from the photosphere. Hot inflated planets likely have a much larger internal flux than Jupiter which would raise the radiative/convective boundary much closer to the photosphere~\citep{Thorngren2019}.      

The vertical mixing in the radiative zone of close-in planets is driven by the strong day-to-night atmospheric circulation induced by the inhomogeneous day/night heating\citep{Parmentier2013,Sainsbury-Martinez2019}, the presence of shear instabilities\citep{Li2010,Fromang2016,Menou2019}, shocks \citep{Heng2012} or gravity waves \citep{Watkins2010}. The strength of most of these mixing processes increase with decreasing pressure~\citep{Menou2019}, meaning that the bottleneck for vertical mixing is likely situated at the radiative/convective boundary itself. An estimate for the mixing timescale accross the radiative zone can therefore be obtained by estimating the time to cross one scale height at the radiative/convective boundary. Whereas estimates from shear instabilities alone lead to values of $K_{zz}$ close to $10$m$^2$/s~\citep{Menou2019} at 100 bars, estimates based on the overturning circulation are much larger, reaching $10^5$m$^2$/s~\citep{Parmentier2013} and possibly higher~\citep{Heng2012,Sainsbury-Martinez2019}. Using equation~\ref{eq::taumix} with a scale height of 500km we obtain that mixing timescale to cross one scale heights should less than $\approx 30$ days for a mixing driven by the overturning circulation but up to $100$ years for a mixing driven by shear instability only, both significantly shorter than the planet evolution timescale. \rev{This comparison indicates that the radiative zone should not be a great barrier to transport and thus that the atmospheric abundances should be a good proxy for the abundances in the convective zone. However these estimates are very uncertain because we do not understand yet the deep atmospheric circulation in hot Jupiter atmospheres}. \rrm{Particularly magnetic drag such as measured by Juno on Jupiter (see \S~\ref{sec:GP_interior}) could efficiently suppress the deep winds~\citep{Rogers2014} and lead to longer mixing timescales.}

Some chemical species expected to be in gaseous phases in the dayside atmosphere can condense on the nightside or in the deep atmospheric layers~\citep{Spiegel2009,Parmentier2013}. Gravitational settling is then in competition with the vertical mixing to determine whether these species are trapped below the photosphere. The efficiency of the gravitational settling depends on the particle size distribution and particularly to the largest possible size of cloud particles. Current models indicate that coagulation play a much smaller role in hot giant exoplanet clouds, due to the low mixing ratio of the cloud condensing materials (e.g. Fe, Mg, Si) compared to the mixing ratio of water in Jupiter for example. As a consequence, clouds in hot exoplanet atmospheres are not believed to grow particles larger than $\approx100\mu m$ \citep{Lee2016,Woitke2020,Powell2019}, which can lead to a significant reduction of the dayside atmospheric abundances compared to local chemical equilibrium~\citep{Spiegel2009,Parmentier2013,Beatty2017a} but not completely deplete the atmosphere of this species \citep{Powell2018}. Such a mechanism has been proposed to explain the apparent lack of TiO in several planets~\citep{Beatty2017a}.

By comparing the atmospheric metallicity measured from the spectra of the planet to the bulk metallicity inferred from its mass, radius and age, one can estimate how heterogeneous is the planet interior~\citep{Thorngren2016,Thorngren2019}. As we show in Figure~\ref{fig:mass_v_met} coreless models with fully mixed interior lead to predicted atmospheric metallicities that are  much larger than the measured one, both in solar system and in exoplanets. Overall, between $50\%$ and $99\%$ of heavy elements must be sequestered below the photosphere, either in the envelope or in the core of the planet, in order to reconcile atmospheric and bulk measurements.

\section{CONSTRAINING GIANT PLANET FORMATION}\label{sec:formation}

Since giant planets are dominated by hydrogen and helium, they must form when the circumstellar gas disk is still present, within a few million years of the formation of the central star. We first present a few theoretical aspects of the formation of giant planets, then examine what we can learn from solar system and giant exoplanets, and provide perspectives for future progress.

The formation mechanism of giant planets is still being investigated. 
There are two main formation models known as "core accretion" and "disk instability" \citep[see][for reviews, and references therein]{2000prpl.conf.1081W, Durisen2007, 2007prpl.conf..591L, 2010exop.book..319D, Helled2014,Nayakshin2017, 2018haex.bookE.140D, 2018haex.bookE.143M,2021arXiv210907790H}. 
In the core accretion model the formation of giant planets begins with the buildup of a heavy-element core. The core is formed via the accretion of solids that can be in the form of planetesimals and/or pebbles. The composition of these solids depends on the formation location of the growing planet due to the existence of various ice lines. 
In the disk instability model, giant planets form via local gravitational instability in the disk which leads to fragmentation. The forming objects must cool down and contract fast enough to remain gravitationally bound. The expected masses and orbital properties of the objects formed via this mechanism are still being debated. Often this formation path is required to explain giant planets around M-stars, massive giant planets at very large radial distances, and massive planets around very young star \citep{Durisen2007,Helled2014}. 
Nevertheless, core accretion is the standard formation  model and the expected mechanism for the formation of the giant planets in the solar system. 


\subsection{The Stages of Planetary Growth}

\begin{figure}[t!]
    \centering
    \includegraphics[width=\linewidth]{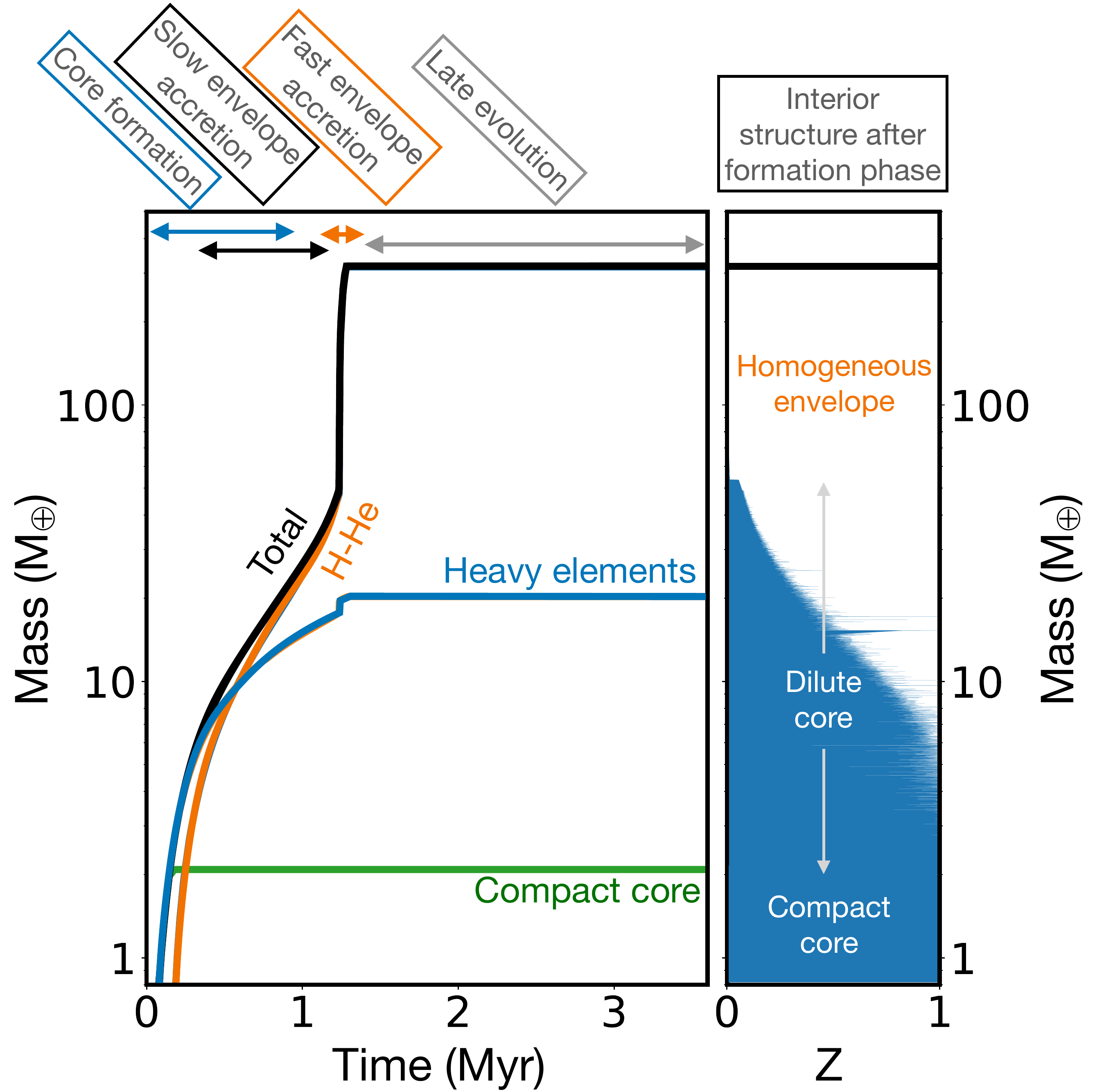}
    \caption{The stages of planetary growth based on the calculations presented by \citet{2017ApJ...836..227L}.  \rev{The left panel indicates the evolution with time of the masses of the compact core, heavy elements and hydrogen and helium in a forming Jupiter-mass planet. The timescales are only illustrative. The right panel shows the interior structure in heavy element mass fraction $Z$ (in blue) as a function of interior planetary mass at the end of the formation phase. Four phases are identified. The core formation phase leads to the formation of first a compact core and then of a dilute core when planetesimals and pebbles dissolve in the envelope. It overlaps with the slow envelope accretion phase.  The fast envelope accretion phase may be extended if the gas supply becomes limited (see text).}}
    \label{fig:formation_phases}
\end{figure}

The different stages of the formation and growth of a giant planet such as Jupiter as originating from the work of \cite{Bodenheimer+86} and \cite{Pollack+1996} are presented in Fig.~\ref{fig:formation_phases}. After a first \rrm{phase} disk phase, a core forms and grows. As \rrm{cooling and contraction proceeds} \rev{the core's gravity becomes strong}, a hydrogen-helium dominated envelope progressively \rev{contracts and} grows. When a cross-over mass of $\sim 20-30\rm\,M_\oplus$ is reached, \rev{the envelope's self-gravity dominates over the core's gravity, triggering a runaway gas accretion limited by the cooling and contraction of the envelope. When the envelope mass reaches $\sim 100 \, \mathrm{M}_\oplus$,} the \rrm{contraction} \rev{envelope growth} becomes limited only by disk supply \rev{\citep[][]{Tanigawa+2002,Tanigawa+2007}}. When gas accretion stops, the giant planets enter a late evolution phase with limited growth (although impacts, and particularly giant impacts may occur), but in which their interior still evolves. The timescales of these different phases are not well-defined and should highly depend on the systems considered. In the solar system, meteoritic constraints indicate that the inner solar system has been separated from the outer system between about 1\,Myr to 4\,Myr after the formation of the first condensates, which is interpreted as marking the times of Jupiter's core formation and rapid envelope accretion phases, respectively \citep{Kruijer+2020}. 
These stages of planetary growth are presented in more details hereafter.  

\subsubsection{The Protoplanetary Disk}

\begin{figure}[t!]
    \centering
    \includegraphics[width=\linewidth]{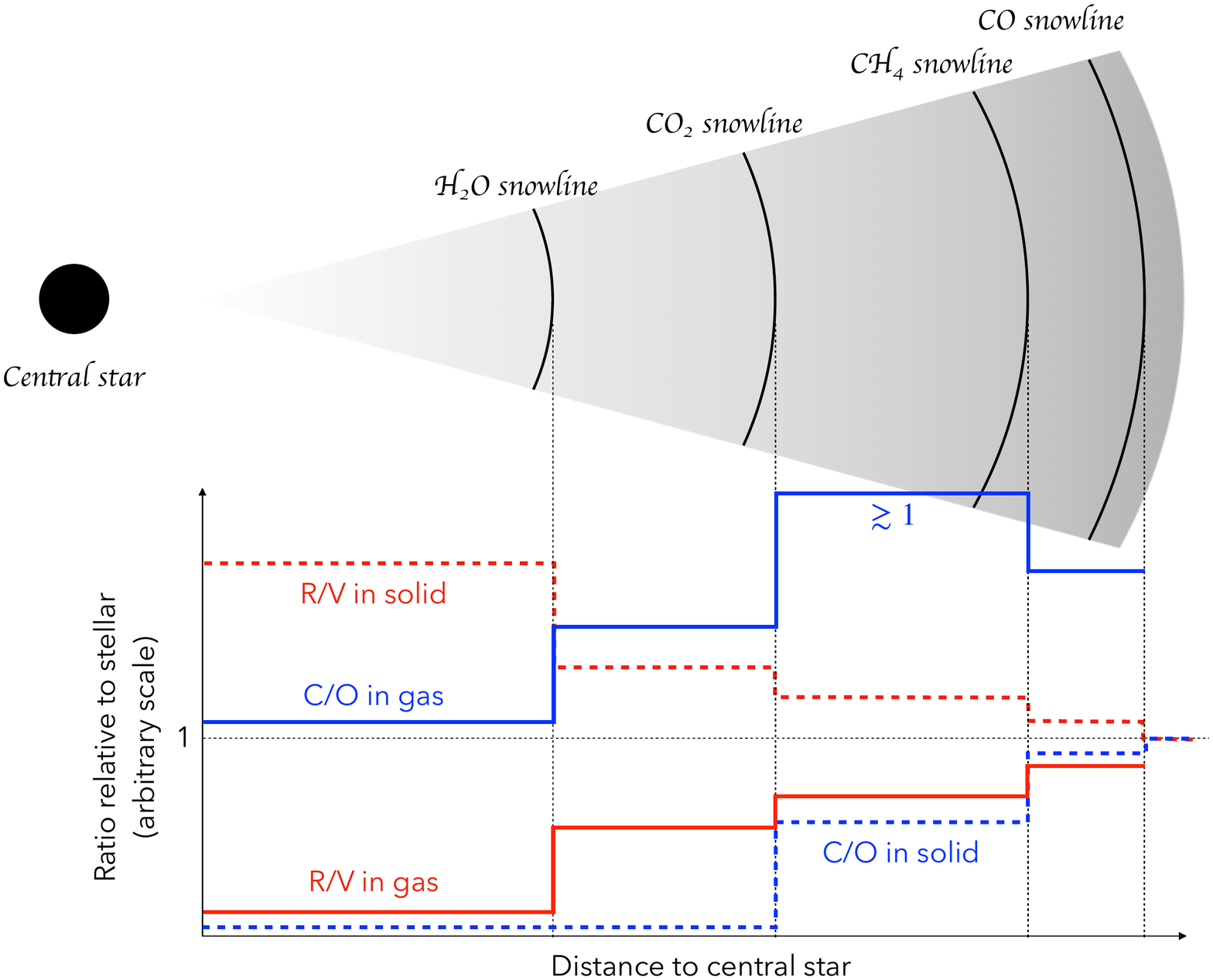}
    \caption{Schematic profiles of elemental abundance in a quiescent initial protoplanetary disk, which are determined solely by the ``snow'' lines for H$_2$O, CO$_2$, CH$_4$, and CO (made following e.g., \citet{Eistrup+2016}). The blue solid and dashed lines represent the carbon-to-oxygen (C/O) ratios in gas and solids, respectively. The red solid and dashed lines represent the refractory-to-volatile (R/V) ratios in gas and solids, respectively. }
    \label{fig:initialdisk}
\end{figure}

Planets form by collecting material from a circumstellar/protoplanetary disk. A simple picture for the radial composition of such a disk is shown in Fig.~\ref{fig:initialdisk}, highlighting the carbon-to-oxygen (C/O) and refractory-to-volatile (R/V) ratios in gas and solids as functions of distance to the central star. 
Here it is assumed that the protoplanetary disk inherits the molecular abundances from the molecular cloud core from which the star and protoplanetary disk originated; the composition of gas and solids is determined almost solely by condensation of the molecules H$_2$O, CO$_2$, CH$_4$ and CO \citep[e.g.,][]{Oberg+2011,Eistrup+2016}. 
Simple to understand is the dependence of the R/V ratios in gas and solids on the radial distance to the central star: The R/V ratio in gas (solids) increases (decreases) with the radial distance because more volatiles condense and are, thus, removed from gas (added to solids) in the outer regions.

However, one has to consider several complications. First, materials are to some extent chemically processed during the gas accretion from molecular clouds onto protoplanetary disks. 
The subsequent chemical reactions in the protoplanetary disk lead to different chemical abundances in gas and solids \citep{Eistrup+2016,Cridland+2019a,Cridland+2019b}. 
\rev{Temperature may not monotonically decrease with radial distance to the central star, and there may be local cold regions in protoplanetary disks \citep[][]{Ohno+2021}.}
In addition, planet formation takes place in an evolving protoplanetary disk \citep[e.g.,][]{Hueso+Guillot2005, Ida+Lin2004}. 
Decline in gas surface density and viscous accretion heating results in a displacement of the ice lines \citep[][]{Min+2011,Oka+2011}.
While kilometer-size and larger planetesimals are expected to remain in place \citep{Weidenschilling1977}, dust and pebbles drift radially, sublimate and recondense near ice lines \citep[e.g.,][]{Ciesla+Cuzzi2006,Hyodo+2021}, causing a redistribution of material and a change in the C/O and R/V ratios \citep[][]{Booth+2017,Booth+2019,Schneider+2021a,Schneider+2021b}. Because giant planets remain efficient gas accretors even after gap opening \citep{Tanigawa+2007,Tanigawa+2016, Durmann+Kley2017}, the fate of the disk gas is also an important, often forgotten, part of the story. The fact that disk photoevaporation takes place in regions devoid of grains leads to a progressive change in the overall metallicity and possibly composition of the gas disk \citep{Guillot+Hueso2006,Monga+Desch2015,Atreya+2018}. 
These complications must be considered to predict and analyze the final elemental abundances in giant planets and their atmospheres.

\subsubsection{Core formation}\label{sec:core_formation}
Traditional core accretion models assumed that giant planet cores have masses of tens of Earth masses in order to initiate rapid gas accretion (i.e., critical-mass cores)  \citep[e.g.,][]{Mizuno1980,Pollack+1996}.  
Recent models, however, estimate that cores first formed are rather small. The smaller the opacity in the envelope, the smaller the critical core mass is \citep[][]{Ikoma+2000,Hubickyj+2005,Valletta+21}. 
Rapid gas accretion with smaller core masses can be a result of various physical processes such as \rev{faster cooling of the protoplanet due to settling of dust grains  leading to lower opacity \citep[][]{Podolak2003,Movshovitz2010,Ormel2014}} 
and the enrichment of the envelope with heavy elements \rev{which increases the mean molecular weight of the envelope leading to faster contraction \citep[e.g.,][]{Hori+11,Venturini+16}}. 
\rrm{In the case of the envelope's opacity, since the radiative regions of the accreting envelope are cool, dust grains contribute mainly to the opacity. While many models assume that dust grains floating in the envelope have similar size and abundance to the insterstellar ones, models considering collisional growth and gravitational sedimentation of dust grains in the envelope demonstrate that the dust abundance and, therefore, the opacity are much lower than previously thought \citep{Podolak2003,Ormel2014}, resulting in small initial cores.}

Second, the critical core mass is strongly dependent on the composition of the envelope and can be of the order of a  few Earth masses.  
Heavy element enrichment (or contamination) of the envelope leads to reducing the critical core mass and the timescale of runaway gas accretion, because the mean molecular weight becomes large, the specific heat is reduced, and endothermic reactions including condensation occur \citep[e.g.,][]{Hori+11, Venturini+15,Venturini+16,2018A&A...611A..65B, Kimura+2020,Valletta+21}. 
The envelope enrichment likely occurs because planetesimals and/or pebbles are captured by the envelope itself \citep[e.g.,][]{Inaba+2003b,Okamura+21,Valletta+21} and subject to ablation and vaporisation  on the way toward the core \citep[e.g.,][]{Podolak+1988,  2020ApJ...900..133V, 2020A&A...634A..15B}.
\rev{Recent simulations of collisional growth from dust grains to gas giant cores show that solids of different sizes contribute to core formation \citep[][]{Kobayashi+2021}.}
Thus, the contamination of the protoplanetary envelope by incoming solids and its feedback on planetary growth now appears to be an essential part of the formation of giant planets. 
Those recent studies put together, the structure of a proto-giant planet in early stages of accretion may be such that there is a relatively small central core (of a few M$_{\oplus}$)  surrounded by an envelope highly contaminated with heavy elements \citep{2017ApJ...836..227L, Helled+Stevenson2017, Valletta_2019}.


\subsubsection{Gas accretion}
As the protoplanet grows in mass, it can accrete gas from the surrounding disk. The composition of the gas is expected to be dominated by H-He, and is often assumed to have a proto-solar composition. However, depending on the planetary formation location, other volatile elements can also be in the gaseous phase,  and therefore  phase and therefore lead to an enrichment of the atmosphere.  
During the early stages, when the proto-planet mass is low ($\sim \le 30$ M$_\oplus$) the gas accretion rate is moderate and is determined by the cooling rate of the protoplanet \citep[see e.g., ][for details]{Bodenheimer+86,Ikoma+2000,Helled2014}. 
When the planet becomes massive enough, \rrm{so the} gas accretion rate is  greater than the amount of gas that the disk can supply \citep[e.g.,][]{Tanigawa+2002}\rrm{, runaway gas accretion begins} (also known as the "detached phase"). 

In this later stage of runaway accretion, the gas accretion rate is given by prescriptions based on hydro-dynamical simulations \citep[e.g., ][]{Lissauer2009}. 
The mechanisms that terminate gas accretion and determine the final planetary mass are still being investigated  \citep[e.g.,][]{Tanigawa+2007,Tanigawa+2016}, with the most common explanations being gap formation and disk dissipation. 



\subsubsection{Late heavy-element accretion}


Most of the heavy-element enrichment of giant planets in the core accretion model occurs during the buildup of the core. However, further enrichment can take place at later stages. 
When planetesimals are embedded in a protoplanetary gaseous disc, the combination of gravitational enhancement and gas drag damping of their random velocities results in widening the separation between the planetesimal and protoplanet \citep[e.g.,][]{Tanaka+1997}. 
An increase in the protoplanet mass via runaway gas accretion leads to widening the protoplanet's feeding zone \rev{which} helps the protoplanet to capture more planetesimals; however, its efficiency is not \rrm{large} \rev{high} enough to collect \rev{tens of Earth masses of} solids \citep[][]{Zhou+2007,Shiraishi+2008,Shibata+Ikoma2019}, unless extremely large amounts of solids are available.

\rev{However, as } originally proposed by \cite{Alibert_2005}, \rev{a much wider feeding zone and thus many more planetesimals are accessible if the proto-giant planet is migrating. A complication arises however, the} trapping of planetesimals in sweeping mean motion resonances with the migrating gas-giant planet \citep[e.g.,][]{Tanaka+1999}. \rev{With dedicated dynamical simulations \citet{Shibata+2020} demonstrate} that the capture of planetesimals takes place only in limited regions of the disc because of the effects of resonant and aerodynamic shepherding. \rev{This so-called sweet spot depends on many parameters including planetesimal size and disc properties, but should be located in the 3-10\,au region, possibly extending to 30\,au for nominal values of the parameters \citep{Shibata+2022}. Thus, planets having crossed that region - in particular those having migrated from large distances would tend to have captured more planetesimals, possibly tens of Earth masses.}
\rrm{Consequently, the amount of heavy elements that a gas giant acquires is determined by how many planetesimals the planet delivers into the sweet spot. 
This indicates that a close-in gas giant with higher metal content may form if the planet was born far from its central star and migrated over a long distance in a denser protoplanetary disc \cite[see also][]{Shibata+2022}. 
Given that more massive gas giants tend to form further away from their central stars, \rev{an anticorrelation} between planet mass and metal content is predicted, \rev{consistently} with the observationally inferred trend \citep{Thorngren2016}.}

\rev{Another mechanism, the accretion of pebbles is strongly suppressed in the giant planet regime} because the planetary gravitational perturbation results in regions with super-Kepler rotation speeds exterior to the orbit of the planet \citep[][]{Lambrechts+2014b}. Instead, if drifting pebbles \rev{drift in}  and vaporise inside the snowline, \rev{the growing gas giant would accrete disc gas that is contaminated with heavy elements, thus possibly ending-up metal rich \citep[][]{Booth+2017}. For this mechanism to work, the planet must be inside the snowline and the metal-poor gas in the outer disk must be lost, possibly by photoevaporation \citep[see][and \S~\ref{sec:C/O}]{Guillot+Hueso2006}.}
\rrm{Gas giants growing by capture of such enriched disc gas would end up metal-rich close-in gas giants \citep[][]{Booth+2017}}. 

\rev{Finally giant planets may also accrete heavy elements after all the gas has dissipated. Calculations for the solar system giant planets indicate that only limited amounts ($0.2\,M_\oplus$ or less) would be accreted \citep{Matter+2009}. However, it is interesting to notice that for a planetesimal in the outer solar system, its ejection probability is much higher than its accretion probability, mostly because the escape speed at Jupiter is much larger than its orbital speed (by a factor $\sqrt{2(M_{\rm p}/M_\star)(R_{\rm p}/a_{\rm p})}\sim 4.6$ where $M_{\rm p}$ and $R_{\rm p}$ are the planetary mass and radius, $M_\star$ the stellar mass and $a_{\rm p}$ the planet orbital distance). The ejection probability becomes negligibly small when the planet escape speed becomes smaller than the orbital speed, i.e. when $M_{\rm p}\le (1/2)(R_{\rm p}/a_{\rm p})M_\star$. For a solar-type star and a Jupiter-radius planet, this corresponds to $M_{\rm p}\sim 17\,M_\oplus$ at 5\,au and to $M_{\rm p}\sim 2.7\,M_{\rm Jup}$ at 0.1\,au \citep{Guillot+Gladman2000}. When not accompanied by other giant planets, hot Jupiters should be efficient accretors of planetesimals and planets.}

\subsection{Evolution of a formed planet}

\subsubsection{Importance of primordial entropy and thermal state}
 Traditional giant planet formation models often over-simplify the treatment of the accretion shock associated with runaway gas accretion. 
 During this late stage most of the planetary mass is accreted and the influence of the accertion shock on the planetary thermal state and consequent evolution is of high importance \citep[e.g.,][]{Marley2007,Marleau2017,Mordasini2017,Berardo2017,BC2017,Cumming2018}. 
 
 Depending on the assumptions made regarding the accretion shock, it was shown that the accreted H-He gas  can have much higher entropy than the interior of the accreting protoplanet \citep{BC2017,Cumming2018}. This results in a large radiative region. This region can become convective after several Myrs as the planet contracts and cools down. The fact that the deep interior is expected to have composition gradients that also act towards inhibiting convection suggest that giant planets starts their contraction having large non-convective regions. As a result, the onset of convective mixing is expected to be delayed. 
 It is therefore clearly desirable to  include the primordial  entropy profiles when studying the evolution and current-state of giant planets. 


Progress can also be achieved through direct observations of young objects. 
Probably the best example is the young ($\sim 21$\,Myr-old) giant planet  $\beta$~Pic~b, which was detected by direct imaging \citep{Lagrange+2010,Bonnefoy+2013} and whose mass, $11\pm2\rm\,M_{Jup}$, was measured via astrometry \citep{Snellen2018}.
The large brightness of $\beta$ Pic b indicates that it was formed "hot", i.e. with a limited loss of entropy during accretion \citep[e.g.,][]{Bonnefoy+2013}. Separately, H$\alpha$ emission has been detected for the young gas giants PDS~70~b and c \citep[$<$ 10~Myr old;][]{Wagner2018,Haffert2019,Hashimoto2020}, indicating the presence of an accretion shock.
Models of the late-stage accretion of gas by forming planets show that explaining the observations requires an accretion flow occurring on a small fraction of the planetary surface and with a significant entropy loss \citep[][]{Aoyama2018,Aoyama2019,Marleau2019,Takasao+2021,Aoyama+2021}. 
The masses of PDS~70~b and c being unknown, it is however too early to understand whether this result is in tension with the $\beta$~Pic~b result or whether it may apply to a different class of planets.

\subsubsection{Re-distribution of Heavy-Elements during the Long-Term Evolution \label{sec:re-distribution}}
In order to link the measured composition and current-state structure to the formation process, the long-term (timescales of $10^9$ years) evolution should be investigated. 
There are several processes that can change the distribution of chemical elements in giant planets such as convective mixing, settling and phase transitions. 

If young giant planets consist of composition gradients, convective mixing can erode the gradient, and mix some of the heavy-elements to the outer convective regions. This in return, leads to an enrichment of the atmosphere with heavy elements. 
The mixing rate depends on the exact local conditions such as the shape of the gradient, the thermal state of the planet, and the mixing parameters. The initial entropy and primordial temperature profile determine the planetary evolution and the efficiency of convective mixing.

The evolution of Jupiter with composition gradients has been presented by 
\citet{Vazan+2018}). In this model an evolutionary path that can lead to the current-state measured properties of Jupiter was consider. As a result, the primordial thermal state and internal structure have been inferred to fit Jupiter today and are not guided by formation models. It was shown that the outer part of the planet, $\sim$ 50\% of Jupiter's radius, becomes homogeneous after several $10^6$ years due to convective mixing. The erosion of the composition gradient enriches the planetary envelope with heavy elements, while Jupiter's deep interior is found to remain stable against convection. It was then concluded that composition gradients which are stable against convection can persist for timescales of Gyr. This in return directly affect the planetary thermal structure and cooling history: in such a configuration the internal temperatures can be significantly higher than in the case of an adiabatic structure and the cooling of the planet is less efficient. 

The primordial thermal state of the young Jupiter considered by \citet{Vazan+2018} corresponds to a rather cold start which seems to be unrealistic according to planet formation models that simulate the  the planetary growth and the accretion shock during the rapid gas accretion state. 
As a result, a thermal evolution model in which Jupiter’s growth from the onset of runaway gas accretion is considered was presented by \citet{Muller2020}. Then the long-term evolution was simulated considering the energy transport and heavy-element mixing. Various formation conditions were investigated including different primordial composition gradients,  heavy-element accretion rates, and shock properties. 
It was found that in all the models after several Myr most of Jupiter's the envelope ($\sim$ 80\% by mass) becomes convective and  homogeneous. \rrm{This is because the formation models predict a much hotter initial state where the composition gradient, even if steep,  cannot inhibit large-scale convection unless the young Jupiter is somewhat unrealistically cold.} Fig.~10 shows the primordial and current-state heavy-element distribution of Jupiter as calculated by \citet{Vazan+2018} and two of the formation models presented by \citet{Muller2020} which are hotter by $\sim50\%$ throughout most of the envelope. 
\rrm{The dilute-core region of \citet{Vazan+2018} is much more extended than the one inferred from formation models, unless efficient planetesimal accretion is assumed. In addition, formation models typically predict high internal temperatures that lead to more efficient convection and  and a more homoegenously mixed. }
\rev{Thus, according to the latter,} composition gradients tend to be relatively compact, even when a substantial amount of heavy elements is accreted at late stages, unless a further mechanism, such as a giant impact takes place \citep{Liu+2019}.

\begin{figure}[htb!]
    \centering
    \includegraphics[width=\linewidth]{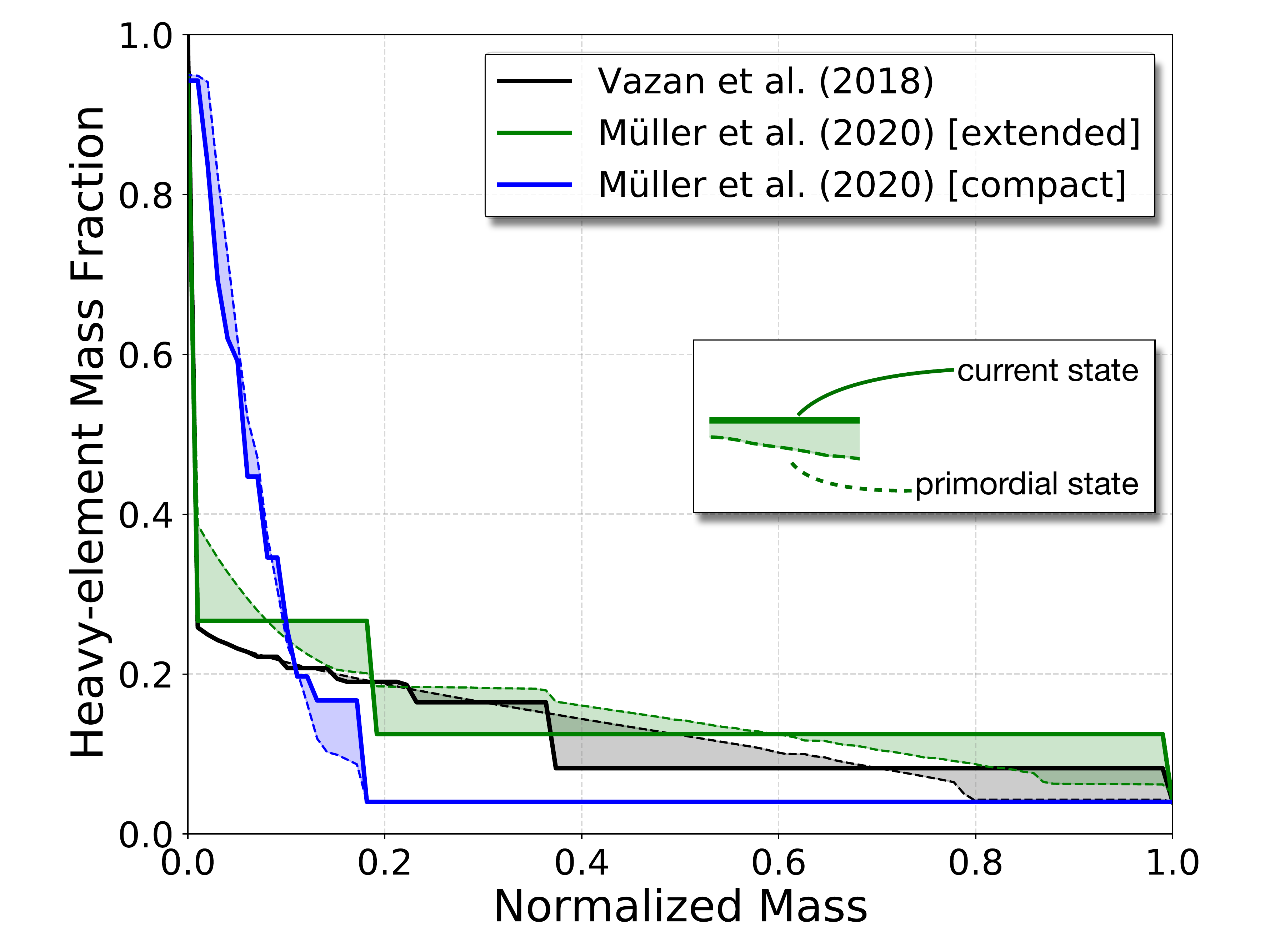}
    \caption{Heavy-element mass fraction vs.~normalised mass of Jupiter shortly after its formation (dashed) and today (plain). The different curves show different models of Jupiter: the \rev{black} lines correspond to the model of \citet{Vazan+2018} while the \rev{green} and \rev{blue} lines to the hot-extended  and hot-compact models of \citet{Muller2020}. 
    It is shown that \rev{4.5\,Gyr} of evolution can significantly change the heavy-element distribution of giant planets, with the efficiency of mixing depending on the \rev{initial distribution of heavy elements} and the thermal state of the young planet (see text for details).}
    \label{fig:mixing}
\end{figure}

\subsubsection{Effect of phase transitions}
Phase transtions can affect the structure and evolution of giant planets. This is in particular relevant for warm/cold giant planets that have internal temperatures that are low enough to experience such transitions as indicated in Fig.~2. The phase  separation  of  helium  in metallic  hydrogen leads to an inhomogenous distribution of helium and an enrichment of helium in the deep interior and possibly the formation of a pure-He shell \citep[e.g.,][]{Mankovich+Fortney2020}. In addition, this process affects the energy transport by the creation of regions where heat is transported via conduction/radiation/semi-convection leading to non-adiabatic temperature profile that prolongs the planetary contraction. 
Other phase transtions linked to elemental solidification and solubility of mixtures  can also play a role in giant planet evolution. 
Indeed, the low observed luminosities of Saturn are often explained by helium rain and composition gradients, respectively.  Compositions gradients and boundary layers could also exist in Jupiter, Uranus, and Neptune. 
It should be noted that the combination of chemical phase transitions and composition gradients makes the modeling of giant planet interiors even more challenging \citep[e.g.,][]{Vazan2016}.  
We suggest that future structure and evolution models should consider phase transtions and their effect on the evolution and internal structure of giant planets. 

\subsection{Confronting Theory \& Observations}

\subsubsection{Accounting for Atmospheric Abundances}

\rev{While giant planet formation models have often focused on accounting for parameters such as core mass and bulk composition}, some models also accounted for the measured atmospheric abundance. \rrm{or more generally, the atmospheric metallicity.} It was shown by \citet{Alibert_2005} that if Jupiter migrated to its current location from $\sim$ 8\,au, it could have accreted sufficient amount of planetesimals to explain Jupiter's measured  atmospheric composition, and predicting the O/H ratio in Jupiter to be six time solar. The authors have also followed the formation of Saturn and predicted the accretion of $\sim$ 13 M$_{\oplus}$ of heavy elements in Saturn with at least 5.4 M$_{\oplus}$ of ices. 

Recent Jupiter formation models that assume core formation via pebble accretion followed by planetesimal accretion can also reproduce Jupiter's atmospheric metallicity of $\sim$ 3 times solar \citep{2018NatAs...2..873A}. \citet{2020A&A...634A..31V} have preformed a large parameter space study to explore under what conditions are needed for Jupiter's heavy-element mass to be between 20 M$_{\oplus}$ and 40 M$_{\oplus}$, and found that Jupiter could have accreted 1--15 M$_{\oplus}$ of heavy elements during runaway gas accretion, depending on the assumed
initial surface density of planetesimals and the heavy-element accretion rate during the final stage of its  formation.  

The atmospheres of Uranus and Neptune are expected to be highly enriched with heavy elements. However, given the large uncertainties linked to their formation histories and bulk compositions \citep[e.g.,][and references therein]{20helled,2020RSPTA.37890474H}, we are still not a stage that atmospheric composition can constrain the planetary origin or vice versa. Future measurements of the atmospheric compositions and gravitational fields of the ice giants can be used to further constrain their structure, and formation path. 

Jupiter's enrichment in noble gases and in nitrogen (see Fig.~\ref{fig:compositions_JSUN}), a major surprise from the Galileo probe, has been seen as a sign of the planet's formation at very low temperatures $\sim 30\,$K \citep{99owen}. Instead, \cite{Guillot+Hueso2006} proposed that it is a consequence of (i) the fact that noble gases stick to grains in the cold outer protoplanetary disk, (ii) that these grains migrate to the mid-plane and to the inner disk where the nobles gases which can be accreted as gas by the growing giant planets and (iii) that photoevaporation in the disk occurs at locations where grains have low abundances and therefore leads to a preferential loss of hydrogen and helium (and potentially neon, depending on the outer disk temperature). The model required a relatively late formation of the planet, in sync with the photoevaporation of the disk, but it has been proposed that grain migration can lead to increase metallicities locally, lifting somewhat this timing requirement \citep{Monga+Desch2015,Mousis_2019}. 

Finally, it should be noted that the observed enrichment could also be a result of post-formation accretion of heavy elements and/or the re-distribution of heavy elements from the deep interior. Therefore in order to link the atmospheric composition with the formation process it is critical to properly model the planetary evolution.

\subsubsection{C/O and Other Abundance Ratios}\label{sec:C/O}
%


In principle, the measurement of abundance ratios are a powerful way to understand how giant planets were assembled. \cite{madhu2014} proposed to retrace the formation mode and migration history of giant exoplanets from their C/O ratio: They proposed that hot Jupiters with high C/O ratios and low O/H abundances should have been formed by disk-free migration. \cite{Ali-Dib+2014} proposed that giant planets with envelopes accreted inside the water snowline should be carbon-rich (with a C/O ratio higher than that of the star). \cite{Mousis+2014} proposed that Saturn's low $\rm^{14}N/^{15}N$ value is indicative of a formation of its building blocks at 45\,K in the solar nebula, predicting an O/H value of $\sim 34.9$ times protosolar and a $\sim 43.1\rm\,M_\oplus$ mass of heavy elements in the planet. Both \cite{Oberg+Wordsworth2019} and \cite{Bosman+2019} interpreted Jupiter's near protosolar C/N ratio as an indication that Jupiter's core assembled exterior to the N$_2$ snowline, echoing a similar conclusion made twenty years earlier \citep{99owen,Owen+2001}. \cite{Cridland+2020} concurred with this result, with an independent conclusion that, owing to its C/O and N/O ratios, Jupiter likely did not form in the inner 5\,au in the solar system. Finally, \cite{Notsu+2020} found that hot Jupiters with $\rm C/O > 1$ can only form between the CO$_2$ and CH$_4$ snowlines, corresponding to locations between 2.6 and 16\,au in a classical minimum-solar nebula.   

These conclusions might be correct. Unfortunately, they result from models based on many underlying hypotheses, most of those untested. Using a planet population synthesis model based on planetesimal accretion, \cite{Mordasini+2016} obtained planets with relatively low C/O ratios (0.1 to 0.5), but wisely noted that {\it "To link a formation history to a specific C/O, a better understanding of the disk chemistry is thus needed"}. \cite{Booth+2017} and \cite{Madhu2019} showed that outcomes of the C/O ratios and globally enrichments in giant exoplanets critically depend on whether giant planet growth occurs via accreting planetesimals or pebbles.  \cite{Cridland+2019b} showed that carbon-chemistry, a still unanswered puzzle in the solar system \citep[e.g.][]{Gail+Trieloff2017}, can strongly modify the resulting planetary C/O ratios depending on the assumption used. \cite{Schneider+2021b} found that Jupiter's nitrogen content can be explained by inward-diffusing nitrogen-rich vapour, so that Jupiter does not need to form close to the N$_2$ snowline, contrary to what inferred by \cite{Oberg+Wordsworth2019} and \cite{Bosman+2019}. Separately we can also see that the new constraint on Saturn's heavy element content \citep[see][and \S~\ref{sec:bulk_comp}]{Mankovich+Fuller2021} rules out the clathrate model proposed by \cite{Mousis+2014}. 

Clearly, we must acknowledge that giant planets result from a combination of many complex and uncertain processes. As we have seen (\S~\ref{sec:formation}), we do not know the relative proportion of dust, pebbles and planetesimals that led to the present-day giant planets. Even assuming a fixed composition for the protoplanetary disk, we do not know the composition of planetesimals that are formed \citep[e.g.,][]{Ida+Guillot2016, Hyodo+2021}. The growth of dust to pebbles leads to large-scale radial motions (a "pebble wave") and an evolving composition of disks \citep[e.g.][]{Booth+2019,Kunitomo+Guillot2021}, probably in a complex time-dependent way \citep[e.g.,][]{Elbakyan+2020}. Most importantly, giant planets are efficient gas accretors. This has two strong implications: heavy elements may be accreted as gas, inside of their snowline, and the photoevaporation of gas in the disk has an essential role in setting the final planetary masses \citep{Tanigawa+2007,Tanigawa+2016} but also the final balance between heavy elements and hydrogen and helium in the planetary envelope \citep{Guillot+Hueso2006, Monga+Desch2015, Atreya+2018}. Finally, as we have demonstrated in \S~\ref{sec:GP_interior}, giant planets may not be fully convective. This implies that the composition of their atmosphere will be modified mostly by the late accretion phases rather than by the formation of their core as often implicitly assumed. 

\subsubsection{\rev{Isotopic Ratios}}\label{sec:isotopes} 
For our solar system giant planets, two isotopic ratios are \rrm{in fact} most easily explained by invoking a partial mixing of the envelope and interior. The first one is D/H, a particularly important tracer of planetary formation because of its low value in the main reservoir H$_2$, and its high value \rev{(160 to 530\,ppm)} in cometary ices \citep{Hersant+2004, Bockelee-Morvan+2015}. \rev{The protosolar D/H is extremely difficult to determine due to its destruction in the Sun. Indirect inference from $\rm ^3He/^4He$ measurements in the solar wind point to a low value $16.7\pm 2.5$\,ppm (Table~\ref{tab:isotopes}), but measurements in the local interstellar medium point to a value $>21$\,ppm \citep[][and references therein]{Asplund+2021}. The most accurate values obtained for Jupiter and Saturn are $22.5\pm 3.5$\,ppm and $21\pm1.3$\,ppm, respectively (see Table~\ref{tab:isotopes}). Assuming full mixing and a $\sim 300$\,ppm D/H ratio in solids, we should expect an enrichment over the the protosolar value of $1$ to $4$\,ppm for Jupiter and 7\,ppm in Saturn \citep[updated from][using \S~\ref{sec:bulk_comp}]{Guillot1999}}, in clear tension with the measurements. If, as expected from the interior structure derived by \cite{Mankovich+Fuller2021}, only the $\sim 2\rm\,M_\oplus$ of heavy elements in the envelope are allowed to exchange their deuterium with the envelope, this value becomes only 0.7\,ppm over the protosolar value, in agreement with the measurement. Similarly, the D/H values measured in H$_2$ by HERSCHEL-PACS, \rev{$44\pm 4$\,ppm on Uranus and $41\pm 4$\,ppm on Neptune are much lower than the $\sim 100$\,ppm that would be obtained from calculations assuming full mixing,} indicating that only a small fraction of the ices in the interior exchanged isotopically with hydrogen \citep{Guillot+Gautier2015, Ali-Dib+Lakhlani2018}. 

The second one is the $\rm^{15}N/^{14}N$ ratio. \rev{As shown in Table~\ref{tab:isotopes}, its protosolar value measured by Genesis is $2.36\pm 0.33$\,\textperthousand\   while its value in comets (where it is carried as mostly NH$_3$)} is centred on $7.37\pm 0.30$\,\textperthousand\ \citep{Shinnaka+2016}. In Jupiter, its value is $2.3\pm 0.3$\,\textperthousand\ and it is less than $2.0$\,\textperthousand\ in Saturn. \rrm{Of course, we have seen that} One interpretation might be that both Jupiter and Saturn accreted extremely cold N$_2$ ices \citep{Owen+2001}. Another interpretation appears more likely: First whatever the nature of the ices that delivered nitrogen in the deep interiors of Jupiter and Saturn, they are not interacting with the envelope, keeping the high $\rm^{15}N/^{14}N$ locked in the dilute core. Second, atmospheric nitrogen could have been delivered during the fast envelope accretion principally as gas, with a protosolar $\rm^{15}N/^{14}N$. The enrichment in nitrogen compared to a protosolar gas seen in Fig.~\ref{fig:compositions_JSUN} could be accounted for with the same mechanism proposed by \cite{Guillot+Hueso2006} to explain the enrichement in noble gases in Jupiter: A progressive enrichement of the inner protoplanetary disk and the photoevaporation of low-metallicity regions of the disk. 

\rev{The third main isotopic ratio measured in Jupiter, Saturn and Neptune is $\rm ^{13}C/^{12}C$. With values of $1.08\pm0.05\%$, $1.09\pm 0.10\%$ and $1.16_{-0.27}^{+0.50}\%$, respectively, these are all comparable to the $1.07\pm 0.01\%$ protosolar value (see Table~\ref{tab:isotopes}). Other values in the solar system generally cluster around the terrestrial standard value of $1.124\%$, with relative departures of at most 10\%, including for interplanetary dust particles and ultracarbonaceous micrometeorites believed to originate from the outer solar system \citep{Rojas+2022}. 
Comets in the solar system are also consistent with these values, but with larger observational uncertainties \citep{Bockelee-Morvan+2015}. 
Altogether, the small but real $5\%$ offset between solar and terrestrial values is attributed to CO self-shielding or inheritence from the parent cloud \citep{Lyons+2018}. 
The measurement of $\rm ^{13}C/^{12}C=1.03\pm 0.12\%$ in an isolated L-type, young ($\sim 125$\,Myr) $12-26\,M_{\rm Jup}$ brown dwarf, 2MASS J03552337+1133437 \citep{Zhang2021bd} thus appears unsurprisingly in line with the solar system measurements. However, the new measurements in the two exoplanets TYC~8998-760-1~b and WASP-77A~b are puzzling, as these yield much higher $\rm ^{13}C/^{12}C$ ratios of $3.2_{-0.7}^{+0.6}\,\%$ ($1\sigma$ error bars) and $6.0\pm 3.7\%$, respectively \citep{Zhang2021,Line+2021}. The first is a young ($\sim 17$\,Myr), accreting, $14\pm 3\,M_{\rm Jup}$ exoplanet/brown dwarf at 160\,au projected separation from its star. The second is a $1.8\,M_{\rm Jup}$ hot Jupiter at 0.024\,au from its star. Explaining this puzzle will thus require isotopic measurements in other exoplanets. This is a highly promising research topic.}

\subsubsection{The Dilute Cores}


\rev{Although overlooked \citep[see however][]{Stevenson1985}, the existence of dilute cores in giant planets is in fact a natural consequence of the formation process. As discussed in \S~\ref{sec:core_formation}, the concurrent accretion of gas and planetesimals/pebbles leads to the formation of a strong gradient in heavy element abundance \citep[see Fig.~\ref{fig:formation_phases} and][]{2017ApJ...836..227L,Helled+Stevenson2017,2020ApJ...900..133V}. The question is really whether the extent of the dilute core resulting from formation models and after 4.57\,Gyr of evolution can be made to agree with the one inferred from interior models, i.e., within the inner $\sim 60$ to $180\,M_\oplus$ for Jupiter and $\sim 50$ to $60\,M_\oplus$ for Saturn (see Fig.~\ref{fig:interiorslices} and \S~\ref{sec:GP_interior}).} 

\rev{The evolution calculations presented in \S~\ref{sec:gp_evolution} show that efficient mixing in Jupiter's envelope leads to a uniform composition except in the inner $60\,M_\oplus$ where some primordial heavy element gradient can remain \citep[see][and Fig.~\ref{fig:mixing}]{Muller2020}. There is thus some tension with Jupiter interior models, calling either for a reduction of the size of dilute core inferred from gravitational soundings, a suppression of mixing or another mechanism. On the other hand, in the case of Saturn, and although specific calculations are not available, we can infer from the Jupiter calculations that Saturn's inner $60\,M_{\oplus}$ mass should not become mixed by convection. Thus, the extent of Saturn's dilute core as inferred by interior models may be broadly consistent with evolution calculations.}

\rev{A further increase of the extent of the dilute core to explain that observed in Jupiter might be due to double-diffusive convection \citep{Moll+2017}, however this seems unlikely as this would require a delayed outward mixing in order to avoid the high-entropy period when the outer envelope cools rapidly. Another explanation that could potentially lead to an extended dilute core (beyond Jupiter's inner $100\,M_\oplus$) would be the head-on impact of an object with a mass of $\sim 10\,M_{\oplus}$ \citep{Liu+2019}. Again however, this should occur late in the planet's evolution and seems thus, at this state, unlikely. Detailed discussions on that topic can be found in \cite{Helled+2022}.}  

\rrm{The existence of dilute cores in giant planets must be explained by formation and evolution models. 
One possibility is that dilute cores are a direct result of the formation process and the buildup of composition gradients in the planetary deep interior \citep[e.g.,][]{2017ApJ...836..227L,Helled+Stevenson2017,2020ApJ...900..133V}. }

\rrm{However, since heavy-elements can be re-distributed due to convective mixing, it is desirable to investigate whether such primordial composition gradients can persist. Such mixing can enrich the atmosphere with heavy-elements, potentially explaining some of the enriched atmospheres of giant exoplanets. 
As discussed above, this is still being investigated, as the mixing efficiency strongly depends on the initial thermal state of the planet and the nature of the gradient which are poorly known.} 

\rrm{The dilute core might also be a result of the erosion of a more compact core and the re-distribution of the heavy elements \citep[e.g.,][]{2017ApJ...849...24M}. Double-diffusive convection can lead to sufficient upward mixing in the presence of composition gradients \citep[e.g.,][]{Leconte+Chabrier2012}. 
However, given the efficiency of the overlaying free convection \citep{Vazan+2018}, it is not clear that double-diffusive layers would be able to extend through a significant fraction of the planetary radius. Evolution models including double-diffusive convection have not yet been calculated. }

\rrm{Finally, another explanation for Jupiter's dilute core is a giant impact post-formation. 
It has been shown by \cite{Liu+2019} that a head-on impact of an object with a mass of 10 M$_{\oplus}$ can mix the material of a primordial compact heavy-element core thanks to the energy associated with the impact. It was shown that such an impact can resulting in an extended fuzzy core similar to the one predicted by recent structure models of Jupiter.}

\subsubsection{Overdense giant planets}

Some of the observed giant exoplanets with measured mass and radius are extremely dense and thus seem require high enrichments in heavy elements or large core masses. 
An extreme example is the hot Jupiter HD~149026~b \citep{Sato+2005}, which is inferred to contain $\sim$~50-80 M$_\oplus$ of heavy elements relative to its total of 110~M$_\oplus$ \citep{Ikoma+2006,Fortney+2006}.
In addition, a few tens of close-in giant planets have so far been known to have tens of Earth masses and more of heavy elements in their deep interiors \citep[][]{Guillot+2006,Miller+2011,Moutou+2013,Thorngren2016}. 

Such massive heavy elements are not easily collected by giant planets even in the core accretion scenario.
The critical core mass is larger for higher solid accretion rate \citep[][]{Ikoma+2000}. For its value to be $\gtrsim$ 30~M$_\oplus$, however, unrealistically high solid accretion rates are needed, provided planetesimals are the carrier of solids. This suggests that further heavy element addition occur during and/or after the onset of runaway gas accretion. 
Not by planetesimal accretion but by pebble accretion, such high accretion rates are possible \citep[][]{Lambrechts+2012,Lambrechts+2014a}; however, once the cores exceed $\sim$ 20~M$_\oplus$, gravitational perturbation the core exerts on the nearby disc gas is large enough to prevent pebbles from accreting \citep{Lambrechts+2014b}. 
Once solid accretion is halted, the runaway gas accretion takes place \citep[][]{Ikoma+2000,Hubickyj+2005}.

Because of rapid mass growth due to runaway gas accretion, the planetary feeding zone expands and engulfs the surrounding planetesimals. However, the efficiency of such an in-situ accretion is not high enough to account for observed massive heavy elements \citep{Zhou+2007,Shiraishi+2008,Shibata+Ikoma2019}. 
Thus some additional processes are needed, including migration \citep{Shibata+2020}, giant impacts \citep{Ikoma+2006,Ginzburg+Chiang2020,Ogihara+2021}, and pebble accretion \citep{Booth+2017,Schneider+2021a}.

\subsubsection{\rev{Failed Cores?}}
\rev{The formation process of intermediate-mass gaseous planets like Uranus and Neptune is still a matter of continuous investigation \citep[see e.g.,][and references therein]{20helled, 2020RSPTA.37890474H}. Since Uranus and Neptune have relatively small amounts of H-He (see \S~\ref{sec:bulk_comp}), it is expected that they never reached the last phase of runaway gas accretion. This is consistent with the fact that they are located rather far from the sun, where the solid surface density in the disc was lower and the growth was slower, preventing the outermost planets from becoming gas giants. While it seems that the core-accretion scenario naturally explains the formation of intermediate-mass gaseous planets, it is yet to be determined whether the planetary growth was dominated by pebble or planetesimal accretion and whether such planets can form at large radial distances. In addition, while recent formation models of Uranus and Neptune clearly predict that the deep interiors of the planets consist of composition gradients, matching the heavy-element to H-He ratio is still challenging. More accurate measurements of the gravitational fields and the atmospheric compositions of Uranus and Neptune combined with detailed characterisation of intermediate-mass exoplanets would reveal important information that can be used to better understand this unique planetary type and further constrain their formation history.}

\rev{Separately, the discovery of many close-in super-Earths \citep[e.g.,][for review]{Batalha2014} have brought a new challenge to the core accretion scenario. Super-Earths are as massive as critical-mass cores unless the envelope opacity and/or solid accretion rate are quite high \citep[e.g.,][]{Ikoma+2000,Hori+11}. The slow growth scenario for Uranus/Neptune described above may not be applied to close-in super-Earths, because core growth is rapid at such short orbital periods. 
Furthermore, this issue is highlighted by the existence of super-Earths with limited atmospheres (often referred to as low-density super-Earths) such as planets orbiting Kepler-11 \citep{Lissauer+2011}, because the observation fact indicates that those super-Earths obviously formed under the presence of disk gas.
Thus, some mechanisms or scenarios are needed for preventing runaway gas accretion. One possible scenario is gas accretion in a dissipating disk after the merger of relatively small cores through giant collisions \citep{Ikoma+2012,Lee+2014,Ogihara+2020}. The huge impact energy deposited in the rocky interior is subsequently transferred to the atmosphere, resulting in preventing the gas accretion and also eroding the atmosphere significantly \citep[sometimes called the core-powered mass loss;][]{Ikoma+2012,Ginzburg+2016}.
This scenario is also consistent with the fact that many close-in super-Earths are slightly out of resonant chains \citep[e.g.,][]{Lissauer2009}. Alternatively, hydrodynamic simulations show that disk gas rotating around a super-Earth at short periods is not gravitationally bound to the planet but brought out (recycled) of the planet's Hill sphere by Keplerian shear flows \citep[][]{Ormel+2015, Lambrechts+Lega2017}. 
The recycling has been proposed for a possible mechanism of preventing the runaway gas accretion.
Its efficiency, however, depends on the treatment of radiative cooling and remains still controversial \citep[][]{D'Angelo+2013,Kurokawa+2018,Moldenhauer+2022}. 
Finally, photo-evaporation of the envelopes of gas giants may be a possible way to explain the existence of close-in super-Earths \citep[e.g.,][]{Valencia+2010}; however, the bimodal size distribution of planets of radii 1-4~$\mathrm{R}_\oplus$ \citep[][]{Fulton+2017,Fulton+2018} implies that complete erosion of gas giant envelopes is unlikely. 
To date, no unified view has been reached on the origin of close-in super-Earths.}

\section{CONCLUSIONS}

Giant planets acquire gas and elements during the early formation stages of planetary systems. Their compositions and interior structures are thus relics of the star and planet formation process itself. However, recent results show that giant planets are not the simple, fully convective and largely uniform worlds that they were once envisioned to be. 

New developments thanks to the Juno and Cassini missions indicate that both Jupiter and Saturn have stable regions and central heavy element cores that are diluted in their envelope. Differential rotation (the observed atmospheric zonal winds and banded structure) extends deep from the atmosphere to the level at which conductivity becomes sufficiently high for the magnetic field to drag the flow into nearly uniform rotation. Separately, observations of the atmospheres of Jupiter, Saturn, Uranus and Neptune show that strong latitudinal inhomogeneities exist and, at least for Jupiter, extend well-below the cloud condensation level of the main condensate, water. 


The growth of the ensemble of known giant exoplanets and brown dwarfs provides us with a wealth of new data. Measurements of masses and radii or luminosity and age combined to evolution models provide us with the possibility to constrain bulk compositions. Spectroscopic observations provide the means to measure atmospheric properties including cloud structure, variability, winds, spin rate, atmospheric composition and abundances. Brown dwarfs have a scatter of C/O values and metallicity that is, to  first order, consistent with that of the stellar population. Giant exoplanets are characterised first by the large diversity of their bulk compositions. The elemental atmospheric abundances, although preliminary, \rrm{clearly} point to a large partitioning between their interior and atmosphere. 

The consequences for our understanding of their formation are multiple. For field brown dwarfs, a formation by processes also responsible for the formation of stars is favoured. For giant planets, many questions remain, in particular, how are heavy elements delivered, what is responsible for ending their growth and how much interior mixing ("core erosion") is responsible for their atmospheric compositions. The strong and multiple indications that the interiors of giant planets --both in our solar system and outside-- are not fully convective imply that inferences between atmospheric composition and formation scenario based on the (often implicit) assumption of full mixing are at best tentative. 

The diversity of the giant planet population, the multiplicity of the mechanisms that led to their formation and their complexity imply that new observational constraints are required to understand their formation. Both extremely accurate measurements of the planetary characteristics for a limited number of objects and moderately accurate ones for a statistically significant population are needed. The possibility to measure accurate atmospheric abundances and bulk compositions on a large ensemble of exoplanets is highly awaited for. The gaps in our understanding of basic physical processes such as convection in the presence of compositional gradients call for an in-depth study of giant planets in our solar system. In that respect, Uranus and Neptune, which have never been visited by an orbiter, hold crucial clues to understand how giant planets and their planetary systems form. 

\bigskip

\noindent\textbf{Acknowledgements} LNF was supported by a European Research Council Consolidator Grant (under the European Union's Horizon 2020 research and innovation programme, grant agreement No 723890) at the University of Leicester. RH acknowledges support from the Swiss National Science Foundation (SNSF) under grant $200020\_188460$. 
MI was supported by Japan Society for the Promotion of Science (JSPS) KAKENHI JP17H01153 and JP18H05439 and JSPS Core-to-core Program ‘International Network of Planetary Sciences.'
\bigskip

\bibliographystyle{pp7}
\bibliography{refs.bib,Abundance_table.bib}


\appendix


\begin{table}[htbp]
\begin{center}
{\caption{Elemental abundances measured in the tropospheres of solar system giant planets}\label{tab:comp}}
\small
\begin{tabular}{llllllc} \hline\hline\vspace{.8ex}
 {\bf Element} & {\bf Carrier} & \multicolumn{2}{c}{{\bf Abundance ratio/H}$^\dagger$} & {\bf Enrichment} & {\bf Method} & {\bf Notes} \\
               &               &  {\bf Planet}$^\dagger$ & {\bf Protosun}$^a$ & $\frac{\mbox{\bf Planet}}{\mbox{\bf Protosun}}$ & & 
\vspace{.3ex}\\ \hline
\multicolumn{7}{l}{\bf Jupiter}\\
He/H     & He       & $(7.88\pm0.16)\times 10^{-2}$ & $(9.64\pm0.29)\times 10^{-2}$ & $0.818\pm0.016$      & Galileo/Nephelometer & $^c$ \\
C/H      & CH$_4$   & $(1.19\pm0.28)\times 10^{-3}$ & $(3.33\pm0.31)\times 10^{-4}$ & $3.56\pm0.86$        & Galileo/GPMS         & $^d$ \\
N/H      & NH$_3$$^\star$ & $(2.04\pm0.13)\times 10^{-4}$ & $(7.81\pm1.26)\times 10^{-5}$ & $2.61\pm0.16$        & Juno/MWR             & $^e$ \\
         & NH$_3$$^\star$ & $(3.32\pm1.27)\times 10^{-4}$ &                           & $4.25\pm1.63$        & Galileo/GPMS         & $^f$ \\
O/H      & H$_2$O$^\star$ & $(1.45_{-0.93}^{+1.28})\times 10^{-3}$ & $(5.66\pm0.52)\times 10^{-4}$ & $2.57_{-1.64}^{+2.26}$ & Keck,IRTF,Juno/MWR   & $^g$ \\
S/H      & H$_2$S$^\star$ & $(4.45\pm1.05)\times 10^{-5}$ & $(1.52\pm0.11)\times 10^{-5}$ & $2.92\pm0.69$        & Galileo/GPMS         & $^c$ \\
Ne/H     & Ne       & $(1.24\pm0.01)\times 10^{-5}$ & $(1.33\pm0.15)\times 10^{-4}$ & $0.093\pm0.001$      & Galileo/GPMS         & $^h$ \\
Ar/H     & Ar       & $(9.10\pm1.80)\times 10^{-6}$ & $(2.77\pm0.64)\times 10^{-6}$ & $3.28\pm0.65$        & Galileo/GPMS         & $^h$ \\
Kr/H     & Kr       & $(4.65\pm0.85)\times 10^{-9}$ & $(1.52\pm0.35)\times 10^{-9}$ & $3.05\pm0.56$        & Galileo/GPMS         & $^h$ \\
Xe/H     & Xe       & $(4.45\pm0.85)\times 10^{-10}$ & $(1.92\pm0.22)\times 10^{-10}$ & $2.32\pm0.44$        & Galileo/GPMS         & $^h$ \\
P/H      & PH$_3$$^\star$ & $(1.10\pm0.06)\times 10^{-6}$ & $(2.97\pm0.21)\times 10^{-7}$ & $3.71\pm0.20$        & Cassini/CIRS         & $^i$ \\
         & PH$_3$$^\star$ & $(4.64_{-1.16}^{+1.16})\times 10^{-7}$ &                           & $1.56_{-0.39}^{+0.39}$ & Juno/JIRAM           & $^j$ \\
Ge/H     & GeH$_4$$^\star$ & $(3.36\pm0.58)\times 10^{-10}$ & $(4.82\pm1.11)\times 10^{-9}$ & $0.070\pm0.012$      & VLT/CRIRES           & $^k$ \\
         & GeH$_4$$^\star$ & $(4.64_{-0.58}^{+0.58})\times 10^{-10}$ &                           & $0.096_{-0.012}^{+0.012}$ & Juno/JIRAM           & $^j$ \\
As/H     & AsH$_3$$^\star$ & $(3.19\pm0.58)\times 10^{-10}$ & $(2.30\pm0.21)\times 10^{-10}$ & $1.38\pm0.25$        & VLT/CRIRES           & $^l$ \\
         & AsH$_3$$^\star$ & $(3.48\pm1.16)\times 10^{-10}$ &                           & $1.51\pm0.50$        & Juno/JIRAM           & $^k$ \\
\smallskip\\
\multicolumn{7}{l}{\bf Saturn}\\
He/H     & He       & $(2.87\pm0.87)\times 10^{-2}$ & $(9.64\pm0.29)\times 10^{-2}$ & $0.30\pm0.09$        & Cassini/CIRS         & $^m$ \\
         & He       & $(6.20\pm1.25)\times 10^{-2}$ &                           & $0.64\pm0.13$        & Cassini/CIRS+UVIS    & $^n$ \\
         & He       & $(2.75_{-0.75}^{+0.50})\times 10^{-2}$ &                           & $0.29_{-0.08}^{+0.05}$ & Cassini/VIMS         & $^o$ \\
C/H      & CH$_4$   & $(2.50\pm0.11)\times 10^{-3}$ & $(3.33\pm0.31)\times 10^{-4}$ & $7.50\pm0.32$        & Cassini/CIRS         & $^p$ \\
N/H      & NH$_3$$^\star$ & $(2.66\pm0.53)\times 10^{-4}$ & $(7.81\pm1.26)\times 10^{-5}$ & $3.40\pm0.68$        & Cassini/VIMS         & $^q$ \\
S/H      & H$_2$S$^\star$ & $(9.40\pm4.70)\times 10^{-5}$ & $(1.52\pm0.11)\times 10^{-5}$ & $6.17\pm3.09$        & VLA                  & $^r$ \\
P/H      & PH$_3$$^\star$ & $(1.64\pm0.16)\times 10^{-6}$ & $(2.97\pm0.21)\times 10^{-7}$ & $5.52\pm0.53$        & Cassini/VIMS         & $^q$ \\
         & PH$_3$$^\star$ & $(4.36\pm0.30)\times 10^{-6}$ &                           & $14.67\pm1.02$       & Cassini/CIRS         & $^i$ \\
Ge/H     & GeH$_4$$^\star$ & $(2.12\pm2.12)\times 10^{-10}$ & $(4.82\pm1.11)\times 10^{-9}$ & $0.04\pm0.04$        & IRTF/FTS             & $^s$ \\
As/H     & AsH$_3$$^\star$ & $(1.17\pm0.16)\times 10^{-9}$ & $(2.30\pm0.21)\times 10^{-10}$ & $5.07\pm0.69$        & Cassini/VIMS         & $^q$ \\
\smallskip\\
\multicolumn{7}{l}{\bf Uranus}\\
He/H     & He       & $(8.96\pm1.95)\times 10^{-2}$ & $(9.64\pm0.29)\times 10^{-2}$ & $0.93\pm0.20$        & Voyager/IRIS+RSS     & $^t$ \\
C/H      & CH$_4$$^\star$ & $(1.95\pm0.49)\times 10^{-2}$ & $(3.33\pm0.31)\times 10^{-4}$ & $58.51\pm14.63$      & HST,Keck,IRTF        & $^u$ \\
N/H      & NH$_3$$^\star$ & $(1.04_{-0.23}^{+0.42})\times 10^{-4}$ & $(7.81\pm1.26)\times 10^{-5}$ & $1.33_{-0.30}^{+0.53}$ & ALMA,VLA             & $^v$ \\
S/H      & H$_2$S$^\star$ & $(5.30_{-0.88}^{+1.84})\times 10^{-4}$ & $(1.52\pm0.11)\times 10^{-5}$ & $34.83_{-5.77}^{+12.11}$ & VLA                  & $^v$ \\
\smallskip\\
\multicolumn{7}{l}{\bf Neptune}\\
He/H     & He       & $(8.75_{-1.29}^{+1.00})\times 10^{-2}$ & $(9.64\pm0.29)\times 10^{-2}$ & $0.91_{-0.13}^{+0.10}$ & ISO/LWS              & $^w$ \\
C/H      & CH$_4$$^\star$ & $(2.45\pm0.61)\times 10^{-2}$ & $(3.33\pm0.31)\times 10^{-4}$ & $73.49\pm18.37$      & HST/STIS             & $^x$ \\
N/H      & NH$_3$$^\star$ & $(2.88_{-2.15}^{+1.48})\times 10^{-4}$ & $(7.81\pm1.26)\times 10^{-5}$ & $3.69_{-2.76}^{+1.90}$ & ALMA,VLA             & $^y$ \\
S/H      & H$_2$S$^\star$ & $(7.97_{-1.75}^{+2.38})\times 10^{-4}$ & $(1.52\pm0.11)\times 10^{-5}$ & $52.32_{-11.47}^{+15.62}$ & ALMA,VLA             & $^y$ \\
\hline\hline
\end{tabular}
\end{center}
\noindent
$^\star$: Species which condense or are in chemical disequilibrium, i.e., with vertical/horizontal variations of their concentration. The global elemental abundances are estimated from the maximum measured mixing ratio, but these may still only be lower limits to the bulk abundance. \\
$^\dagger$: Abundance ratios $r$ are measured with respect to atomic hydrogen. In these atmospheres dominated by molecular hydrogen and helium, mole fractions $f$ are found by $f\simeq 2r/(1+r_{\rm He})$ where $r_{\rm He}$ is the He/H abundance ratio. \\
$^a$\citep{Asplund+2021}; $^b$ Enrichment uncertainties do not include the uncertainties on the protosolar values; $^c$ \citet{Niemann+1998}; $^d$ \citet{Wong+2004}; $^e$ \citet{Li+2020}; $^f$ \citet{Wong+2004}; $^g$ \citet{Bjoraker+2018, Li+2020}; $^h$ \citet{Mahaffy+2000, Atreya+2018}; $^i$ \citet{Fletcher+2009ph3}; $^j$ \citet{Grassi+2020}; $^k$ \citet{Giles+2017}; $^l$ \citet{Giles+2017}; $^m$ \citet{Achterberg+Flasar2020}; $^n$ \citet{Koskinen+Guerlet2018}; $^o$ \citet{Sromovsky+2016}; $^p$ \citet{Fletcher+2009ch4}; $^q$ \citet{Fletcher+2011}; $^r$ \citet{Briggs+Sackett1989} w/ ad hoc uncertainties; $^s$ \citet{Noll+Larson1991}; $^t$ \citet{Conrath+1987}; $^u$ \citet{Sromovsky+2019}; $^v$ \citet{Molter+2021}; $^w$ \citet{Burgdorf+2003}; $^x$ \citet{Karkoschka+Tomasko2011}; $^y$ \citet{Tollefson+2021}.
\end{table}

\begin{table}[htbp]
\begin{center}
{\caption{Selected isotopic ratios measured in solar system giant planets$^\star$}\label{tab:isotopes}}
\small
\begin{tabular}{llllllc} \hline\hline\vspace{.8ex}
 {\bf Ratio} & {\bf Carrier} & \multicolumn{2}{c}{{\bf Abundance ratio}} & {\bf Enrichment}$^b$ & {\bf Method} & {\bf Notes} \\
               &               &  {\bf Planet}$^\dagger$ & {\bf Protosun}$^a$ & $\frac{\mbox{\bf Planet}}{\mbox{\bf Protosun}}$ & & 
\vspace{.3ex}\\ \hline
\multicolumn{7}{l}{\bf Jupiter}\\
D/H      & H$_2$    & $(2.25\pm0.35)\times 10^{-5}$ & $(1.67\pm0.25)\times 10^{-5}$ & $1.35\pm0.21$        & ISO/SWS              & $^c$ \\
         & H$_2$    & $(2.95\pm0.55)\times 10^{-5}$ &                           & $1.77\pm0.33$        & Cassini/CIRS         & $^d$ \\
         & H$_2$    & $(2.60\pm0.70)\times 10^{-5}$ &                           & $1.56\pm0.42$        & Galileo/GPMS         & $^e$ \\
$\rm ^{13}C/^{12}C$ & CH$_4$   & $(1.08\pm0.05)\times 10^{-2}$ & $(1.07\pm0.01)\times 10^{-2}$ & $1.01\pm0.05$        & Galileo/GPMS         & $^f$ \\
$\rm ^{15}N/^{14}N$ & NH$_3$   & $(2.30\pm0.30)\times 10^{-3}$ & $(2.26\pm0.33)\times 10^{-3}$ & $1.02\pm0.13$        & Galileo/GPMS         & $^g$ \\
\smallskip\\
\multicolumn{7}{l}{\bf Saturn}\\
D/H      & H$_2$    & $(2.10\pm0.13)\times 10^{-5}$ & $(1.67\pm0.25)\times 10^{-5}$ & $1.26\pm0.08$        & Cassini/CIRS         & $^d$ \\
         & H$_2$    & $(1.70_{-0.45}^{+0.75})\times 10^{-5}$ &                           & $1.02_{-0.27}^{+0.45}$ & ISO/SWS              & $^c$ \\
$\rm ^{13}C/^{12}C$ & CH$_4$   & $(1.09\pm0.10)\times 10^{-2}$ & $(1.07\pm0.01)\times 10^{-2}$ & $1.02\pm0.09$        & Cassini/CIRS         & $^h$ \\
\smallskip\\
\multicolumn{7}{l}{\bf Uranus}\\
D/H      & H$_2$    & $(4.40\pm0.40)\times 10^{-5}$ & $(1.67\pm0.25)\times 10^{-5}$ & $2.63\pm0.24$        & Herschel/PACS        & $^i$ \\
\smallskip\\
\multicolumn{7}{l}{\bf Neptune}\\
D/H      & H$_2$    & $(4.10\pm0.40)\times 10^{-5}$ & $(1.67\pm0.25)\times 10^{-5}$ & $2.46\pm0.24$        & Herschel/PACS        & $^i$ \\
$\rm ^{13}C/^{12}C$ & C$_2$H$_6$ & $(1.16_{-0.27}^{+0.50})\times 10^{-2}$ & $(1.07\pm0.01)\times 10^{-2}$ & $1.09_{-0.25}^{+0.47}$ & NASA ITF             & $^j$ \\
\hline\hline
\end{tabular}
\end{center}
\noindent
$^\star$: Isotopic ratios of noble gases in Jupiter, not listed here, are globally consistent with the solar value \citep[see][]{Guillot+Gautier2015, Atreya+2018}. \\
$^a$ From \citet{Asplund+2021} w/ centered errors except for $^{13}$C/$^{12}$C from \citet{Lyons+2018}; $^b$ Enrichment uncertainties do not include the uncertainties on the protosolar values; $^c$ \citet{Lellouch+2001}; $^d$ \citet{Pierel+2017}; $^e$ \citet{Mahaffy+1998}; $^f$ \citet{Niemann+1998}; $^g$ \citet{Owen+2001}; $^h$ \citet{Fletcher+2009ch4}; $^i$ \citet{Feuchtgruber+2013}; $^j$ \citet{Sada+1996}.
\end{table}

\begin{table}[htb!]
    \centering
        \caption{Chemical species detected in exoplanet atmospheres as of March 2022. } \label{Table:Detection_table_full}
    \includegraphics[trim=70 70 70 70,clip,width=0.95\linewidth]{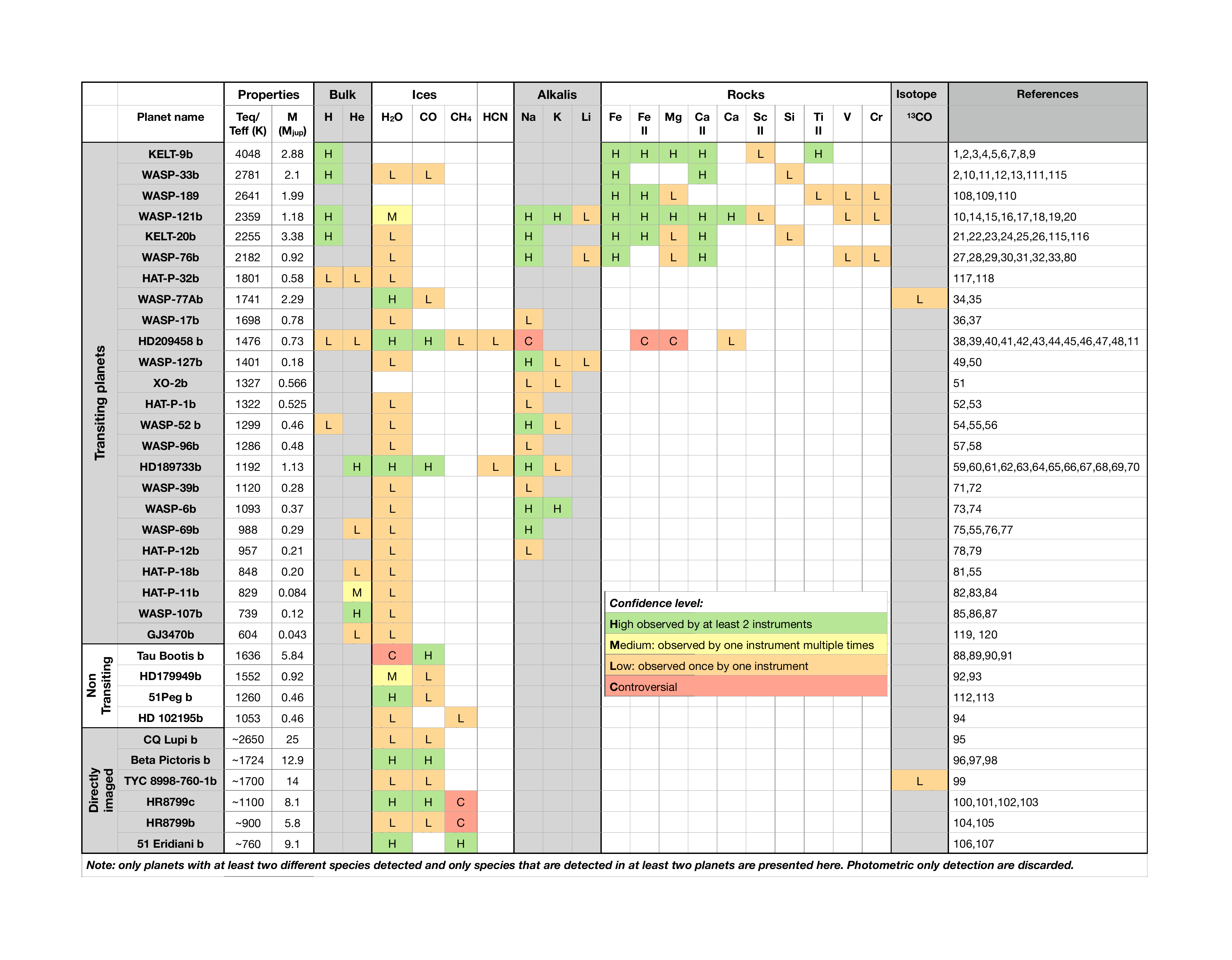}\\
        \noindent\parbox{16.5cm}{\small
    {\bf Note}: We included only planets where more than 2 species have been detected and species that have been detected in at least two planets. {\bf References}: 
        (1)~\cite{Wyttenbach2020} (2)~\cite{Yan2019} 
        (3)~\cite{Turner2020} (4)~\cite{Hoeijmakers2018} (5)~\cite{Cauley2019} 
        (6)~\cite{Pino2020} 
        (7)~\cite{Kasper2021} (8)~\cite{Hoeijmakers2019} (9)~\cite{Yan2018} 
        (10)~\cite{Borsa2021} (11)~\cite{Nugroho2021} (12)~\cite{Nugroho2020} (13)~\cite{Yan2021} 
        (14)~\cite{Cabot2020} (15)~\cite{Evans2018} (16)~\cite{Mikal-Evans2020} (17)~\cite{Hoeijmakers2020} (18)~\cite{Merritt2021} (19)~\cite{Gibson2022} (20)~\cite{Sing2019} (21)~\cite{Casasayas-Barris2018} (22)~\cite{Yan2019} (23)~\cite{Casasayas-Barris2019} (24)~\cite{Nugroho2020a} (25)~\cite{Hoeijmakers2020} (26)~\cite{Stangret2020} (27)~\cite{Fu2021} 
        (28)~\cite{Seidel2019} (29)~\cite{Seidel2021} (30)~\cite{Kesseli2022} (31)~\cite{Kesseli2021} (32)~\cite{Ehrenreich2020} (33)~\cite{Casasayas-Barris2021} (34)~\cite{Mansfield2022} (35)~\cite{Line+2021}
        (36)~\cite{Mandell2013} (37)~\cite{Sing2016} (38)~\cite{Vidal-Madjar2003} (39)~\cite{Alonso-Floriano2019} (40)~\cite{Deming2013} (41)~\cite{Giacobbe2021} (42)~\cite{Line2016} (43)~\cite{Snellen2010} (44)~\cite{Charbonneau2002} (45)~\cite{Sing2008} (46)~\cite{Casasayas-Barris2021a} (47)~\cite{Cubillos2020} (48)~\cite{Vidal-Madjar2013} (48)~\cite{Astudillo-Defru2013} (49)~\cite{Spake2020} 
        (50)~\cite{Chen2018} 
        (51)~\cite{Sing2012} (52)~\cite{Wakeford2013} (53)~\cite{Nikolov2014} (54)~\cite{Chen2020} (55)~\cite{Tsiaras2018} (56)~\cite{Chen2017} 
        (57)~\cite{Yip2021} (58)~\cite{Nikolov2018} (59)~\cite{Salz2018} (60)~\cite{Guilluy2020} (61)~\cite{McCullough2014} (62)~\cite{Birkby2013} (63)~\cite{Brogi2016} (64)~\cite{Boucher2021}
        (65)~\cite{deKok2013} (66)~\cite{Rodler2013} (67)~\cite{Huitson2012} (68)~\cite{Wyttenbach2015} (69)~\cite{Khalafinejad2017} (70)~\cite{Keles2019} (71)~\cite{Wakeford2017b} (72)~\cite{Nikolov2016} (73)~\cite{Nikolov2015} (74)~\cite{Carter2020} (75)~\cite{Nortmann2018} (76)~\cite{Khalafinejad2021} (77)~\cite{Casasayas-Barris2017} (78)~\cite{Line2013} (79)~\cite{Deibert2019} (80)~\cite{Deibert2021} (81)~\cite{Paragas2021} (82)~\cite{Mansfield2018} (83)~\cite{Allart2018} (84)~\cite{Fraine2014} (85)~\cite{Spake2018} (86)~\cite{Allart2018} (87)~\cite{Kreidberg2018} (88)~\cite{Lockwood2014} (89)~\cite{Pelletier2021} (90)~\cite{Brogi2012} (91)~\cite{Rodler2012} (92)~\cite{Brogi2014} 
        (93)~\cite{Webb2020} (94)~\cite{Guilluy2019} (95)~\cite{Schwarz2016} (96)~\cite{Chilcote2017} (97)~\cite{GRAVITYCollaboration2020} (98)~\cite{Hoeijmakers2018} (99)~\cite{Zhang2021} (100)~\cite{Wang2020} (101)~\cite{Wang2018} (102)~\cite{Konopacky2013} (103)~\cite{Lavie2017} (104)~\cite{Barman2015} 
        (105)~\cite{PetitditdelaRoche2018} (106)~\cite{Macintosh2015} (107)~\cite{Samland2017} (108)~\cite{Yan2020} (109)~\cite{Stangret2021} (110)~\cite{Prinoth2022} (111)~\cite{vanSluijs2022} (112)~\cite{Birkby2017} (113)~\cite{Brogi2013} (114)~\cite{Hawker2018} (115)~\cite{Cont2022} 
        (116)~\cite{Fu2022} (117)~\cite{Czesla2022} (118)~\cite{Damiano2017} (119)~\cite{Benneke2019} (120)~\cite{Palle2020}
        }
        \end{table}
\end{document}